\documentclass[twoside,11pt]{article}

\usepackage{jmlr2e}
\usepackage{tikz}
\usetikzlibrary{fit,positioning}
\usepackage{amsmath}
\usepackage{graphicx} 
\usepackage{subfigure}
\usepackage{enumerate}
\usepackage{bm}

\usepackage{algorithmic}
\usepackage[lined,boxed]{algorithm2e}

\jmlrheading{1}{2000}{1-48}{4/00}{10/00}{Kai Fan, Allison E. Aiello and Katherine A. Heller}

\ShortHeadings{Learning Heterogeneous Personalized Health Data}{Fan, Aiello and Heller}
\firstpageno{1}

\begin{document}

\title{Bayesian Models for Heterogeneous Personalized Health Data}

\author{\name Kai Fan \email kai.fan@stat.duke.edu \\
       \addr Computational Biology and Bioinformatics\\
       Duke University\\
       Durham, NC 27705, USA
       \AND
       \name Allison E.\ Aiello \email aaiello@unc.edu \\
       \addr Gillings School of Global Public Health\\
       University of North Carolina-Chapel Hill\\ 
       Chapel Hill, NC 27599, USA
       \AND
       \name Katherine A.\ Heller \email kheller@stat.duke.edu \\
       \addr Department of Statistical Science\\
       Duke University\\
       Durham, NC 27705, USA}

\editor{}

\maketitle

\begin{abstract}
The purpose of this study is to leverage modern technology (such as mobile or web apps in \cite{beckman2014isis}) to enrich epidemiology data and infer the transmission of disease. Homogeneity related research on population level has been intensively studied in previous work. In contrast, we develop hierarchical Graph-Coupled Hidden Markov Models (hGCHMMs) to simultaneously track the spread of infection in a small cell phone community and capture person-specific infection parameters by leveraging a link prior that incorporates additional covariates. We also reexamine the model evolution of the hGCHMM from simple HMMs and LDA, elucidating additional flexibility and interpretability. Due to the non-conjugacy of sparsely coupled HMMs, we design a new approximate distribution, allowing our approach to be more applicable to other application areas. Additionally, we investigate two common link functions, the beta-exponential prior and sigmoid function, both of which allow the development of a principled Bayesian hierarchical framework for disease transmission. The results of our model allow us to predict the probability of infection for each person on each day, and also to infer personal physical vulnerability and the relevant association with covariates. We demonstrate our approach experimentally on both simulation data and real epidemiological records.
\end{abstract}


\section{Introduction}
Disease diffusion modeling is an important topic in medical informatics \cite{mould2012models}. Recently, much emphasis has been placed on personalized medical treatment and advice, encouraging prediction of disease on an individual level. A notable example is Apple Inc., who recently released the apple watch and built-in Health apps. However, the majority of research on detecting patterns and discovering the risks of infectious disease has been at the population level, such as Google flu trend prediction. This paper aims to incorporate sensor and app collected data from mobile devices for individualized prediction of infectious disease risk and transmission. The data may contain, or allow us to learn, information including personal health condition, physical location and social contacts. 

From dynamic social network information, recorded via a smartphone, we can capture transmission paths at the individual level. The use of social networks provides a framework for identifying interactions and transmission and has been used to populate various complex systematic models and personalized risk models. However, the preponderance of available social network data relies primarily on reported network connections, resulting in a missing data problem and reducing the robustness of inferences that can be made. In this study, we sought to overcome these problems by utilizing a novel cell phone bluetooth network contact app to infer dynamic social network interactions, infection probability and transmission. Location and time based information from this app allow us to track personal daily contacts between participants. Proximity can be measured by the contact duration within a certain range. Similar social experiments have been conducted and mentioned in \cite{beckman2014isis,dong2012graph,dong2011modeling}, but these prior models assumed homogeneous individuals or global parameter sharing within the networks, and did not include data on potential modifying factors, such as personal health habits and demographic features of individuals in the network.   

\subsection{Related Works}
Hidden Markov models are widely used in simulating the state space based disease progression \cite{ohlsson2001using,jackson2003multistate,sukkar2012disease,dong2012graph}, since HMMs actually mimic the behavior of an SIS model with discretized timestamps and are convenient in terms of the development of inference algorithm for different variants. In our work, we also adopt this basic framework to design our model. Other work like \cite{zhou2012modeling, zhou2013modeling} proposed a group lasso formulation or multi-task learning framework using optimization, and is far from our domain. A recent work by \cite{wang2014unsupervised} learned a continuous time series model looking at trajectories for different patients, but ignoring interactions between chains. Our work attempts to unify social network modeling of individuals using a hierarchical structure. \cite{christley2005infection} utilized a fixed social network analysis on susceptible-infectious-recovered (SIR) models to identify high-risk individuals. \cite{salathe2010high}'s work on close proximity interactions (CPIs) of dynamic social networks at a high school indicated immunization strategies are more credible if extra contact data is provided. 

The idea of using hierarchies to improve model flexibility is extensively studied in topic modeling, in models such as latent Dirichlet allocation (LDA)  \cite{chang2010hierarchical,blei2010nested,paisley2012nested}. \cite{chang2010hierarchical} used a sigmoid link function, introduced in Relational Topic Model to learn fixed networks of documents. These, and further works have exemplified a trend in data-driven machine learning applications -- hierarchical modeling is used to make inferences when the data structure is complex. Our work can be considered as a hierarchical extension of either GCHMMs \cite{dong2012graph} or topic HMMs \cite{gruber2007hidden} with a nested transition function. From this viewpoint, we can also interpret our graphical model in terms of hierarchical LDA modeling.

\subsection{Notation}

\begin{table}[t]
\caption{Notations}
\begin{tabular}{cl} \hline
\textbf{} & \textbf{Description} \\ \hline
$n\in[N]$ & index for participants  \\ \hline
$t\in[T]$ & index for timestamp \\ \hline
$s\in[S]$ & index for observed symptoms\\ \hline
$\mathbf{z}_{n}$ & covariates indicating personal features\\ \hline
$G_{t-1}$ & dynamic social networks between $t-1$ and $t$\\ \hline
$\gamma(\gamma_n)$ & recovery probability if infectious at previous timestamp  \\ \hline
$\alpha(\alpha_n)$ &  probability of being infected from some one outside networks \\ \hline
$\beta(\beta_n)$ &  probability of being infected from some one inside networks \\ \hline
$\pi$ & infection probability of initial hidden state $x_{n,1}$ \\ \hline
$x_{n,t} \in \mathcal{X}=\{0,1\}$ & latent variable indicates whether infected \\ \hline
$y_{n,t} \in \mathcal{Y}=\{0,1\}^S$ & reported symptoms during each time interval \\ \hline
$\theta_{x_{n,t},s}$ & emission probability of symptom $s$ onset given $x_{n,t}$  \\ 
\hline
\end{tabular}
\label{tab:notation}
\end{table}

Let $N$ be the number of participants in the social community, and $T$ be the days being tracked. The health record for each participant can be simulated as an HMM with $T$ timestamps. Let $\mathcal{Y}$ be the observation space of a Markov chain with hidden state space $\mathcal{X}$, and initial probabilities $\pi$. We also refer to the infection rate related parameters as $\gamma$, $\alpha$ and $\beta$. In particular, $\gamma$ gives the probability that a previously-infectious individual recovers and again becomes susceptible. $\alpha$ represents the probability that an infectious person from outside the community infects a previously-susceptible person within the community. $\beta$ represents the probability that an infectious person from the community infects a previously-susceptible person. These parameters are used to construct the transition probability of HMMs, whereas the emission probability $\theta_X$ merely depends on the hidden state.  

Additionally, we introduce the notation associated with the mobile app survey or sensor logs. $S$ is the number of symptoms in self report record. Temporal features in our survey are denoted by $z$. Since our discussion of HMMs varies between parameter sharing and the inhomogeneous setting, we temporarily did not include any subscripts to avoid ambiguity. Clarification will be given in the subsequent sections. However, the overall notation description is concisely summarized in Table \ref{tab:notation}. 

\subsection{Network Data Imputation}
Our experiments are mainly conducted using both actively reported symptoms via mobile apps and passively detected location through bluetooth sensors. Occasionally, professional flu diagnoses will be given if symptoms imply a high risk of infection. Contact between any two agents can be inferred by the sensor collected location information. However, various uncertainties may cause the electronic data to be missing, such as dead batteries or a weak signal. This necessitates data preprocessing to recover the dynamic social network $G_t$, which will be used for the next stage of our analysis. Relational modeling is not our current focus, so we use a nonparametric Bayesian dynamic network learning algorithm from \cite{durante2014nonparametric}.

Given $G_t=(g_{ij}(t))_{N\times N}$ being a symmetric binary matrix to represent the dynamic social networks at timestamp $t$, let $\Xi(t)$ be the indicated symmetric probability matrix, i.e. 
\begin{align*}
\Pr(g_{ij}(t)=1)=\xi_{ij}(t), \text{ or equivalently} \quad g_{ij}(t)\sim\text{Bernoulli}(\xi_{ij}(t)) .
\end{align*}
In \cite{durante2014nonparametric}, they applied a $[0,1]$ scaled transformation on a deliberately constructed latent space to obtain the time varying matrix $\Xi(t)$. A link function $l:\mathbb{R}\rightarrow[0,1]$ or a prior can be imposed on the latent space. For simplicity, the commonly used sigmoid link function is introduced.
\begin{align*}
\xi_{ij}(t) = \sigma\left(s_{ij}(t)\right) .
\end{align*}
where $\sigma(x)=(1+e^{-x})^{-1}$. In order to reduce the problem complexity and avoid modeling $\frac{N(N-1)}{2}$ stochastic processes, the dynamic latent space can be expressed as 
\begin{align*}
s_{ij}(t)=\mu(t)+\mathbf{h}_i(t)^\top \mathbf{h}_j(t),
\end{align*} 
where $\mu(t)$ is a scalar indicating the baseline relation intensity, and $\mathbf{h}_i(t)=(h_{i1},h_{i2},\dots,h_{iH})'$ is a vector in latent space with dimensionality $H$. This design also allows borrowing information to exploit the underlying process inducing similarities among the units. The prior imposed on the latent space variable is Gaussian process to capture the time dependency. 
\begin{align*}
h_{ih}(\cdot)\sim\mathcal{GP}(0,\tau_h^{-1}c_H),\tau_h=\prod_{k=1}^h \theta_k, \theta_1\sim\text{Gamma}(a_1,1),\theta_k\sim\text{Gamma}(a_2,1), k\geq2
\end{align*}
where $c_H$ is a squared exponential (SE) correlation function $c_H(t,t')=\exp\left(-\kappa_H\|t-t'\|^2/2\right)$ with parameter $\kappa_H$, and the prior of $\tau_h$ has a shrinkage effect. Similarly, the prior of baseline variables is $\mu(\cdot)\sim\mathcal{GP}(0,c_\mu)$ with SE function $c_\mu$.

This model allows missing data to appear in the observed $G_t$, since it can be easily sampled from the posterior of $\Xi_t$. The latent space sampling greatly depends on a Polya-Gamma distribution, thus allowing the posterior computation to be performed with fully Bayesian inference. In this paper, we merely describe the generative process of their model, whereas the inference method involving Gibbs sampling using a data-augmentation trick \cite{polson2013bayesian} is beyond our scope and will be omitted. We refer the interested readers to these previous publications for details on the methodology.

\subsection{Contributions of the Paper}

We review the homogenous Graph-coupled HMM based flu infection model within both the frameworks of Bayesian statistics and factor graph theory, and analyze various connections, for example, to the prominent work of \cite{dong2012graph}. We also extend this model based on covariates, introducing associated parameter heterogeneity, and apply it to personalized health data given via wearable equipment. To the best of our knowledge, both the model extension and the application have never been investigated in previous work. We also provide a description of the relationship to topic modeling, showing that our model can be derived from standard LDA or HMMs. Using two available real-world datasets, we discover several reasonable and interesting phenomena from our experimental evaluation, that lead to intuitive interpretations of our approach. The practical results offer new possibilities for applications in future personal health research. Specifically, this paper
\begin{enumerate}[(I)]
\item designs a more accurate approximation for the distribution of auxiliary variables to track infection source, especially when the values of $\alpha,\beta,\gamma$ violate the near-zero assumption that only holds in the domain of epidemiology, allowing our model to potentially fit other application areas. 
\item generalizes the standard Baum-Welch algorithm to GCHMMs under the constraint of sparse networks. The sparsity assumption imposed on daily social networks is reasonable in any real-world setting.
\item characterizes person-specific infection parameters by imposing a covariate dependent hierarchical structure on GCHMMs. The link function is associated with \textit{covariates} relates to temporal personal features, such as gender, weight, hygiene habits, and diet habits. We also adopt the hypothesis that better personal habits should result in lower susceptibility to influenza.
\item explores both deterministic and probabilistic link functions in our model. In particular, the sigmoid link function and beta-exponential prior can be generalized to the softmax and Dirichlet-exponential respectively.
\item develops an efficient parameter estimation algorithm for the non-conjugate prior. Inspired by \cite{delyon1999convergence}, our proposed solver performs Gibbs sampling and optimization iteratively. A faster version of our EM-like algorithm for binary hidden variables is primarily used for our experimental tests, which significantly accelerates the computational speed without a significant impact on accuracy.
\end{enumerate} 

\subsection{Organization of the Rest of the Paper}

In Section \ref{sec:gchmm}, we describe the basic idea of GCHMMs and discussion the Gibbs sampling and message passing methods used, where a connection and comparison will be presented through an illustrative example. Meanwhile, a new decomposition trick for the auxiliary variable distribution is visually compared with previous work \cite{dong2012graph}. In Section \ref{sec:hgchmm}, we test our proposed hGCHMMs by extending the GCHMMs to be both heterogeneous and hierarchical, explicitly discussing the two different link schemes and relevant parameter interpretation. A new model evolution from LDA to hGCHMMs is also depicted in this context. In Section \ref{sec:inference}, we focus on how to modify the EM algorithm to a Gibbs sampling version when the expectation is intractable with respect to hidden or latent variable. In Section \ref{sec:experiment}, we report empirical results, applying it to semi-synthetic and real-world datasets. Conclusions and future research directions are discussed in Section \ref{sec:conclusion}.

\section{Graph-coupled Hidden Markov Model}
\label{sec:gchmm}

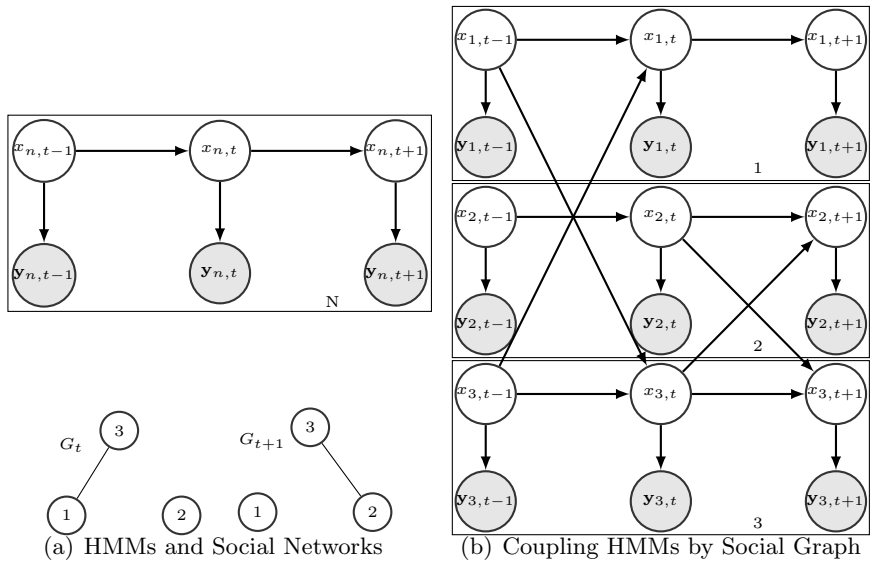
\begin{figure}
\tiny
\centering
\subfigure[HMMs and Social Networks]{
\begin{tikzpicture}
\tikzstyle{main}=[circle, minimum size = 8mm, thick, draw =black!80, node distance = 15mm,inner sep=0pt]
\tikzstyle{connect}=[-latex, thick]
\tikzstyle{box}=[rectangle, draw=black!100]
  \node[main, fill = white!100] (X1t0) {$x_{n,t-1}$};
  \node[main] (X1t) [right=of X1t0] {$x_{n,t}$};
  \node[main] (X1t1) [right=of X1t] {$x_{n,t+1}$};
  \node[main, fill = black!10] (Y1t0) [below=of X1t0,yshift=7mm] {$\mathbf{y}_{n,t-1}$};
  \node[main, fill = black!10] (Y1t) [below=of X1t,yshift=7mm] {$\mathbf{y}_{n,t}$};
  \node[main, fill = black!10] (Y1t1) [below=of X1t1,yshift=7mm] {$\mathbf{y}_{n,t+1}$};
  \node[main, minimum size=5mm] (1) [below=of Y1t0, xshift=3mm, yshift=-10mm] {1};
  \node[main, minimum size=5mm] (2) [right=of 1, xshift=-5mm] {2};
  \node[main, minimum size=5mm] (3) [above=of 1, xshift=7mm, yshift=-9mm] {3};
  \node[main, minimum size=5mm] (11) [below=of Y1t, xshift=5mm, yshift=-10mm] {1};
  \node[main, minimum size=5mm] (22) [right=of 11, xshift=-5mm] {2};
  \node[main, minimum size=5mm] (33) [above=of 11, xshift=7mm, yshift=-9mm] {3};
  
  \path (X1t0) edge [connect] (X1t)
        (X1t) edge [connect] (X1t1)
        (X1t0) edge [connect] (Y1t0)
        (X1t) edge [connect] (Y1t)
        (X1t1) edge [connect] (Y1t1)
        (1) edge (3)
        (22) edge (33);

\node[rectangle, inner sep=0.5mm,draw=black!100, fit= (X1t0) (X1t) (X1t1) (Y1t0) (Y1t) (Y1t1)] {};
\node[rectangle, inner sep=0.5mm, fit= (X1t0) (X1t) (X1t1) (Y1t0) (Y1t) (Y1t1),label=below right:N, yshift=3mm] {};
\node[rectangle, inner sep=0mm, fit= (1) (2) (3),xshift=-7mm,yshift=3mm] {$G_t$};
\node[rectangle, inner sep=0mm, fit= (11) (22) (33),xshift=-7mm,yshift=3mm] {$G_{t+1}$};
\end{tikzpicture}
}
\subfigure[Coupling HMMs by Social Graph]{
\begin{tikzpicture}
\tikzstyle{main}=[circle, minimum size = 8mm, thick, draw =black!80, node distance = 15mm,inner sep=0pt]
\tikzstyle{connect}=[-latex, thick]
\tikzstyle{box}=[rectangle, draw=black!100]
  \node[main, fill = white!100] (X1t0) {$x_{1,t-1}$};
  \node[main] (X1t) [right=of X1t0] {$x_{1,t}$};
  \node[main] (X1t1) [right=of X1t] {$x_{1,t+1}$};
  \node[main, fill = black!10] (Y1t0) [below=of X1t0,yshift=9mm] {$\mathbf{y}_{1,t-1}$};
  \node[main, fill = black!10] (Y1t) [below=of X1t,yshift=9mm] {$\mathbf{y}_{1,t}$};
  \node[main, fill = black!10] (Y1t1) [below=of X1t1,yshift=9mm] {$\mathbf{y}_{1,t+1}$};
  \node[main] (X2t0) [below=of Y1t0,yshift=14mm] {$x_{2,t-1}$};
  \node[main] (X2t) [below=of Y1t,yshift=14mm] {$x_{2,t}$};
  \node[main] (X2t1) [below=of Y1t1,yshift=14mm] {$x_{2,t+1}$};
  \node[main, fill = black!10] (Y2t0) [below=of X2t0,yshift=9mm] {$\mathbf{y}_{2,t-1}$};
  \node[main, fill = black!10] (Y2t) [below=of X2t,yshift=9mm] {$\mathbf{y}_{2,t}$};
  \node[main, fill = black!10] (Y2t1) [below=of X2t1,yshift=9mm] {$\mathbf{y}_{2,t+1}$};
  \node[main] (X3t0) [below=of Y2t0,yshift=14mm] {$x_{3,t-1}$};
  \node[main] (X3t) [below=of Y2t,yshift=14mm] {$x_{3,t}$};
  \node[main] (X3t1) [below=of Y2t1,yshift=14mm] {$x_{3,t+1}$};
  \node[main, fill = black!10] (Y3t0) [below=of X3t0,yshift=9mm] {$\mathbf{y}_{3,t-1}$};
  \node[main, fill = black!10] (Y3t) [below=of X3t,yshift=9mm] {$\mathbf{y}_{3,t}$};
  \node[main, fill = black!10] (Y3t1) [below=of X3t1,yshift=9mm] {$\mathbf{y}_{3,t+1}$};
  
  \path (X1t0) edge [connect] (X1t)
        (X1t) edge [connect] (X1t1)
        (X1t0) edge [connect] (Y1t0)
        (X1t) edge [connect] (Y1t)
        (X1t1) edge [connect] (Y1t1)
        (X2t0) edge [connect] (X2t)
        (X2t) edge [connect] (X2t1)
        (X2t0) edge [connect] (Y2t0)
        (X2t) edge [connect] (Y2t)
        (X2t1) edge [connect] (Y2t1)
        (X3t0) edge [connect] (X3t)
        (X3t) edge [connect] (X3t1)
        (X3t0) edge [connect] (Y3t0)
        (X3t) edge [connect] (Y3t)
        (X3t1) edge [connect] (Y3t1)
        (X1t0) edge [connect] (X3t)
        (X3t0) edge [connect] (X1t)
        (X2t) edge [connect] (X3t1)
        (X3t) edge [connect] (X2t1);

\node[rectangle, inner sep=0.3mm,draw=black!100, fit= (X1t0) (X1t) (X1t1) (Y1t0) (Y1t) (Y1t1)] {};
\node[rectangle, inner sep=0mm, fit= (X1t0) (X1t) (X1t1) (Y1t0) (Y1t) (Y1t1),label=below right:1, yshift=3mm] {};
\node[rectangle, inner sep=0.3mm,draw=black!100, fit= (X2t0) (X2t) (X2t1) (Y2t0) (Y2t) (Y2t1)] {};
\node[rectangle, inner sep=0mm, fit= (X2t0) (X2t) (X2t1) (Y2t0) (Y2t) (Y2t1),label=below right:2, yshift=3mm] {};
\node[rectangle, inner sep=0.3mm,draw=black!100, fit= (X3t0) (X3t) (X3t1) (Y3t0) (Y3t) (Y3t1)] {};
\node[rectangle, inner sep=0mm, fit= (X3t0) (X3t) (X3t1) (Y3t0) (Y3t) (Y3t1),label=below right:3, yshift=3mm] {};
\end{tikzpicture}
}
\caption{Left, top, an HMM graphical model. Left, bottom, a dynamic social network, Right, illustration of the formation of GCHMMs involving 3 people. 1 and 3 have social contact between $t-1$ and $t$. The infection states of 1 and 3 at $t$ are then both influenced by each others' infection states at time $t-1$.}
\label{fig:gchmm}
\end{figure}

We first briefly introduce the graph-coupled hidden Markov model (GCHMM), evolving from coupled hidden Markov model (CHMM) \cite{brand1997coupled}, a dynamic representation for analyzing the discrete-time series data by considering the interactions between Markov chains (see Figure \ref{fig:gchmm} for an example, where filled nodes are observed). The standard CHMM is typically fully connected, between hidden nodes at successive timestamps, whereas the intrinsic sparsity of a dynamic social network can couple multiple HMMs with the possibility for fast inference. The number of parameters needed to be inferred will decrease dramatically from $\mathcal{O}(N^N)$ to $\mathcal{O}(N^{n_{\max}})$ where $n_{\max}$ is the maximum degree of hidden nodes. This advantage will further benefit our message passing algorithm and hGCHMMs in the subsequent section.

\subsection{Generative Modeling}

Let $G_t=(N,E_t)$ be a network structure snapshot between timestamps $t-1$ and $t$, where each agent or participant is represented by a node $n\in \mathcal{N}=[N]$\footnote{[N] means a set including integers from 1 to $N$.} in graph $G_t$, and $E_t$ is a set of undirected edges in $G_t$, where unordered pair $(n_i,n_j)\in E_t$ if two participants $n_i$ and $n_j$ have a valid contact during the time interval $(t-1,t]$. The bottom in Figure \ref{fig:gchmm}(a) illustrates an example of the dynamic social networks. Assuming that each participant $n$ is represented a HMM  with binary hidden state $x_{n,t}$ shown as the top in Figure \ref{fig:gchmm}(a), where 0 and 1 indicate susceptible and infectious respectively. The observed node $\mathbf{y}_{n,t}$ is an $S$ dimensional binary vector $(y_{n,t,1},y_{n,t,2},\dots,y_{n,t,S})$ as an indicator for $S$ symptoms. Thus, the generative model of GCHMMs is given in a fully bayesian way.
\begin{equation}\label{eq:generative}
\begin{split}
\pi&\sim\text{Beta}(a_\pi,b_\pi) \\
\alpha\sim\text{Beta}(a_\alpha,b_\alpha) \quad
\beta&\sim\text{Beta}(a_\beta,b_\beta) \quad  \gamma\sim\text{Beta}(a_\gamma,b_\gamma)\\
\theta_{0,s}&\sim\text{Beta}(a_0,b_0) \quad \theta_{1,s}\sim\text{Beta}(a_1,b_1)\\
x_{n,0}&\sim\text{Bernoulli}(\pi) \\
x_{n,t} &\sim\text{Bernoulli}\left(\phi_{n,x_{n'}:(n,n')\in G_t}(\alpha,\beta,\gamma)\right)\\
y_{n,t,s}&\sim\text{Bernoulli}(\theta_{x_{n,t},s})
\end{split}
\end{equation}
where the transition probability $\phi_{n,x_{n'}:(n,n')\in G_t}(\alpha,\beta,\gamma)$ is a function of the infection parameters and the dynamic graph structure. This homogenous setting means all HMMs share the same parameters set or similar transition function. The difference is reflected as Figure \ref{fig:gchmm} indicated, the transition of the hidden state is not only dependent on the previous state of its own Markov chain but also may be influenced by states from other HMMs that have edges connected to it. One undirected edge in $G_t$ indicates a valid contact in time interval $(t-1,t]$, thus leading to a directed edge in GCHMMs. Recalling the definition in terms of $\gamma,\alpha,\beta$ (See Table \ref{tab:notation}), it is natural to construct the transition probability function as follows:
\begin{equation}\label{eq:trans}
\begin{split}
\phi_{n,x_{n'}:(n,n')\in G_t}(\gamma,\alpha,\beta)=
\left\{ \begin{matrix} 
      \gamma &  x_{n,t}=1,x_{n,t+1}=0; \\
      1-\gamma &  x_{n,t}=1,x_{n,t+1}=1; \\
      1-(1-\alpha)(1-\beta)^{C_{n,t}} & x_{n,t}=0,x_{n,t+1}=1;\\
      (1-\alpha)(1-\beta)^{C_{n,t}} & x_{n,t}=0,x_{n,t+1}=0.
   \end{matrix}\right.
\end{split}
\end{equation} 
where $\mathbb{I}_{\{\}}$ is the indicator function and $C_{n,t}=\sum_{n':(n',n)\in E_t}\mathbb{I}_{\{x_{n',t}=1\}}$ is the count of possible infectious sources for node $n$ in $G_t$, in other words, it means that besides participant $n$, the number of other infectious nodes that have social contacts with $n$ at the previous day. The epidemiological intuition is very simple, the more infectious people one comes in contact with, the more probable one is to be contaminated. This Bayesian formulation of GCHMMs can be applied to fit homogenous susceptible-infectious-susceptible (SIS) epidemic dynamics. Additionally, the Bayesian inference of this model (e.g. \cite{dong2012graph}) can be reduced to a special case from our heterogenous model by getting rid of link hierarchy and personalization of infection parameters. We do not go into details here, but we generalize the Baum-Welch algorithm to sparse-coupled HMMs and relax the assumption of near-zero parameters, i.e. $\alpha$, $\beta$ and $\gamma$ are all $\approx 0$.

\subsection{Generalized Baum-Welch Algorithm}

HMMs usually model independent sequenced or discrete time-series data, and represent long-range dependencies between observations for each data point, mediated via latent variables. We are inspired by the fact (\cite{loeliger2004introduction}) that the forward-backward, Viterbi and EM for HMMs (see \cite{murphy2012machine}'s review) learning algorithms have their equivalent message passing formulation in factor graph representation. The independency property allows the three algorithms to be efficient and effective, while the blownup parameter space of standard coupled HMMs makes analogous treatment impossible and impractical. Thus different assumptions have been imposed on CHMM, such as \cite{zhong2001new}. Fortunately, in our sparse coupling setting, a Generalized Baum-Welch Algorithm can be developed via a similar factor graph representation and enables efficient forward-backward, Viterbi and approximate EM algorithms.

\begin{figure}[t]
\tiny
\centering
\begin{tikzpicture}
\tikzstyle{main}=[circle, minimum size = 8mm, thick, draw =black!80, node distance = 15mm,inner sep=0pt]
\tikzstyle{connect}=[-latex, thick]
\tikzstyle{box}=[rectangle, draw=black!100]
  \node[main, fill = white!100] (X1t0) {$x_{1,t-1}$};
  \node[rectangle,draw=black!100] (p101) [right=of X1t0, xshift=10mm] {$\phi_{1,t-1,t}$};
  \node[main] (X1t) [right=of p101] {$x_{1,t}$};
  \node[rectangle,draw=black!100] (p111) [right=of X1t, xshift=10mm] {$\phi_{1,t,t+1}$};
  \node[main] (X1t1) [right=of p111] {$x_{1,t+1}$};
  \node[rectangle, fill = black!10] (Y1t0) [below=of X1t0,yshift=3mm] {$\phi_{1,t-1,y|x}$};
  \node[rectangle, fill = black!10] (Y1t) [below=of X1t,yshift=3mm] {$\phi_{1,t,y|x}$};
  \node[rectangle, fill = black!10] (Y1t1) [below=of X1t1,yshift=3mm] {$\phi_{1,t+1,y|x}$};
  \node[main] (X2t0) [below=of Y1t0,yshift=10mm] {$x_{2,t-1}$};
  \node[rectangle,draw=black!100] (p201) [right=of X2t0, xshift=10mm] {$\phi_{2,t-1,t}$};
  \node[main] (X2t) [right=of p201] {$x_{2,t}$};
  \node[rectangle,draw=black!100] (p211) [right=of X2t, xshift=10mm] {$\phi_{2,t,t+1}$};
  \node[main] (X2t1) [right=of p211] {$x_{2,t+1}$};
  \node[rectangle, fill = black!10] (Y2t0) [below=of X2t0,yshift=3mm] {$\phi_{2,t-1,y|x}$};
  \node[rectangle, fill = black!10] (Y2t) [below=of X2t,yshift=3mm] {$\phi_{2,t,y|x}$};
  \node[rectangle, fill = black!10] (Y2t1) [below=of X2t1,yshift=3mm] {$\phi_{2,t+1,y|x}$};
  \node[main] (X3t0) [below=of Y2t0,yshift=10mm] {$x_{3,t-1}$};
  \node[rectangle,draw=black!100] (p301) [right=of X3t0, xshift=10mm] {$\phi_{3,t-1,t}$};
  \node[main] (X3t) [right=of p301] {$x_{3,t}$};
  \node[rectangle,draw=black!100] (p311) [right=of X3t, xshift=10mm] {$\phi_{3,t,t+1}$};
  \node[main] (X3t1) [right=of p311] {$x_{3,t+1}$};
  \node[rectangle, fill = black!10] (Y3t0) [below=of X3t0,yshift=3mm] {$\phi_{3,t-1,y|x}$};
  \node[rectangle, fill = black!10] (Y3t) [below=of X3t,yshift=3mm] {$\phi_{3,t,y|x}$};
  \node[rectangle, fill = black!10] (Y3t1) [below=of X3t1,yshift=3mm] {$\phi_{3,t+1,y|x}$};
  
  \path (X1t0) edge (p101)
        (p101) edge (X1t)
        (X1t) edge (p111)
        (p111) edge (X1t1)
        (X1t0) edge (Y1t0)
        (X1t) edge (Y1t)
        (X1t1) edge (Y1t1)
        (X2t0) edge (p201)
        (p201) edge (X2t)
        (X2t) edge (p211)
        (p211) edge (X2t1)
        (X2t0) edge (Y2t0)
        (X2t) edge (Y2t)
        (X2t1) edge (Y2t1)
        (X3t0) edge (p301)
        (p301) edge (X3t)
        (X3t) edge (p311)
        (p311) edge (X3t1)
        (X3t0) edge (Y3t0)
        (X3t) edge (Y3t)
        (X3t1) edge (Y3t1)
        (X1t0) edge (p301)
        (X3t0) edge (p101)
        (X2t) edge (p311)
        (X3t) edge (p211);

  \node[rectangle, inner sep=2mm,draw=black!100, fit= (X1t0) (X1t) (X1t1) (Y1t0) (Y1t) (Y1t1)] {};
  \node[rectangle, inner sep=0mm, fit= (X1t0) (X1t) (X1t1) (Y1t0) (Y1t) (Y1t1),label=below right:1, xshift=15mm, yshift=3mm] {};
  \node[rectangle, inner sep=2mm,draw=black!100, fit= (X2t0) (X2t) (X2t1) (Y2t0) (Y2t) (Y2t1)] {};
  \node[rectangle, inner sep=0mm, fit= (X2t0) (X2t) (X2t1) (Y2t0) (Y2t) (Y2t1),label=below right:2, xshift=15mm, yshift=3mm] {};
  \node[rectangle, inner sep=2mm,draw=black!100, fit= (X3t0) (X3t) (X3t1) (Y3t0) (Y3t) (Y3t1)] {};
  \node[rectangle, inner sep=0mm, fit= (X3t0) (X3t) (X3t1) (Y3t0) (Y3t) (Y3t1),label=below right:3, xshift=15mm, yshift=3mm] {};
  
\end{tikzpicture}
\caption{Factor Graph of Figure \ref{fig:gchmm}(b)}
\label{fig:fgchmm}
\end{figure}
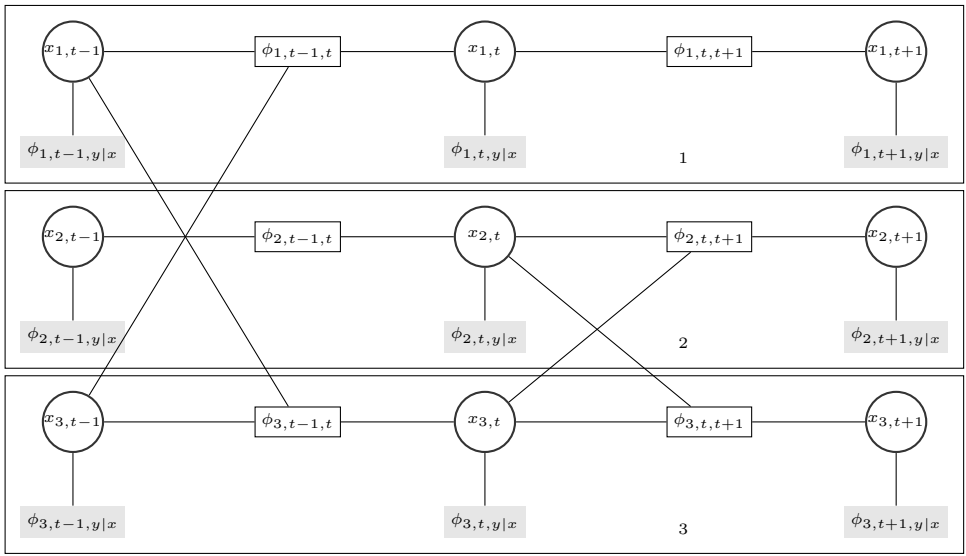

First, we describe the factor graph in Figure \ref{fig:fgchmm} derived from the Bayesian network in Figure \ref{fig:gchmm}(b). Commonly, factor graphs contain two types of nodes, variable nodes and factor nodes. In our model, the factor nodes can be categorized as emission probability functions and transition probability functions,
\begin{align}
&\phi_{n,t,y|x,s}(y_{n,t,s}|x_{n,t})=p(y_{n,t,s}|x_{n,t})=\theta_{x_{n,t},s}^{y_{n,t,s}}(1-\theta_{x_{n,t},s})^{y_{n,t,s}}\\
&\phi_{n,t-1,t}(x_{n,t-1},x_{n',t-1:(n',n)\in E_t},x_{n,t})=p(x_{n,t}|x_{n,t-1},x_{n',t-1:(n',n)\in E_t})
\end{align}
where the second equation defines the same function (\ref{eq:trans}). Notice that the factor graph is undirected and direction information is comprised in the factor node. The factor graph is still shown with plates; each plate represents a participant while the interaction is captured by transition factor. In fact, during the running of the EM algorithm, all parameters are unknown, so $\alpha,\beta,\gamma$  can go at the top of the current graph, adding edges to all transition factors. For simplicity, we did not include them in our factor graph, though this widely used trick is introduced in independent HMMs (\cite{loeliger2004introduction}).

\subsection{Forward-backward Algorithm}

Notice that even if the Bayesian network is not exactly a directed polytree (cyclic paths exist if omitting edge direction), marginal inference on each hidden node can still be approximated with belief propagation on the factor graph for the sake of efficiency. In this section, we will derive the single node belief propagation rules (\cite{pearl2014probabilistic}) on Figure \ref{fig:fgchmm}. Denote the message passing to the child, i.e. from $x_{m,t-1}$ to $x_{n,t}$ as $\pi_{x_{m,t-1}}(x_{n,t})$, and the message to parent, i.e. from $x_{k,t+1}$ to $x_{n,t}$ as $\lambda_{x_{k,t+1}}(x_{n,t})$. Then, the belief or probability distribution passed with all evidence shown is denoted as $\text{BEL}(x_{n,t})=P(x_{n,t}|Y)$. Our derivation is based on Pearl's belief propagation algorithm. All $\propto$ in the following imply the term should be normalized to 1 as a valid probability distribution. For notation simplicity, we further denote $\pi_{x_{n,t}}(x_{\cdot,t-1})=p(x_{n,t}|x_{\cdot,t-1})$ and $\lambda_{x_{\cdot,t+1}}(x_{n,t})=p(x_{\cdot,t+1}|x_{n,t})$. The principal part of forward-backward algorithm can be summarized as follows, where the detailed update for $\lambda$ and $\pi$ is derived in Appendix A.
\begin{align}
\pi^{(i)}(x_{n,t})&=\sum_{x_{n,t-1},x_{n',t-1:(n'n)\in E_t}}\phi_{n,t-1,t}\prod_{n\cup\{n':(n',n)\in E_{t-1}\}}\pi_{x_{n,t}}^{(i)}(x_{\cdot,t-1})\\
\lambda^{(i)}(x_{n,t})&=\prod_{s=1}^S\lambda_{y_{n,t,s}}(x_{n,t})\prod_{n\cup\{n':(n,n')\in E_t\}}\lambda_{x_{\cdot,t+1}}^{(i)}(x_{n,t})\\
\text{BEL}^{(i)}(x_{n,t})&\propto \pi^{(i)}(x_{n,t})\lambda^{(i)}(x_{n,t})
\end{align}
Even if it has been proven that belief propagation on standard HMMs is equivalent to the forward-backward algorithm, the update step for message passing in our case is a little more complicated since the single node dependence is generalized to multi-nodes. Let the maximum degree of all $G_t$s be $M$, then the update of sum-product for message can be computed with complexity $\mathcal{O}(2^M)$ at each iteration. This is the reason why the sparsity assumption is required in our algorithm. In addition, the initialization for all messages from variable nodes to factor nodes, such as $\pi_{x_\cdot}(x_\cdot),\lambda_{x_\cdot}(x_\cdot)$, can be set to all 1s. 

\paragraph{Viterbi} algorithm can be derived in a straightforward way, if the sum-product $\sum\prod$ in forward-backward is substitute by max-product $\max\prod$. Since the message updating step is almost the same, it won't be discussed here.

\subsection{Approximate EM Algorithm}

In this section, we will put forward a parameter learning scheme via the generalized Baum-Welch Algorithm. It is straightforward to derive the expected complete data log-likelihood given by
\begin{align}
\hspace{-2cm}
Q(\Theta,\Theta^{old})&=\sum_{X}\sum_{n=1}^N\left\{x_{n,0}\log\pi+(1-x_{n,0})\log(1-\pi)+\sum_{t=1}^T\log\phi_{n,t-1,t}+\sum_{t=1}^T\sum_{s=1}^S\log\phi_{n,t,y|x,s}\right\}\Pr(\mathbf{X}|\mathbf{Y},\Theta^{old})
\end{align}
This is exactly the E-step in EM algorithm, and the non-approximate M-step can be optimized for parameters $\pi$, $\theta_X$ and $\gamma$. In fact, due to the conjugacy of these parameters, their posterior distribution can also be analytically computed. Taking the partial derivative on $Q$, i.e.
\begin{align}
\frac{\partial Q(\Theta,\Theta^{old})}{\partial\xi}=0 \quad \frac{\partial Q(\Theta,\Theta^{old})}{\partial \theta_{0,s}}=0 \quad \frac{\partial Q(\Theta,\Theta^{old})}{\partial \theta_{1,s}}=0 \quad \frac{\partial Q(\Theta,\Theta^{old})}{\partial \gamma}=0
\end{align}
By solving above equations, we obtain the update formula for corresponding parameters.
\begin{align}
\pi&=\frac{\sum_{n=1}^N\mathbb{E}[x_{n,0}]}{N}\\
\theta_{i,s}&=\frac{\sum_{x_{1:N,1:T}\in\{0,1\}^{N^2}}\left(\sum_{n=1}^N\sum_{t=1}^Ty_{n,t,s}\mathbb{I}_{x_{n,t=i}}\right)p(\mathbf{X}|\mathbf{Y},\Theta^{old})}{\sum_{x_{1:N,1:T}\in\{0,1\}^{N^2}}\left(\sum_{n=1}^N\sum_{t=1}^T\mathbb{I}_{x_{n,t=i}}\right)p(\mathbf{X}|\mathbf{Y},\Theta^{old})}, i=0,1\nonumber\\
\gamma&=\frac{\sum_{x_{1:N,1:T}\in\mathcal{X}^{N^2}}\left(\sum_{n=1}^N\sum_{t=2}^T\mathbb{I}_{x_{n,t-1}=1,x_{n,t}=0}\right)p(\mathbf{X}|\mathbf{Y},\Theta^{old})}{\sum_{x_{1:N,1:T}\in\mathcal{X}^{N^2}}\left(\sum_{n=1}^N\sum_{t=2}^T\mathbb{I}_{x_{n,t-1}=1}\right)p(\mathbf{X}|\mathbf{Y},\Theta^{old})} \nonumber
\end{align}

Notice that it does not matter if we change the $p(\mathbf{X}|\mathbf{Y},\Theta^{old})$ to $p(\mathbf{X},\mathbf{Y}|\Theta^{old})$, and $\frac{\partial Q(\Theta,\Theta^{old})}{\partial\alpha}=0$, $\frac{\partial Q(\Theta,\Theta^{old})}{\partial\beta}=0$ can result in an analogous computation as for $\gamma$ for the iteration. However, except for $\pi$, the exact computational complexity of the iteration step for other parameters is intractable, exponentially increasing with $N$ or $N^2$. Since we did not assume near-zero parameters, the induced non-conjugacy requires further approximation in the M-step. If we approximate $P(\mathbf{X}|\mathbf{Y},\Theta^{old})=\prod_{n,t}p(x_{n,t}|\mathbf{Y},\Theta^{old})$ in a fully factorized form, then all the optimized results for $\theta$ would update analytically, because $p(x_{n,t}|\mathbf{Y},\Theta^{old})$ (i.e. $\text{BEL}(x_{n,t})$) can be computed by the \textit{forward-backward} algorithm as derived before.
\begin{align}
\theta_{i,s}&=\frac{\sum_{n=1}^N\sum_{t=1}^Ty_{n,t,s}\mathbb{E}[\mathbb{I}_{i=0}(1-x_{n,t})+\mathbb{I}_{i=1}x_{n,t}]}{\sum_{n=1}^N\sum_{t=1}^T\mathbb{E}[\mathbb{I}_{i=0}(1-x_{n,t})+\mathbb{I}_{i=1}x_{n,t}]}, i=0,1
\end{align}

Updating $\gamma,\alpha,\beta$ is a little tricky. First we introduce the approximation for $\gamma$, which will make the other two updates more understandable. Even if we use full factorization in the approximation, the update of $\gamma$ is associated with variables $x_{n,t-1}$ and $x_{n,t}$. A natural idea is to use Monte Carlo methods to sample $\{\tilde{x}_{1:N,1:T}\}\in \mathcal{X}^{N^2}$ from $p(\mathbf{X}|\mathbf{Y},\Theta^{old})=\prod_{n,t}p(x_{n,t}|\mathbf{Y},\Theta^{old})$. Then we can count the number of times that event $x_{n,t-1}=1,x_{n,t}=0$ happens. For simplicity, we can directly assign the simulated sample by Bayesian decision strategy according to each $p(x_{n,t}|\mathbf{Y},\Theta^{old})$ instead of sampling. That is to say, we only need to set the sample $x_{n,t}=\arg\max_{\tilde{x}_{n,t}=\{0,1\}}p(x_{n,t}|\mathbf{Y},\Theta^{old})$ and give the following result.
\begin{align}
\gamma&=\frac{\sum_{n=1}^N\sum_{t=1}^{T}\mathbb{I}_{\tilde{x}_{n,t-1}=1,\tilde{x}_{n,t}=0}}{\sum_{n=1}^N\sum_{t=1}^{T}\mathbb{E}[x_{n,t-1}]}\approx\frac{\sum_{n=1}^N\sum_{t=1}^{T}\mathbb{I}_{\tilde{x}_{n,t-1}=1,\tilde{x}_{n,t}=0}}{\sum_{n=1}^N\sum_{t=1}^{T}\mathbb{I}_{\tilde{x}_{n,t-1}=1}}
\end{align}

The same trick can be applied to update $\alpha,\beta$. However, along with this approximation trick, we also need variable substitution. Let $\tau_j=(1-\alpha)(1-\beta)^j$, then $\alpha=1-\tau_0$. Therefore the update of $\alpha$ and $\tau_i,i=1,...,M$  is analogous to $\gamma$.
\begin{align}
\alpha=\frac{\sum_{n=1}^N\sum_{t=1}^{T}\mathbb{I}_{\tilde{x}_{n,t-1}=0,\tilde{x}_{n,t}=1}\mathbb{I}_{\sum_{n':(n,n')\in E_{t-1}}\tilde{x}_{n',t-1}=0}}{\sum_{n=1}^N\sum_{t=1}^{T}\mathbb{E}[1-x_{n,t-1}]}
\\
\tau_i = \frac{\sum_{n=1}^N\sum_{t=1}^{T}\mathbb{I}_{\tilde{x}_{n,t-1}=0,\tilde{x}_{n,t}=0}\mathbb{I}_{\sum_{n':(n,n')\in E_{t-1}}\tilde{x}_{n',t-1}=j}}{\sum_{n=1}^N\sum_{t=1}^{T}\mathbb{E}[1-x_{n,t-1}]}
\end{align}
$\beta=1-\left(\frac{\tau_j}{1-\alpha}\right)^{\frac{1}{j}}$, thus meaning we have $M$ estimations for $\beta$s. How to combine these $\beta$s to obtain a better estimation may vary for different applications. We suggest one possibility using averaging that can be adjusted under various circumstances. In summary, we organize our generalized Baum-Welch Algorithm for GCHMMs in the following framework \ref{alg:GBW}.
\begin{align}
\beta=1-\frac{1}{M}\sum_{j=1}^M\left(\frac{\tau_i}{1-\alpha}\right)^{\frac{1}{j}}
\end{align} 

\begin{algorithm}[t]
 \KwData{$\mathbf{Y}$, $\mathbf{G}$}
 \KwResult{parameter set $\Theta=\{\pi,\theta,\alpha,\beta,\gamma\}$ and hidden matrix $\mathbf{X}$}
 \textbf{Initialize} coefficient parameter $\Theta^{(0)}$\;
 \Repeat{$\Theta^{new}$ Convergence}{
 Run one iteration forward-backward algorithm on $\Theta^{old}$ to obtain $p(x_{n,t}|\mathbf{Y},\Theta^{old})$\;
  Update $\Theta^{new}$ based on equations from (10) to (15)\;
 }
 \caption{Generalized Baum-Welch Algorithm}\label{alg:GBW}
\end{algorithm}

\begin{figure}[t]
\centering
\subfigure{
\includegraphics[width=70mm]{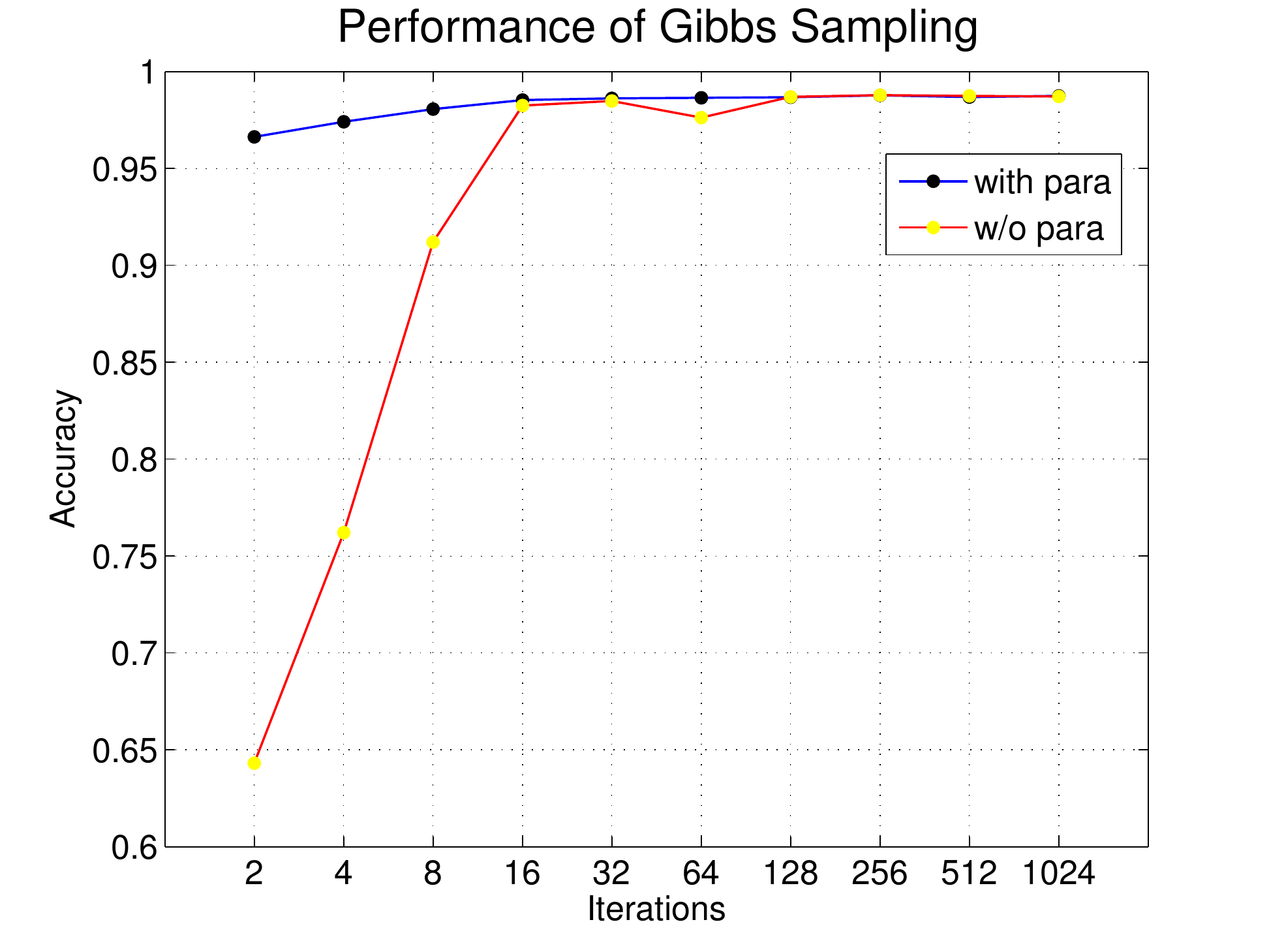}
}
\subfigure{
\includegraphics[width=70mm]{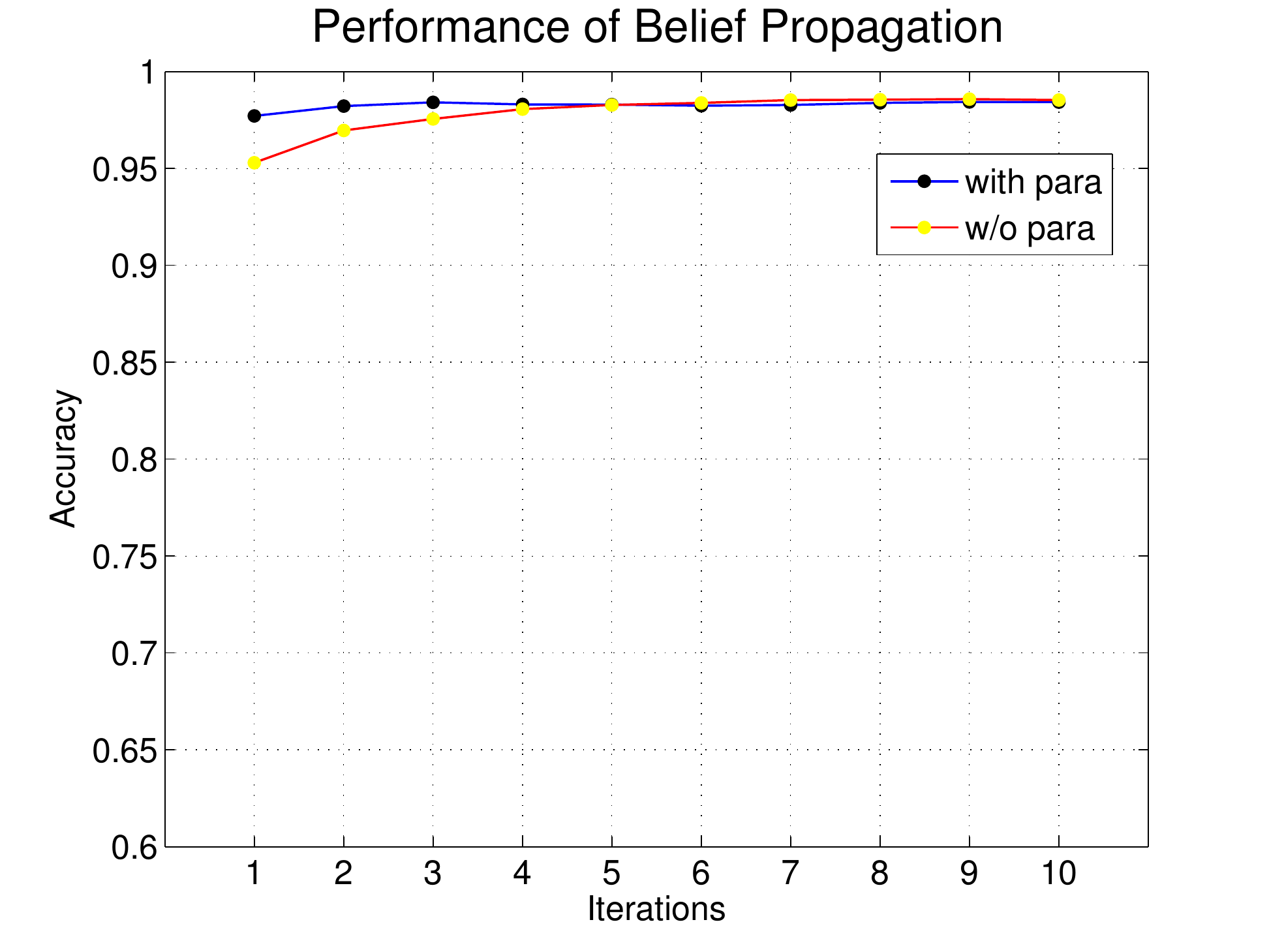}
}
\centering
\caption{The axis of Gibbs sampling is scaled by logarithm. Both algorithms can achieve an accuracy above 98.5\%.}
\label{fig:iter}
\end{figure}

\subsection{A Simulation Study}

\paragraph{Synthesized data based on Real Social Networks} We leveraged a dynamic social network dataset with 84 nodes over 107 days in \cite{madan2012sensing} denoted as $\mathbf{G}_{84\times84\times107}$, we modified it to make its maximum degree be bounded by constant $M=11$. Based on the generative model, we simulated infection states $\mathbf{X}_{84\times108}$ (including $X_{\cdot,0}$ without emission) and observed symptoms data $\mathbf{Y}_{84\times107\times6}$. We ran inference using two algorithms Generalized BW (GBW) and the Gibbs sampling developed by \cite{dong2012graph} in two different settings, when parameters are known and unknown. Notice that GBW with known parameters is reduced to forward-backward belief propagation. Both algorithms won't take $\mathbf{X}$ as arguments but predict an $\mathbf{X}$. Gibbs sampling and GBW are run for 500 and 15 iterations respectively. The Gibbs sampling implemented in the experiments is burns-in half of the total iterations. Since $\mathbf{X}$ is a binary matrix, a threshold of 0.5 is used for prediction.

In order to observe the performance with respect to the number of iterations, we conduct experiments testing the predictive accuracy of $\mathbf{X}$ with results shown in Figure \ref{fig:iter}. With parameters known, both algorithms can achieve good performance within a few iterations, while in the unknown case, the generalized BW algorithm gives better performance than Gibbs sampling over fewer iterations. However, the excellent performance of GBW is dependent on the initialization of parameters. If they are chosen inappropriately, GBW may not perform well, a common disadvantage of EM algorithm. In addition, with more iterations, Gibbs sampling will give more robust prediction results.

\section{Extending GCHMMs to Hierarchies}
\label{sec:hgchmm}

Though we successfully design a generalized EM algorithm for GCHMMs, we still overlook temporal covariates $\mathbf{z}_n$. In epidemics, it is commonly assumed that personal health features (covariates) are relevant to influenza vulnerability. In our model, $\mathbf{z}_n\in \mathcal{R}^K$, where $K$ is the dimension of the covariate feature space. Without loss of generalization, the feature is a constant of 1 in the feature space we are looking at. This relevance is captured by the mapping $f:\mathcal{R}^K\rightarrow[0,1]$ or the transformation from the feature space to infection parameters. In this section, we propose two constructions using two different link functions. A natural way to go is to extend the beta prior of the standard GCHMM to a beta-exponential link.

\paragraph{Beta-Exponential link} 
\begin{align}
&\bm\eta_{\cdot,\cdot} \sim \mathcal{N}(\bm\mu,\Sigma)  \\
&\gamma_n \sim \text{Beta}(\exp(\mathbf{z}_n^\top\bm\eta_{r,1}),\exp(\mathbf{z}_n^\top\bm\eta_{r,2})) \nonumber \\
&\alpha_n \sim \text{Beta}(\exp(\mathbf{z}_n^\top\bm\eta_{a,1}),\exp(\mathbf{z}_n^\top\bm\eta_{a,2})) \nonumber \\
&\beta_n \sim \text{Beta}(\exp(\mathbf{z}_n^\top\bm\eta_{b,1}),\exp(\mathbf{z}_n^\top\bm\eta_{b,2})) \nonumber
\end{align}
where $\bm\eta_{\cdot,\cdot}$ is distributed as a multivariate Gaussian, playing the role of the regression coefficients, since the expectation $\frac{1}{1+e^{\mathbf{z}_n^\top(\bm\eta_{\cdot,1}-\bm\eta_{\cdot,2})}}$ can be considered as an approximation for logistic regression with coefficients $-(\bm\eta_{\cdot,1}-\bm\eta_{\cdot,2})$. This link also enables the exponential term $\exp(\mathbf{z}_n^\top\bm\eta_{\cdot,\cdot})$ to take the place of the hyper-parameter of the beta prior. The usual count update to the hyper-parameter will implicitly update $\bm\eta$ via our EM algorithm.

Once $\gamma,\alpha,\beta$ are allowed to be indexed by $n$, a new \textit{transmission function} merely needs an index modification of the arguments in (\ref{eq:trans}), but otherwise remains the same, i.e. $\phi_{n,x_{n'}:(n,n')\in G_t}(\gamma_n,\alpha_n,\beta_n)$. The advantage of this setting is that it allows for the approximate Gibbs sampling of infection parameters in a way that still holds for GCHMMs, except for $\eta$, so that the original Gibbs sampling scheme to update the beta distribution by event counts is the same in the later E-step. Another advantage is, when $\mathbf{X}$ is generalized to categorical variables, a similar construction also works. We use an individual level distribution with the transition but not with the emission matrix because it makes more sense that everyone has the same probability of a physical behavior given an infection state. Patients should have corresponding symptoms, such as cough or throat pain, or the flu cannot be discovered or diagnosed. Instead of the Beta-Exponential distribution, we can introduce a deterministic  link.

\paragraph{Sigmoid link} 
\begin{align}\label{eq:sigmoid}
\bm\eta_\cdot \sim \mathcal{N}(\bm\mu,\Sigma),\quad\gamma_n = \sigma(\mathbf{z}_n^\top\bm\eta_r),\quad \alpha_n = \sigma(\mathbf{z}_n^\top\bm\eta_a),\quad\beta_n = \sigma(\mathbf{z}_n^\top\bm\eta_b) 
\end{align} 
In this generative process, less $\bm\eta$s are present, thus leading to a simpler model. Instead of sampling, the equation (\ref{eq:sigmoid}) will actually make infection parameters vanish in the model. In other words, $\phi_{n,x_{n'}:(n,n')\in G_t}(\gamma_n,\alpha_n,\beta_n)$ is replaced by $\phi_{n,x_{n'}:(n,n')\in G_t}(z_n,\eta_{\cdot},\sigma(\cdot))$. From an implementation perspective, the EM derivation will be easier, and the experimental results imply its performance is more competitive. Additionally, for both link constructions, the generative model (\ref{eq:generative}) is also individually indexed by $n$.

{\bf Another interpretation of $\beta_n$} In above two extensions, it is implicitly assumed that $\beta_n$ means the individual infection probability from another person within the network, is as given in Equation (\ref{eq:ip1}). From a biological perspective, the contagiousness of the infected person varies, meaning that $\beta_n$ can be interpreted as the probability of spreading illness to any other person in the social network. This heterogenous inconsistency will not appear in the previously discussed homogenous setting. However, the second interpretation results in a slightly complicated mathematical calculation (details in inference section), since both the total count of infectious contacts $C_{n,t}$ and their individual features are required. Thus, the probability of infection has two different definitions. 
\begin{align} 
P(x_{n,t+1}=1|x_{n,t}=0)&=1-(1-\alpha_n)(1-\beta_{n})^{C_{n,t}} \label{eq:ip1} \\
P(x_{n,t+1}=1|x_{n,t}=0)&=1-(1-\alpha_n)\prod_{n'\in S_{n,t}}(1-\beta_{n'}) \label{eq:ip2}
\end{align}
where the node set of infectious contacts is defined as $S_{n,t}=\left\{n'\in[N]:(n,n')\in E_t, x_{n',t}=1\right\}$.

\subsection{Model Evolution}

\begin{figure}
\centering
\tiny
\begin{tikzpicture}
\tikzstyle{main}=[circle, minimum size = 8mm, thick, draw =black!80, node distance = 15mm,inner sep=0pt]
\tikzstyle{connect}=[-latex, thick]
\tikzstyle{box}=[rectangle, draw=black!100]
  \node[main, fill = white!100] (theta) {$\theta_X$};
  \node[main] (h) [above=of theta,yshift=-5mm] {$h$};
  \node[main] (X1t0) [right=of theta,xshift=-5mm, yshift=30mm] {$x_{1,t-1}$};
  \node[main] (X1t) [right=of X1t0] {$x_{1,t}$};
  \node[main] (X1t1) [right=of X1t] {$x_{1,t+1}$};
  \node[main, fill = black!10] (Y1t0) [below=of X1t0,yshift=10mm] {$y_{1,t-1}$};
  \node[main, fill = black!10] (Y1t) [below=of X1t,yshift=10mm] {$y_{1,t}$};
  \node[main, fill = black!10] (Y1t1) [below=of X1t1,yshift=10mm] {$y_{1,t+1}$};
  \node[rectangle,draw=black!100] (rab1) [below=of X1t1, xshift=15mm, yshift=5mm] {$\gamma_1,\alpha_1,\beta_1$};
  \node[main, fill = black!10] (Z1) [above=of rab1, yshift=-10mm] {$z_1$};
  \node[main] (X2t0) [below=of Y1t0,yshift=14mm] {$x_{2,t-1}$};
  \node[main] (X2t) [below=of Y1t,yshift=14mm] {$x_{2,t}$};
  \node[main] (X2t1) [below=of Y1t1,yshift=14mm] {$x_{2,t+1}$};
  \node[main, fill = black!10] (Y2t0) [below=of X2t0,yshift=10mm] {$y_{2,t-1}$};
  \node[main, fill = black!10] (Y2t) [below=of X2t,yshift=10mm] {$y_{2,t}$};
  \node[main, fill = black!10] (Y2t1) [below=of X2t1,yshift=10mm] {$y_{2,t+1}$};
  \node[rectangle,draw=black!100] (rab2) [below=of X2t1, xshift=15mm, yshift=5mm] {$\gamma_2,\alpha_2,\beta_2$};
  \node[main, fill = black!10] (Z2) [above=of rab2, yshift=-10mm] {$z_2$};
  \node[main] (eta) [right=of Z2, xshift=-5mm, yshift=5mm] {$\eta$};
  \node[main] (normal) [above=of eta,yshift=-5mm] {$\mu,\Sigma$};
  \node[main] (X3t0) [below=of Y2t0,yshift=14mm] {$x_{3,t-1}$};
  \node[main] (X3t) [below=of Y2t,yshift=14mm] {$x_{3,t}$};
  \node[main] (X3t1) [below=of Y2t1,yshift=14mm] {$x_{3,t+1}$};
  
  \path (h) edge [connect] (theta)
  		(X1t0) edge [connect] (X1t)
        (X1t) edge [connect] (X1t1)
        (X1t0) edge [connect] (Y1t0)
        (X1t) edge [connect] (Y1t)
        (X1t1) edge [connect] (Y1t1)
        (theta) edge [connect] (Y1t0)
        (rab1) edge [connect] (X1t0)
        (rab1) edge [connect] (X1t)
        (rab1) edge [connect] (X1t1)
        (Z1) edge [connect] (rab1)
        (X2t0) edge [connect] (X2t)
        (X2t) edge [connect] (X2t1)
        (X2t0) edge [connect] (Y2t0)
        (X2t) edge [connect] (Y2t)
        (X2t1) edge [connect] (Y2t1)
        (theta) edge [connect] (Y2t0)
        (rab2) edge [connect] (X2t0)
        (rab2) edge [connect] (X2t)
        (rab2) edge [connect] (X2t1)
        (Z2) edge [connect] (rab2)
        (eta) edge [connect] (rab1)
        (eta) edge [connect] (rab2)
        (X3t0) edge [connect] (X3t)
        (X3t) edge [connect] (X3t1)
        (X1t0) edge [connect] (X3t)
        (X3t0) edge [connect] (X1t)
        (X2t) edge [connect] (X3t1)
        (X3t) edge [connect] (X2t1)
        (normal) edge [connect] (eta);

  \node[rectangle, inner sep=0mm,draw=black!100, fit= (X1t0) (X1t) (X1t1) (Y1t0) (Y1t) (Y1t1) (rab1) (Z1)] {};
  \node[rectangle, inner sep=0mm, fit= (X1t0) (X1t) (X1t1) (Y1t0) (Y1t) (Y1t1) (rab1) (Z1),label=below right:1, xshift=15mm, yshift=3mm] {};
  \node[rectangle, inner sep=0mm,draw=black!100, fit= (X2t0) (X2t) (X2t1) (Y2t0) (Y2t) (Y2t1) (rab2) (Z2)] {};
  \node[rectangle, inner sep=0mm, fit= (X2t0) (X2t) (X2t1) (Y2t0) (Y2t) (Y2t1) (rab2) (Z2),label=below right:2, xshift=15mm, yshift=3mm] {};

\end{tikzpicture}
\caption{The hGCHMM extended from Figure \ref{fig:gchmm}(b): for plotting simplification, the edges from $\theta_X$ to other observed nodes are blanked out; the two step evolution, GCHMMs$\rightarrow$heterogenous GCHMMs$\rightarrow$hierarchical GCHMMs.}
\label{fig:hgchmm}
\end{figure}
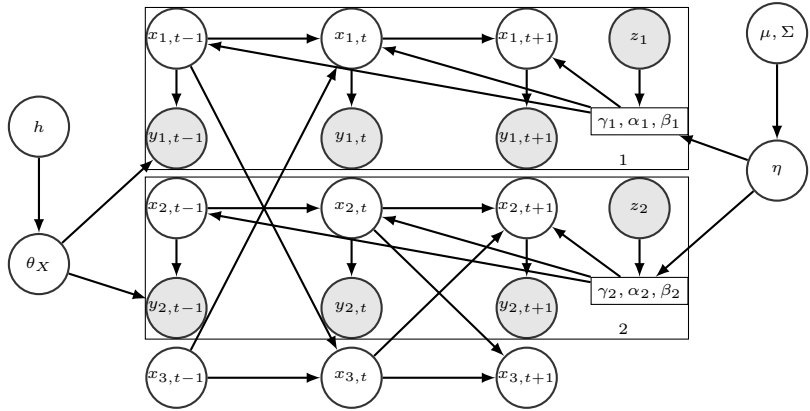

\paragraph{HMM} Along the direction of graphical representation in Figure \ref{fig:gchmm}, the hierarchical GCHMMs shown in Figure \ref{fig:hgchmm} can be seen as a two-step evolution. First, each HMM is allowed to contain its own infection parameters, thus evolving to a heterogenous GCHMMs. Second, the link introduced previously can be used to associate with covariates. This scheme also inspires another corresponding two-separated-segment training: Gibbs sampling to estimate the posterior mean of $(\gamma_n,\alpha_n,\beta_n)$ for heterogenous GCHMMs, and fitting a standard logistic regression between the posterior mean and covariates.

However, there are some things that we need to notice. \textbf{(1)} The beta-exponential generative model is not well defined for unique dataset simulation, because we need only one sample from this generative model, which actually uses a set of fixed parameters $(\alpha_n^f,\beta_n^f,\gamma_n^f)_{n=1}^N$ to simulate $\mathbf{X}_{N\times T}$ and $Y_{N\times T\times S}$. Take $\gamma_n$ for example, $\gamma_n^f\neq \mathbb{E}[\gamma_n]$, since $\gamma_n^f$ is one realization of the generative model. What our algorithm aims to learn is a generative distribution $\text{Beta}(e^{\mathbf{z}_n^\top\bm\eta_{r,1}},e^{\mathbf{z}_n^\top\bm\eta_{r,2}})$ with expectation equal to $\gamma_n^f$, rather than the real $\mathbb{E}[\gamma_n]$. \textbf{(2)} Therefore, in experiments our simulation dataset is always generated from the sigmoid link model. However, it is reasonable to use the beta-exponential link model for inference to eliminate the inconsistency. It has been mentioned previously the expectation of the beta-exponential is basically an approximate logistic regression. An EM like algorithm can perform point estimation well for this expectation, which in turn would be an estimator of the sigmoid link. \textbf{(3)} Another way to make the beta-exponential generative and the inference process work is to sample $\alpha_n,\beta_n,\gamma_n$ both individually and dynamically, i.e. $\gamma_{n,t},\alpha_{n,t},\beta_{n,t}$. A number of samples are sufficient to learn the true generative distribution, though the new inference algorithm will necessarily become more difficult.

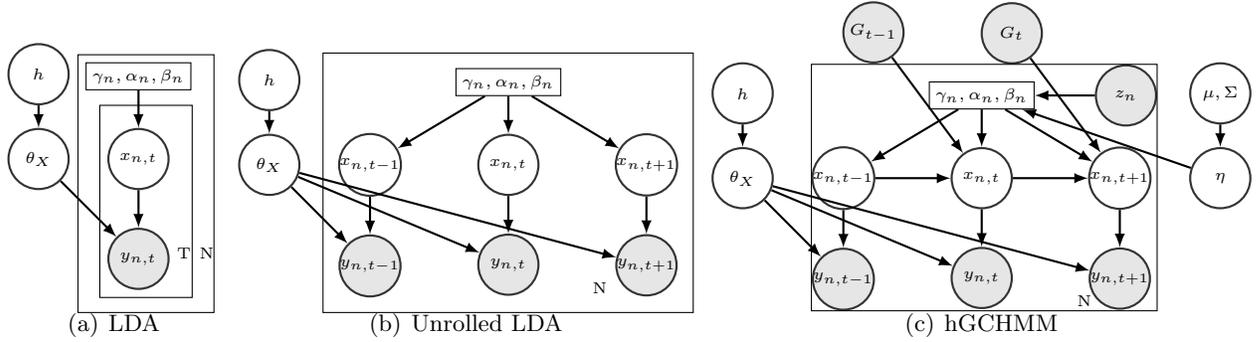
\begin{figure}
\tiny
\subfigure[LDA]{
\centering
\begin{tikzpicture}
\tikzstyle{main}=[circle, minimum size = 8mm, thick, draw =black!80, node distance = 10mm,inner sep=0pt]
\tikzstyle{connect}=[-latex, thick]
\tikzstyle{box}=[rectangle, draw=black!100]

  \node[main, fill = white!100] (theta) {$\theta_X$};
  \node[main] (h) [above=of theta,yshift=-7mm] {$h$};
  \node[main] (Xnt) [right=of theta,xshift=-5mm] {$x_{n,t}$};
  \node[main, fill = black!10] (Ynt) [below=of Xnt,yshift=5mm] {$y_{n,t}$};
  \node[rectangle,draw=black!100] (rab) [above=of Xnt, yshift=-5mm] {$\gamma_n,\alpha_n,\beta_n$};
  
  \path (h) edge [connect] (theta)
        (Xnt) edge [connect] (Ynt)
        (rab) edge [connect] (Xnt)
        (theta) edge [connect] (Ynt);
                
\node[rectangle, inner sep=2mm,draw=black!100, fit= (Xnt) (Ynt) (rab), xshift=1mm,yshift=-1mm] {};
\node[rectangle, inner sep=0mm, fit= (Xnt) (Ynt), label=below right:N,xshift=3mm] {};
\node[rectangle, inner sep=2mm,draw=black!100, fit= (Xnt) (Ynt), xshift=1mm,yshift=1mm] {};
\node[rectangle, inner sep=0mm, fit= (Xnt) (Ynt), label=below right:T,xshift=0mm] {};

\end{tikzpicture}
}
\subfigure[Unrolled LDA]{
\centering
\begin{tikzpicture}
\tikzstyle{main}=[circle, minimum size = 8mm, thick, draw =black!80, node distance = 10mm,inner sep=0pt]
\tikzstyle{connect}=[-latex, thick]
\tikzstyle{box}=[rectangle, draw=black!100]

  \node[main, fill = white!100] (theta) {$\theta_X$};
  \node[main] (h) [above=of theta,yshift=-7mm] {$h$};
  \node[main] (Xnt0) [right=of theta,xshift=-5mm] {$x_{n,t-1}$};
  \node[main] (Xnt) [right=of Xnt0] {$x_{n,t}$};
  \node[main] (Xnt1) [right=of Xnt] {$x_{n,t+1}$};
  \node[main, fill = black!10] (Ynt0) [below=of Xnt0,yshift=5mm] {$y_{n,t-1}$};
  \node[main, fill = black!10] (Ynt) [below=of Xnt,yshift=5mm] {$y_{n,t}$};
  \node[main, fill = black!10] (Ynt1) [below=of Xnt1,yshift=5mm] {$y_{n,t+1}$};
  \node[rectangle,draw=black!100] (rab) [above=of Xnt, yshift=-5mm] {$\gamma_n,\alpha_n,\beta_n$};
  
  \path (h) edge [connect] (theta)
        (Xnt0) edge [connect] (Ynt0)
        (Xnt) edge [connect] (Ynt)
        (Xnt1) edge [connect] (Ynt1)
        (theta) edge [connect] (Ynt0)
        (theta) edge [connect] (Ynt)
        (theta) edge [connect] (Ynt1)
        (rab) edge [connect] (Xnt0)
        (rab) edge [connect] (Xnt)
        (rab) edge [connect] (Xnt1);
        
\node[rectangle, inner sep=2mm,draw=black!100, fit= (Xnt0) (Xnt) (Xnt1) (Ynt0) (Ynt) (Ynt1) (rab)] {};
\node[rectangle, inner sep=0mm, fit= (Xnt0) (Xnt) (Xnt1) (Ynt0) (Ynt) (Ynt1) (rab),label=below right:N, xshift=-5mm, yshift=3mm] {};
\end{tikzpicture}
}
\subfigure[hGCHMM]{
\centering
\begin{tikzpicture}
\tikzstyle{main}=[circle, minimum size = 8mm, thick, draw =black!80, node distance = 10mm,inner sep=0pt]
\tikzstyle{connect}=[-latex, thick]
\tikzstyle{box}=[rectangle, draw=black!100]

  \node[main, fill = white!100] (theta) {$\theta_X$};
  \node[main] (h) [above=of theta,yshift=-7mm] {$h$};
  \node[main] (Xnt0) [right=of theta,xshift=-5mm] {$x_{n,t-1}$};
  \node[main] (Xnt) [right=of Xnt0] {$x_{n,t}$};
  \node[main] (Xnt1) [right=of Xnt] {$x_{n,t+1}$};
  \node[main, fill = black!10] (Ynt0) [below=of Xnt0,yshift=5mm] {$y_{n,t-1}$};
  \node[main, fill = black!10] (Ynt) [below=of Xnt,yshift=5mm] {$y_{n,t}$};
  \node[main, fill = black!10] (Ynt1) [below=of Xnt1,yshift=5mm] {$y_{n,t+1}$};
  \node[rectangle,draw=black!100] (rab) [above=of Xnt, yshift=-5mm] {$\gamma_n,\alpha_n,\beta_n$};
  \node[main, fill = black!10] (Zn) [right=of rab, xshift=-2mm] {$z_n$};
  \node[main] (eta) [right=of Xnt1, xshift=-5mm] {$\eta$};
  \node[main,fill = black!10] (Gt0) [above=of Xnt0, xshift=4mm,yshift=1mm] {$G_{t-1}$};
  \node[main,fill = black!10] (Gt) [above=of Xnt, xshift=4mm,yshift=1mm] {$G_{t}$};
  \node[main] (normal) [above=of eta,yshift=-7mm] {$\mu,\Sigma$};

  \path (h) edge [connect] (theta)
  		(Xnt0) edge [connect] (Xnt)
        (Xnt) edge [connect] (Xnt1)
        (Xnt0) edge [connect] (Ynt0)
        (Xnt) edge [connect] (Ynt)
        (Xnt1) edge [connect] (Ynt1)
        (theta) edge [connect] (Ynt0)
        (theta) edge [connect] (Ynt)
        (theta) edge [connect] (Ynt1)
        (rab) edge [connect] (Xnt0)
        (rab) edge [connect] (Xnt)
        (rab) edge [connect] (Xnt1)
        (Zn) edge [connect] (rab)
        (eta) edge [connect] (rab)
        (Gt0) edge [connect] (Xnt)
        (Gt) edge [connect] (Xnt1)
        (normal) edge [connect] (eta);
        
\node[rectangle, inner sep=0mm,draw=black!100, fit= (Xnt0) (Xnt) (Xnt1) (Ynt0) (Ynt) (Ynt1) (rab) (Zn)] {};
\node[rectangle, inner sep=0mm, fit= (Xnt0) (Xnt) (Xnt1) (Ynt0) (Ynt) (Ynt1) (rab) (Zn),label=below right:N, xshift=-5mm, yshift=3mm] {};

\end{tikzpicture}
}
\caption{(a) is the commonly represented LDA model; (b) is an equivalent but different graphical representation of LDA; (c) is our proposed model, adding topic dependency, document dependency and document-specific features.}
\label{fig:ldahgchmm}
\end{figure}

\paragraph{LDA} As shown in Figure \ref{fig:ldahgchmm}(b), LDA can be trivially represented as an unrolled graph for $T$ documents. If we impose a Markov dependency to model topics changing, an embedded HMM appears with respect to the latent topics. Furthermore, we construct the topic changing function relying on the document relationship and a link associated with document-specific covariates, thus resulting our hGCHMMs. A similar variant is LDA-HMM \cite{griffiths2004integrating}, which requires extra hidden nodes essentially to model syntactic or function words, such as "and" or "however", introducing a sentence-level dependency, while the HMM imposed on topic nodes gives a word-level dependency.

\section{Inference for hGCHMMs}
\label{sec:inference}

\subsection{Approximate Conjugacy} 
The inference process is designed to invert the generative model and to discover the $\bm\eta$ and $\mathbf{X}$ that best explain $\mathbf{G}$ and $\mathbf{Y}$. In our hierarchical extension, however, a fully conjugate prior is not present and knowing what the right prior is can be difficult. Thus an approximate conjugacy is developed by introducing the auxiliary variable $R_{n,t}$, representing the non-specific infection source (inside or outside networks). The idea is to decompose \textit{infection probability} $I_{n,t}\triangleq1-(1-\alpha_n)(1-\beta_n)^{C_{n,t}}$ into the summation of three terms, $\alpha_n(1-\beta_n)^{C_{n,t}}$, $(1-\alpha_n)(1-(1-\beta_n)^{C_{n,t}})$ and $\alpha_n(1-(1-\beta_n)^{C_{n,t}})$, indicating infection from outside, inside and both respectively, thus following a categorical distribution:
\begin{equation}\label{eq:approxR}
\begin{split}
P(R_{n,t})=\left\{ 
  \begin{array}{l l}
    \frac{\alpha_n(1-\beta_n)^{C_{n,t}}}{1-(1-\alpha_n)(1-\beta_n)^{C_{n,t}}},&\text{if outside infection, $R_{n,t}=1$}\\
    \frac{(1-\alpha_n)(1-(1-\beta_n)^{C_{n,t}})}{1-(1-\alpha_n)(1-\beta_n)^{C_{n,t}}},&\text{if inside infection, $R_{n,t}=2$}\\
    \frac{\alpha(1-(1-\beta_n)^{C_{n,t}})}{1-(1-\alpha_n)(1-\beta_n)^{C_{n,t}}},& \text{if both, $R_{n,t}=3$, }\\
  \end{array} \right.
\end{split}
\end{equation}
The exact expression still does not have Beta-Bernoulli conjugacy except for the case where $R_{n,t}=1$. However, using a taylor expansion we have $P(R_{n,t}=2)P(x_{n,t+1}=1|x_{n,t}=0) \approx C_{n,t}(1-\alpha_n)\beta_n$ and $P(R_{n,t}=3)P(x_{n,t+1}=1|x_{n,t}=0) \approx C_{n,t}\alpha_n\beta_n$. The two approximations have the property that local full conditionals can be analytically obtained by discarding $\eta$ temporarily. In practice the term involving $P(R_{n,t}=3)$ can be approximated as 0 for Gibbs sampling. Because of the biological application, $\alpha_n$ and $\beta_n$ are both a positive real value close to 0, resulting in their product being quite small. Even if this probability is taken into consideration in Gibbs sampling, there is a very small chance that $R_{n,t}=3$. This approximation allows the posterior distribution of $\alpha_n,\beta_n$ to be much easier to compute given the current value of $\eta$.

\begin{figure}[t]
\centering
\subfigure[Approximate $I$ w.r.t $\alpha$]{
\includegraphics[width=47mm,clip,trim=69 175 70 65mm]{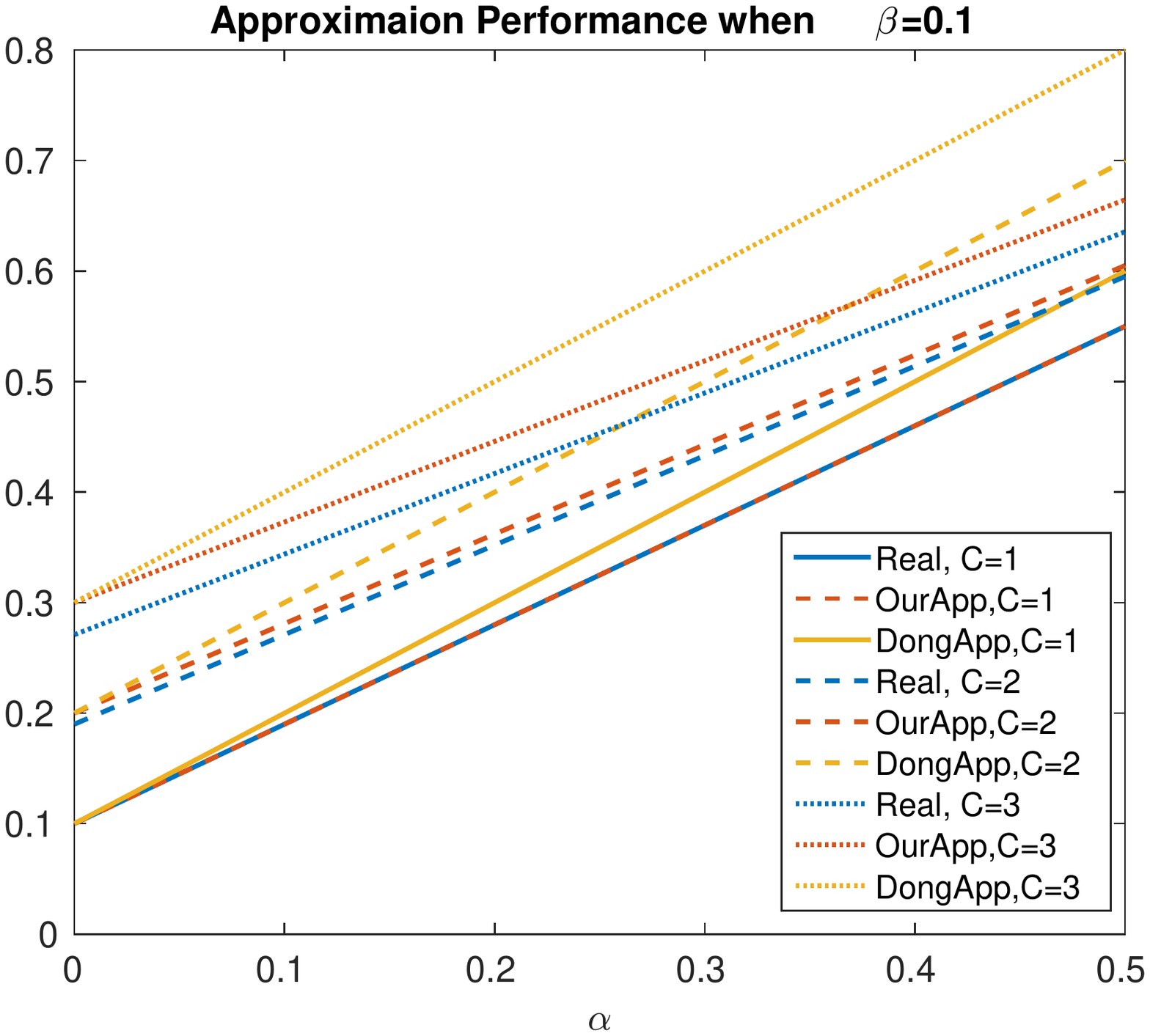}
}
\subfigure[Approximate $I$ w.r.t $\beta$]{
\includegraphics[width=47mm,clip,trim=69 175 70 65mm]{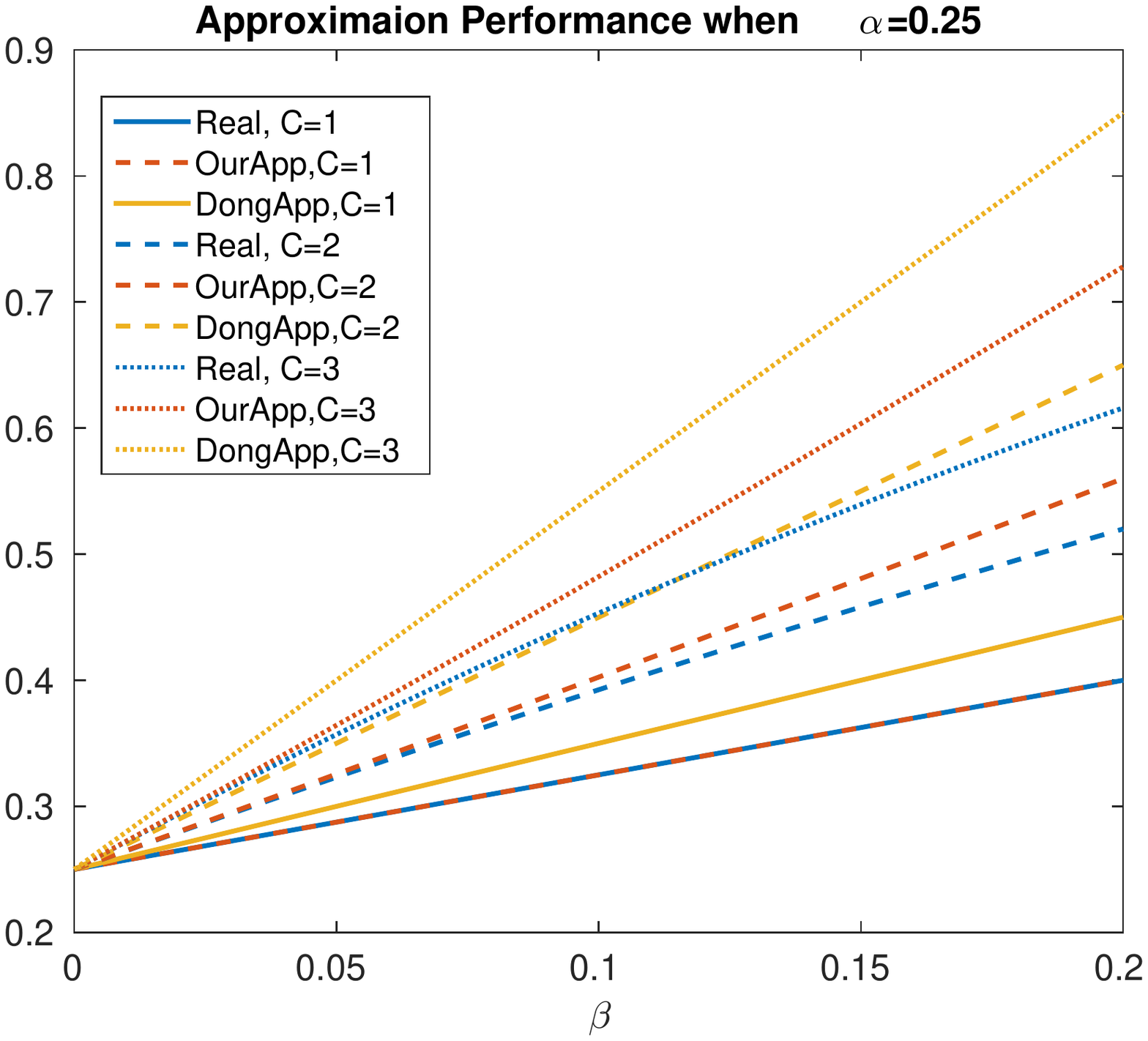}
}
\subfigure[Approximate $p(R)$ w.r.t $\alpha$]{
\includegraphics[width=47mm,clip,trim=69 175 73 65mm]{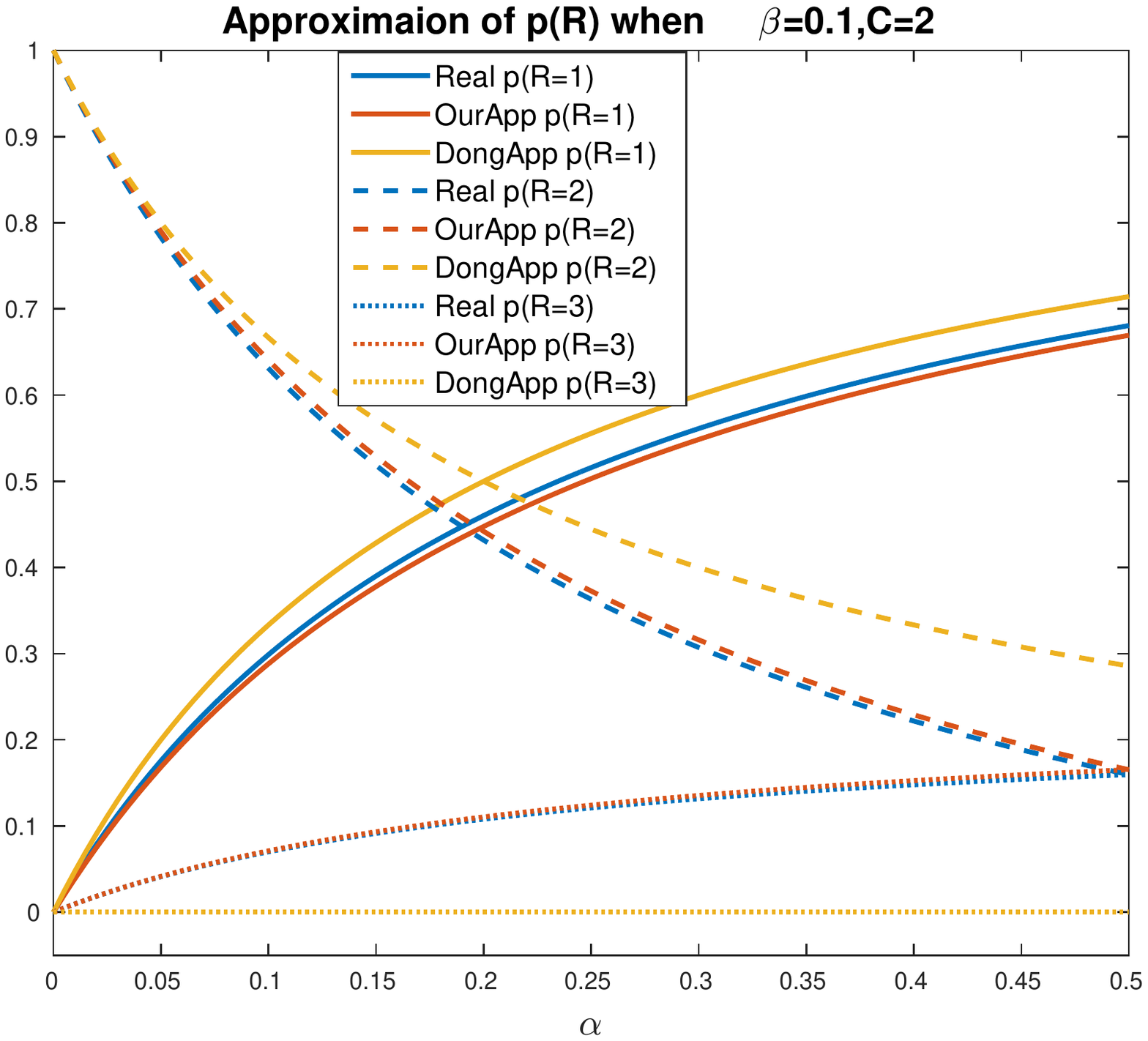}
}
\centering
\caption{Comparison between different approximation methods}
\label{fig:approxR}
\end{figure}

In addition, our approximation works better than the proposed decomposition in \cite{dong2012graph}, i.e. $I_{n,t}=\alpha_n + C_{n,t}\beta_n$. In Figure \ref{fig:approxR}, a quantitative comparison between our proposed approximation and previous work indicates less error achieved by our decomposition. Specifically, if $C_{n,t}=1$, our decomposition recovers $I$ exactly (blue line and red line are overlapped); with $C_{n,t}$ increasing, both of the two approximations are biased, but our approach has constant error regardless of varying $\alpha$, and shows less error for relatively bigger $\beta$; moreover, the induced approximate distribution through our three terms is almost in line with the true distribution (\ref{eq:approxR}). The main advantage of the novel approximation is the possibility of deriving the fully conjugate posterior. Specifically, we have the following posteriors, which will benefit from the EM algorithm described later:
\begin{align*}
\alpha_n&\sim\text{Beta}\left(e^{\mathbf{z}_n^\top\bm\eta_{a,1}}+C_{n,R_n=1,3},e^{\mathbf{z}_n^\top\bm\eta_{a,2}}+C_{n,R_n=2}+C_{n,0\rightarrow0}\right)\\
\beta_n&\sim\text{Beta}\left(e^{\mathbf{z}_n^\top\bm\eta_{b,1}}+C_{n,R_n=2,3},e^{\mathbf{z}_n^\top\bm\eta_{b,2}}+C_{n,R\neq2,3}\right)\\
\gamma_n&\sim\text{Beta}(e^{\mathbf{z}_n^\top\bm\eta_{r,1}}+C_{n,1\rightarrow0},e^{\mathbf{z}_n^\top\bm\eta_{r,2}}+C_{n,1\rightarrow1})
\end{align*}
where the count notations are defined as follows.
\begin{table}[htb]
\centering
\begin{tabular}{l} \hline
$C_{n,i\rightarrow j}=\sum_t\mathbb{I}_{\{x_{n,t}=i,x_{n,t+1}=j\}}, i,j\in\{0,1\}$ \\ \hline
$C_{n,R_n=1,3}=\sum_t\mathbb{I}_{\{R_{n,t}=1,3\}}\approx C_{n,R_n=1}=\sum_t\mathbb{I}_{\{R_{n,t}=1\}}$ \\ \hline
$C_{n,R_n=2,3}=\sum_t\mathbb{I}_{\{R_{n,t}=2,3\}}\approx C_{n,R_n=2}=\sum_t\mathbb{I}_{\{R_{n,t}=2\}}$\\ \hline
$C_{n,R\neq2,3}=\sum_tC_{n,t}\left[\mathbb{I}_{\{R_{n,t}=1\}}+\mathbb{I}_{\{x_{n,t}=0,x_{n,t+1}=0\}}\right]$ \\
\hline\end{tabular}
\label{tab:note}
\end{table}

Note that the auxiliary variable did not appear in the posterior of $\gamma_n$, which can be exactly computed due to conjugacy. Utilizing these approximate posteriors, the complete likelihood $P(\mathbf{X},\mathbf{R},\bm\eta|\mathbf{Z})$ is obtained by integrating out the infection parameters.
\begin{align}
&\int P(\mathbf{R}|\mathbf{X},\alpha_{n=1}^N,\beta_{n=1}^N)P(\mathbf{X}|\gamma_{n=1}^N,\alpha_{n=1}^N,\beta_{n=1}^N)P(\gamma_{n=1}^N,\alpha_{n=1}^N,\beta_{n=1}^N|\bm\eta,\mathbf{Z})P(\bm\eta)\mathrm{d}\gamma_{n=1}^N\mathrm{d}\alpha_{n=1}^N\mathrm{d}\beta_{n=1}^N \nonumber \\
=& P(\bm\eta)\prod_n\left(\frac{B(e^{\mathbf{z}_n^\top\bm\eta_{r,1}}+C_{n,1\rightarrow0},e^{\mathbf{z}_n^\top\bm\eta_{r,2}}+C_{n,1\rightarrow1})}{B(e^{\mathbf{z}_n^\top\bm\eta_{r,1}},e^{\mathbf{z}_n^\top\bm\eta_{r,2}})} \cdot \frac{B(e^{\mathbf{z}_n^\top\bm\eta_{a,1}}+C_{n,R_n=1,3},e^{\mathbf{z}_n^\top\bm\eta_{a,2}}+C_{n,R_n=2}+C_{n,0\rightarrow0})}{B(e^{\mathbf{z}_n^\top\bm\eta_{a,1}},e^{\mathbf{z}_n^\top\bm\eta_{a,2}})} \right. \nonumber \\
&\cdot\left.\frac{B(e^{\mathbf{z}_n^\top\bm\eta_{b,1}}+C_{n,R_n=2,3},e^{\mathbf{z}_n^\top\bm\eta_{b,2}}+C_{n,R\neq2,3})}{B(e^{\mathbf{z}_n^\top\bm\eta_{b,1}},e^{\mathbf{z}_n^\top\bm\eta_{b,2}})}\right) \label{eq:like}
\end{align}
where $B(\cdot)$ is the beta function, and $P(\bm\eta)$ is a multinomial Gaussian distribution. The integral result enables the analytical computation of the gradient $\nabla_\eta$ and $2^{\text{nd}}$ derivative $\partial^2$ of the log-likelihood, used in optimization via Newton's method. Under such $\beta_n$ interpretation, the derivatives of the log likelihood (\ref{eq:like}) are straightforward.

\subsection{Gibbs Sampling for the Conjugate Part}

{\bf Sampling Infection States} Given $\alpha_n,\beta_n,\gamma_n$, the generative model implies a conjugate prior for $x_{n,t}$. The unnormalized posterior probability of $x_{n,t}=i$ can be represented as $p_{n,t}^i,i=0,1$.
\begin{align}\label{eq:posX}
p_{n,t}^0 &\propto \gamma_n^{\mathbb{I}_{x_{n,t-1}=1}} 
I_{n,t}^{\mathbb{I}_{x_{n,t+1}=1}} (1-\alpha_n)^{\mathbb{I}_{\{x_{n,t-1}=0,x_{n,t+1}=0\}}} (1-\beta_n)^{C_{n,t-1}\mathbb{I}_{x_{n,t-1}=0}+C_{n,t}\mathbb{I}_{x_{n,t+1}=0}}\\
&\times \prod_s\theta_{0,s}^{\mathbb{I}_{y_{n,t,s}=1}}(1-\theta_{0,s})^{\mathbb{I}_{y_{n,t,s}=0}} \nonumber\\
p_{n,t}^1&\propto\gamma_n^{\mathbb{I}_{x_{n,t+1}=0}}\cdot(1-\gamma_n)^{\mathbb{I}_{\{x_{n,t-1}=1,x_{n,t+1}=1\}}}I_{n,t}^{\mathbb{I}_{x_{n,t-1}=0}} \prod_s\theta_{1,s}^{\mathbb{I}_{y_{n,t,s}=1}}(1-\theta_{1,s})^{\mathbb{I}_{y_{n,t,s}=0}}
\end{align}
where the normalized posterior of $p(x_{n,t}=1)$ is $\frac{p_{n,t}^1}{p_{n,t}^0+p_{n,1}^1}$. We need to be careful of the boundary condition since $x_{n,1}$ and $x_{n,T}$ do not have this form. $x_{n,1}$ is generated by $x_{n,1}\sim\text{Bernoulli}(\pi)$, where $\pi\sim\text{Beta}(a_\pi,b_\pi)$. The full conditional depends on the initial event occurrence rate $\pi$, further requiring some mild modification. The full conditional of $\pi$ can be efficiently derived.
\begin{align*}
\pi|\mathbf{X} \sim\text{Beta}\left(a_\pi+\sum_n \mathbb{I}_{(x_{n,1}=1)},b_\pi+ N -  \sum_n \mathbb{I}_{(x_{n,1}=1)}\right)
\end{align*}
For the state $x_{n,T}$, the posterior is easily computed since terms associated with $t+1$ cancel out immediately.

{\bf Sampling Missing Observations} For real world data, a missing value problem commonly arises because of underreporting in data collection. Bayesian schemes can successfully fill in missing values by drawing $y_{n,t,s}$ according to the distribution $\text{Bernoulli}(\theta_{x_{n,t},s})$, if they are \texttt{NA}. Given $y_{n,t,s}$, the posterior of $\theta_{i,s}$, $(i=0,1)$ is from a beta distribution.
\begin{align*}
\theta_{i,s}|\mathbf{X},\mathbf{Y}\sim \text{Beta}\left(a_i+\sum_{n,t}\mathbb{I}_{\{y_{n,t,s}=1,x_{n,t}=i\}},b_i+\sum_{n,t}\mathbb{I}_{\{y_{n,t,s}=0,x_{n,t}=i\}}\right)
\end{align*}

\subsection{Burn-in Gibbs EM Algorithm}
Previous works on CHMMs or GCHMMs seldom include the parameter $\bm\eta$, let alone a sampling scheme for inference. In hGCHMMs, the Gaussian prior makes the posterior of $\bm\eta$ not conjugate. One possible solution is the Metropolis Hastings (MH) algorithm due to the approximate likelihood ({\ref{eq:like}}); however, the transition kernel is difficult to choose for MH, and running large numbers of iterations is usually required to achieve good mixing. Another thing to try may be an augmentation trick based on the Poyla-Gamma distribution \cite{polson2013bayesian}, which is mentioned for the network imputation in the introduction. The drawback of this scheme is that it can be straightforward to fit the sigmoid link, while the beta-exponential prior may need further generalization. Variational Bayesian inference (see \cite{beal2003variational} for a detailed introduction) is commonly used for approximate inference by optimizing a lower bound. If the readers are familiar with variational methods, it is obvious that for the conjugate-exponential family the update for parameters (in our model, $\pi$, $\theta_X$,$\gamma$) can be written out analytically, equivalent to the posterior derivation. However, for other parameters associatiated with non-conjugacy, a gradient based method is the first option to explore, such as stochastic variational inference (SVI, \cite{hoffman2013stochastic}), unless another lower bound with respect to previous lower bound can be found (e.g. as an illustrative example in Figure \ref{fig:approxR}(c), $p(R=2)$ is approximated by lower bound but $p(R=1)$ by upper bound. Strictly speaking, this approximation cannot be used as the lower bound of lower bound).

In this section, we propose a fast algorithm based on expectation-maximization. In hGCHMMs, expected sufficient statistics are computationally intractable since there is no closed form for integrating out the latent variables. Stochastic Approximation (SA) or Monte Carlo (MC) EM by \cite{delyon1999convergence,wei1990monte} is an alternative introduced to simulate the expectation, and it is able to obtain convergence to a local minimum with a theoretical guarantee under mild conditions. The basic idea is to use a Monte Carlo sampling approximation; however, we replace this step with Gibbs sampling by utilizing the approximate conjugacy property. 

{\bf E-step}: Sampling $\{\mathbf{X}^{(j)}\}_{j=1}^J$ and $\{\mathbf{R}^{(j)}\}_{j=1}^J$ follows
\begin{align}\label{eq:gibbs}
\alpha_n,\beta_n,\gamma_n&|\mathbf{Z},\bm\eta^{(k-1)}\\
\mathbf{X}&|\alpha_n,\beta_n,\gamma_n,\mathbf{Y} \nonumber \\
\mathbf{R}&|\mathbf{X},\alpha_n,\beta_n,\gamma_n,\mathbf{G} \nonumber
\end{align}
The true expectation or intractable integration $Q^{(k)}(\bm\eta)$ is approximately calculated by a stochastic averaging in a burn-in representation $\hat{Q}^{(k)}(\bm\eta)$ defined as (\ref{eq:Q}), taking advantage of Gibbs sampling. During each E-step, infection parameters are in fact always updated at each inner iteration of Gibbs Sampling, thus making the latent variables $\mathbf{X}$, $\mathbf{R}$ update based on \textit{different} posterior distributions at each sampling, which disagrees with SAEM or MCEM, sampling latent variable from a fixed distribution based on estimated parameter by previous M-step. Therefore the samples at later Gibbs sampling iterations are closer to the true posterior given current $\bm\eta^{(k-1)}$. From this perspective, the Gibbs sampling in E-step may essentially accelerate the convergence rate in the next maximization step. 

{\bf M-step}: Maximizing with respect to $\bm\eta$, i.e. $\arg\max\hat{Q}^{(k)}(\bm\eta)$. However, directly optimizing $\hat{Q}^{(k)}(\bm\eta)$ will suffer from the same drawback as in standard EM. Pathological surfaces of the log-likelihood may be present via saddle points and local optima, meaning that the algorithm is sensitive to initialization. \cite{delyon1999convergence} argued that the augmented objective function $Q_{\bm\eta}^{(k)} \triangleq (1-\delta^{(k)})Q_{\bm\eta}^{(k-1)}+\delta^{(k)}\hat{Q}^{(k)}(\eta)$ can avoid this problem partially, where $\hat{Q}^{(k)}(\bm\eta)$ usually takes few samples to introduce a stochastic property, and $\delta^{(k)}$ is a small positive step size, essentially requiring the conditions in (\ref{eq:stepsize})
\begin{align}\label{eq:stepsize}
\lim_{k\rightarrow\infty}\delta^{(k)}=0, \lim_{k\rightarrow\infty}\delta^{(k)}/\delta^{(k+1)}=1,\sum_k\delta^{(k)}=\infty 
\end{align}
The intuition to solve this intractable objective lies in \cite{celeux1995stochastic}, showing that this optimization can be updated by $\bm\eta^{(k+1)}=(1-\delta^{(k+1)})\bm\eta_{\text{EM}}^{(k+1)}+\delta^{(k+1)}\bm\eta_{\text{SEM}}^{(k+1)}$, where $\bm\eta_{\text{EM}}^{(k+1)}$ is the true EM result approximated by MCEM with large sampling size, and $\bm\eta_{\text{SEM}}^{(k+1)}$ is the special case of MCEM in very few samples or even unique one sometimes.

Generalizing this scheme to the Gibbs sampling setting, we formalize Algorithm \ref{alg:bGEM}, where $\hat{Q}_{\text{bGEM}}$ takes the sample average of the Gibbs algorithm and $\hat{Q}_{\text{SEM}}$ takes the last sample. $\hat{Q}_{\text{SEM}}$ is a stochastic perturbation of EM, and is expected to search more stable points. The algorithm starts by optimizing $\hat{Q}_{\text{SEM}}$ with $\delta^{(1)}=1$, making the search area large for the first few steps. Then it focuses more weight on optimizing $\hat{Q}_{\text{bGEM}}$. A theoretical guarantee for this algorithm can be illustrated by using two convergence bounds; Birkhoff Ergodic theory \cite{durrett2010probability} and Theorem 7 in \cite{delyon1999convergence}. 

\begin{algorithm}[tb]
 \KwData{$\mathbf{Z}$, $\mathbf{Y}$, $\mathbf{G}$, sampling size $J$, burn-in iteration $B$, step size series $\{\delta^{(k)}\}_{k=1}^{\infty}$}
 \KwResult{$\bm\eta$ and $\mathbf{X}$}
 \textbf{Initialize} coefficient parameter $\bm\eta^{(0)}$\;
 \Repeat{$\bm\eta^{(k)}$ Convergence}{
  \For{$i\leftarrow1$ \KwTo $J$}{
  sampling $\{\mathbf{X}^{(j)},\mathbf{R}^{(j)}\}_{j=1}^J$ according to (\ref{eq:gibbs})\;
  }
  \textbf{Compute }
  \begin{align}\label{eq:Q}
   \hat{Q}_{\text{bGEM}}^{(k)}(\bm\eta)&=\frac{1}{J-B}\sum_{j=B+1}^{J}\log\left(P(\mathbf{X}^{(j)},\mathbf{R}^{(j)},\bm\eta|\mathbf{Z},\bm\eta^{(k-1)})\right) \\
   \hat{Q}_{\text{SEM}}^{(k)}(\bm\eta)&=\log\left(P(\mathbf{X}^{(J)},\mathbf{R}^{(J)},\bm\eta|\mathbf{Z},\bm\eta^{(k-1)})\right)
   \end{align}
 \textbf{Optimization} 
   \begin{align*}
   \bm\eta_{\text{bGEM}}^{(k)}=\arg\max\hat{Q}_{\text{bGEM}}^{(k)}(\bm\eta)\\	
   \bm\eta_{\text{SEM}}^{(k)}=\arg\max\hat{Q}_{\text{SEM}}^{(k)}(\bm\eta);
   \end{align*}
   \textbf{Combination} $\bm\eta^{(k)}=(1-\delta^{(k)})\bm\eta_{\text{bGEM}}^{(k)}+\delta^{(k)}\bm\eta_{\text{SEM}}^{(k)}$\;
 }
 \caption{burn-in Gibbs EM Algorithm}\label{alg:bGEM}
\end{algorithm}

{\bf Faster version for binary latent variables} Because taking the first order derivative with respect to $\bm\eta$ and setting it equal to 0 will obtain a non-analytical root, gradient descent based optimization is necessary, and we adopt Newton's Method by taking the advantage of curvature information. Taking the inverse of the Hessian matrix usually requires algorithmic complexity $\mathcal{O}(K^3)$. The dimensionality of $\bm\eta$ is $K$ which is independent of HMMs scale $N$, and a PCA preprocessing will reduce it significantly, where the necessity is illustrated in experiments section. Though $K$ representing the number of temporal health feature is less scalable in most application, there may still be a high cost to computing the Hessian with $\mathcal{O}(JK^2)$ complexity due to matrix addition, unless there is a parallelized implementation with reduce operation \cite{dean2008mapreduce}. Thus we need to seek for more efficient algorithm. For Gaussian variable, \cite{price1958useful} prove a theorem to address the exchangeablity of the derivatives and expectations. \cite{rezende2014stochastic} implemented this idea in a non-Gaussian posterior likelihood and obtained good performance by approximating expectation with unique delicately designed sample. An improved SAEM coupled with MCMC is discussed in \cite{kuhn2004coupling}, which also argues that only one sample is required in the E-step if an appropriate Markov transition kernel is also used. 

Consequently, we mimic these two ideas to design our MC integration with a single sample. Technically, if we omit the probability of tracking a source belonging to both inside and outside the network, latent variable $R_{n,t}$ can be considered as binary variable as well. Then, we can use the posterior mean of the latent variable as the sample we have been looking for. Therefore, at the $k$th iteration of EM, the pseudo-sample is constructed via a Bayesian decision rule based on the burn-in posterior mean in Gibbs sampling, i.e. $\hat{x}_{n,t}=\mathbb{I}{\{\frac{1}{J-B}\sum_{j=B+1}^Jx_{n,t}^{(j)}>0.5\}}$ and $\hat{R}_{n,t}=\mathbb{I}{\{\frac{1}{J-B}\sum_{j=B+1}^JR_{n,t}^{(j)}>0.5\}}$. This means that a unique set $(\hat{\mathbf{X}},\hat{\mathbf{R}})$ is sufficient to approximate $\hat{Q}^{(k)}(\bm\eta)$, that is to say, $\log(P(\hat{\mathbf{X}},\hat{\mathbf{R}},\bm\eta)|\mathbf{Z},\bm\eta^{(k-1)})$ substitutes for $\hat{Q}_{\text{bGEM}}^{(k)}$. This trick applied on non-Gaussian variables is not theoretically guaranteed but has been broadly used in EM or other optimization problems, by assuming a fully factorized joint distribution. In our binary variable case we found that it made no significant difference on accuracy whenever this trick is applied, in practice.

{\bf Optimization} To optimize $\bm\eta_{\text{*EM}}^{(k)}$ at the $k$th M-step, the update formula by the Newton-Raphson Method is briefly outlined in this paragraph, excluding the analytical gradient $G$ and Hessian $H$ computation. For efficiency, we update parameters as follows, with a few iterations. \[\bm\eta_{\text{*EM}:new}^{(k)}=\bm\eta_{\text{*EM}:old}^{(k)}-\delta H^{-1}G\] where *EM varies according to different estimators, bGEM or SEM. It is unnecessary for there to be complete convergence in order to guarantee $Q(\bm\eta^{(k)})>Q(\bm\eta^{(k-1)})$. A similar idea with a single iteration is mentioned in \cite{lange1995gradient}. The step size $\delta$ ensures that the Wolfe conditions (\cite{nocedal2006numerical}) are satisfied. The intuition in adding in step size here is, compared with gradient descent, is that Newton's Method tends to make more progress in the right direction of the local optima, due to the property of affine invariance. This probably leads to an update where the step size is too large, so it is better for stochastic algorithms to enlarge the search domain at first then shrink later.

\subsection{Short Discussion on Sigmoid link}
A sigmoid link function benefits from model simplicity and hiding the infection parameters without the necessity to integrate them out. The likelihood $P(\bm\eta,\mathbf{X}|\mathbf{Z})$ shown in (\ref{eq:sig}) can thus be exactly computed. It means that we can estimate parameters throughout either standard SAEM by getting rid of latent variable $\mathbf{R}$ immediately, or bGEM by introducing $\mathbf{R}$ in E-step and a faster M-step by keeping $\hat{\mathbf{X}}$ alone.
\begin{align}\label{eq:sig}
&P(\bm\eta)\prod_{n=1}^NP(x_{n,1}) \times \prod_{n=1}^N\prod_{t=1}^{T-1} \sigma(\mathbf{z}_n^\top\bm\eta_r)^{\mathbb{I}_{\{x_{n,t}=1,x_{n,t+1}=0\}}}\left(1-\sigma(\mathbf{z}_n^\top\bm\eta_r)\right)^{\mathbb{I}_{\{x_{n,t}=1,x_{n,t+1}=1\}}} \\
&\cdot \left(1-(1-\sigma(z_n^\top\eta_a))(1-\sigma(\mathbf{z}_n^\top\bm\eta_b))^{C_{n,t}}\right)^{\mathbb{I}_{\{x_{n,t}=0,x_{n,t+1}=1\}}} \cdot \left((1-\sigma(\mathbf{z}_n^\top\bm\eta_a))(1-\sigma(\mathbf{z}_n^\top\bm\eta_b))^{C_{n,t}}\right)^{\mathbb{I}_{\{x_{n,t}=0,x_{n,t+1}=0\}}} . \nonumber
\end{align} 

\subsection{Further Discussion on $\beta_n$}
In the second biological interpretation of $\beta_n$ (the probability of infecting others), transition function $\phi_{n,x_{n'}:(n,n')\in G_t}$ will become dependent on a parameter set $\{\beta_{n'}:n'\in S_{n,t}\}$. Consequently, the posterior of each $\beta_n$ requires both a count number and source tracking (conceptually, this is like a "pointer" in the C programming language). However, the likelihood of the beta-exponential model can be simplified to integrate out these parameters due to the auxiliary variable $R_{n,t}$ as well, corresponding to a new approximate categorical distribution, though $P(R_{n,t})$ in previous (\ref{eq:approxR}) actually aggregates the probability with respect to all equal $\beta_n$s. The new categorical distribution and its induced completed likelihood can be represented as follows.
\begin{align}\label{eq:approxR2}
&P(R_{n,t})\approx \text{Categorical}\left(\frac{\alpha_n\prod_{n'\in S_{n,t}}(1-\beta_{n'})}{1-(1-\alpha_n)\prod_{n'\in S_{n,t}}(1-\beta_{n'})},\frac{(1-\alpha_n)\beta_{n'}}{1-(1-\alpha_n)\prod_{n'\in S_{n,t}}(1-\beta_{n'})},\ldots\right) \\
&P(\mathbf{X},\mathbf{R},\bm\eta|\mathbf{Z})=P(\bm\eta)\prod_n\left(\frac{B(e^{\mathbf{z}_n^\top\bm\eta_{r,1}}+C_{n,1\rightarrow0},e^{\mathbf{z}_n^\top\bm\eta_{r,2}}+C_{n,1\rightarrow1}}{B(e^{\mathbf{z}_n^\top\bm\eta_{r,1}},e^{\mathbf{z}_n^\top\bm\eta_{r,2}})}\right. \label{eq:PR2}\\
&\left. \cdot\frac{B(e^{\mathbf{z}_n^\top\bm\eta_{a,1}}+C_{n,R_n=0},e^{\mathbf{z}_n^\top\bm\eta_{a,2}}+C_{n,R_n\neq0}+C_{n,0\rightarrow0}}{B(e^{\mathbf{z}_n^\top\bm\eta_{a,1}},e^{\mathbf{z}_n^\top\bm\eta_{a,2}})} 
\cdot \frac{B(e^{\mathbf{z}_n^\top\bm\eta_{b,1}}+C_{n,R=n},e^{\mathbf{z}_n^\top\bm\eta_{b,2}}+C_{n,R\neq n}}{B(e^{\mathbf{z}_n^\top\bm\eta_{b,1}},e^{\mathbf{z}_n^\top\bm\eta_{b,2}})}\right) \nonumber
\end{align}
where $R_{n,t}$ takes the value $\{0,1,...,C_{n,t}\}$, and 0 means there is an outside network source and other integers refer to specific infection in-network sources. The categorical distribution makes the beta prior for the infection parameters conjugate in the posterior. However, the integral for the likelihood is actually difficult and needs some tricks, especially for $\beta_n$ because of the source tracking (see Appendix B for details). Note the likelihood for sigmoid link can be derived analogously. The new count notations are listed below.
\begin{table}[htb]
\centering
\begin{tabular}{l} \hline
$C_{n,R_n=0}=\sum_t\mathbb{I}_{\{R_{n,t}=0\}}$ \\ \hline
$C_{n,R_n\neq0}=\sum_t\sum_{n'\in S_{n,t}}\mathbb{I}_{\{R_{n,t}=n'\}}$ \\ \hline
$C_{n,R=n}=\sum_{n',t:n\in S_{n',t}}\mathbb{I}_{\{R_{n',t}=n\}}$\\ \hline
$C_{n,R\neq n}=\sum_{n',t:n\in S_{n',t}}[\mathbb{I}_{\{R_{n',t}=0\}}+\mathbb{I}_{\{x_{n',t}=0,x_{n',t+1}=0\}}]$ \\
\hline\end{tabular}
\label{tab:addnote}
\end{table}

\section{Experimental Results}
\label{sec:experiment}

In this section we illustrate, on simulated data, the performance of our approach, hGCHMMs and the burn-in Gibbs EM algorithm on three datasets for the purposes of predicting the hidden infectious state matrix $\mathbf{X}$, filling in missing data -- observations $\mathbf{Y}$, and inferring an individual's physical condition based on parameter estimation. Further application on the public real world Social Evolution Dataset \cite{madan2012sensing} and our mobile apps survey dataset are also shown. 

\subsection{Semi-Simulation Dataset}

\subsubsection{Data Generation}
Differing from completely simulated data or a totally artificial setting, we employed a generative model to synthesize $\mathbf{X}$,$\mathbf{Y}$ based on the real dynamic social network $G_t$ and covariates $Z$ from the real Social Evolution dataset. The predefined $\mathbf{X}$ then plays the role of ground truth, making evaluation for all above points possible.

\begin{figure}[t]
\centering
\subfigure[Synthesized $\mathbf{X}$]{
\includegraphics[width=45mm,height=30mm,clip,trim=20 30 20 0mm]{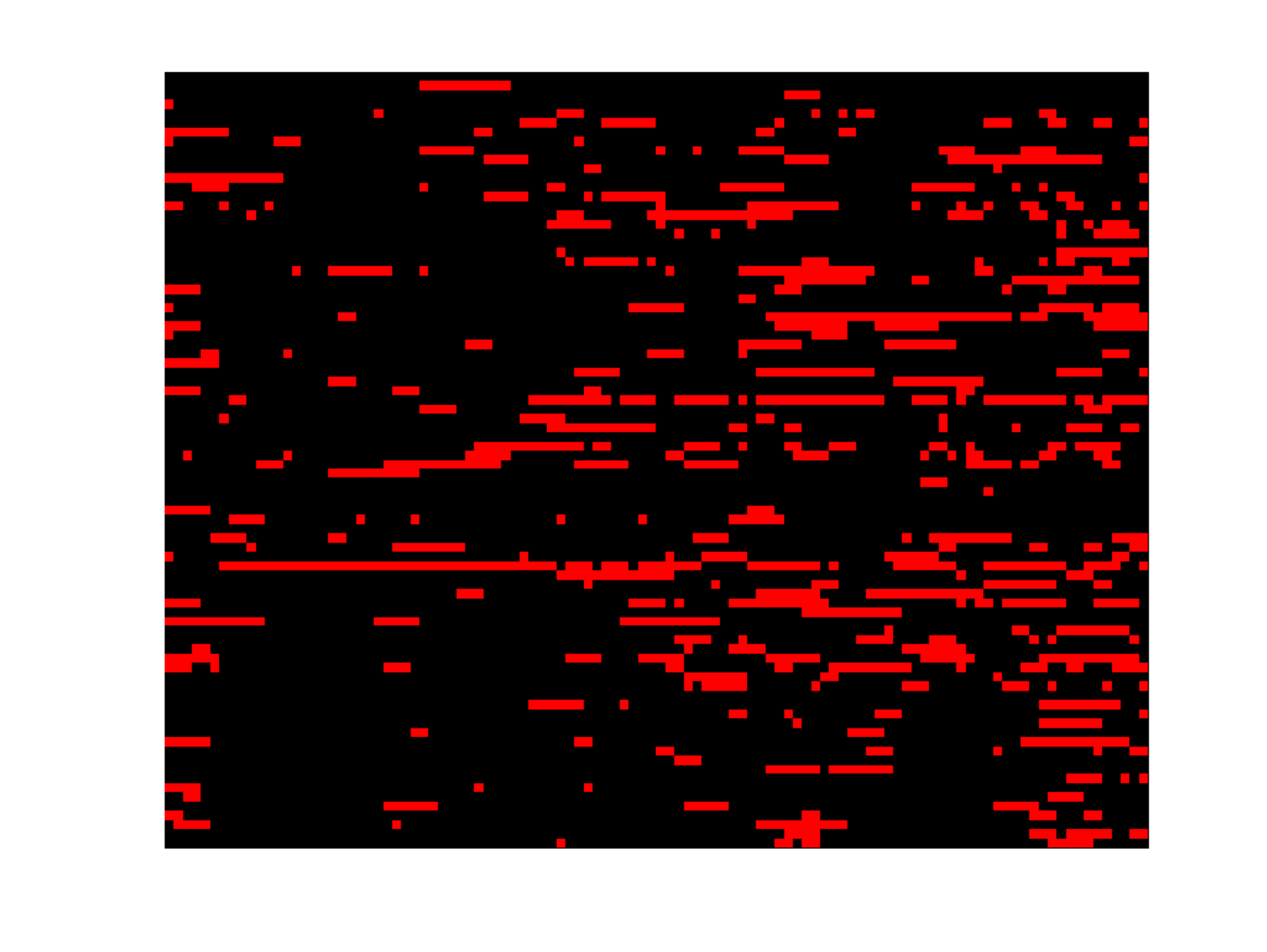}
}
\subfigure[$\hat{\mathbf{X}}$ by Sigmoid link]{
\includegraphics[width=45mm,height=30mm,clip,trim=20 30 20 0mm]{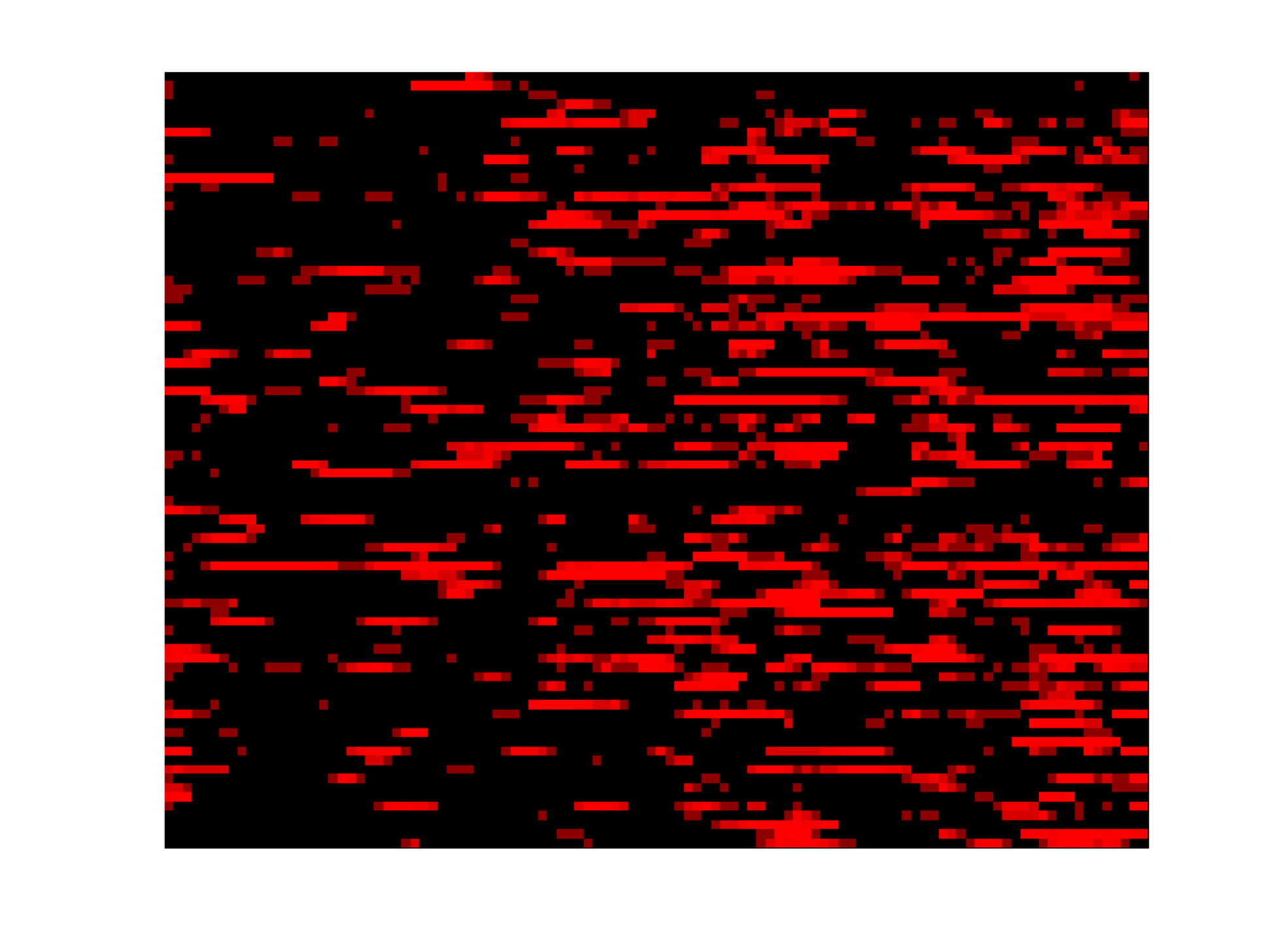}
}
\subfigure[$\hat{\mathbf{X}}$ by Beta-exp link]{
\includegraphics[width=45mm,height=30mm,clip,trim=20 30 20 0mm]{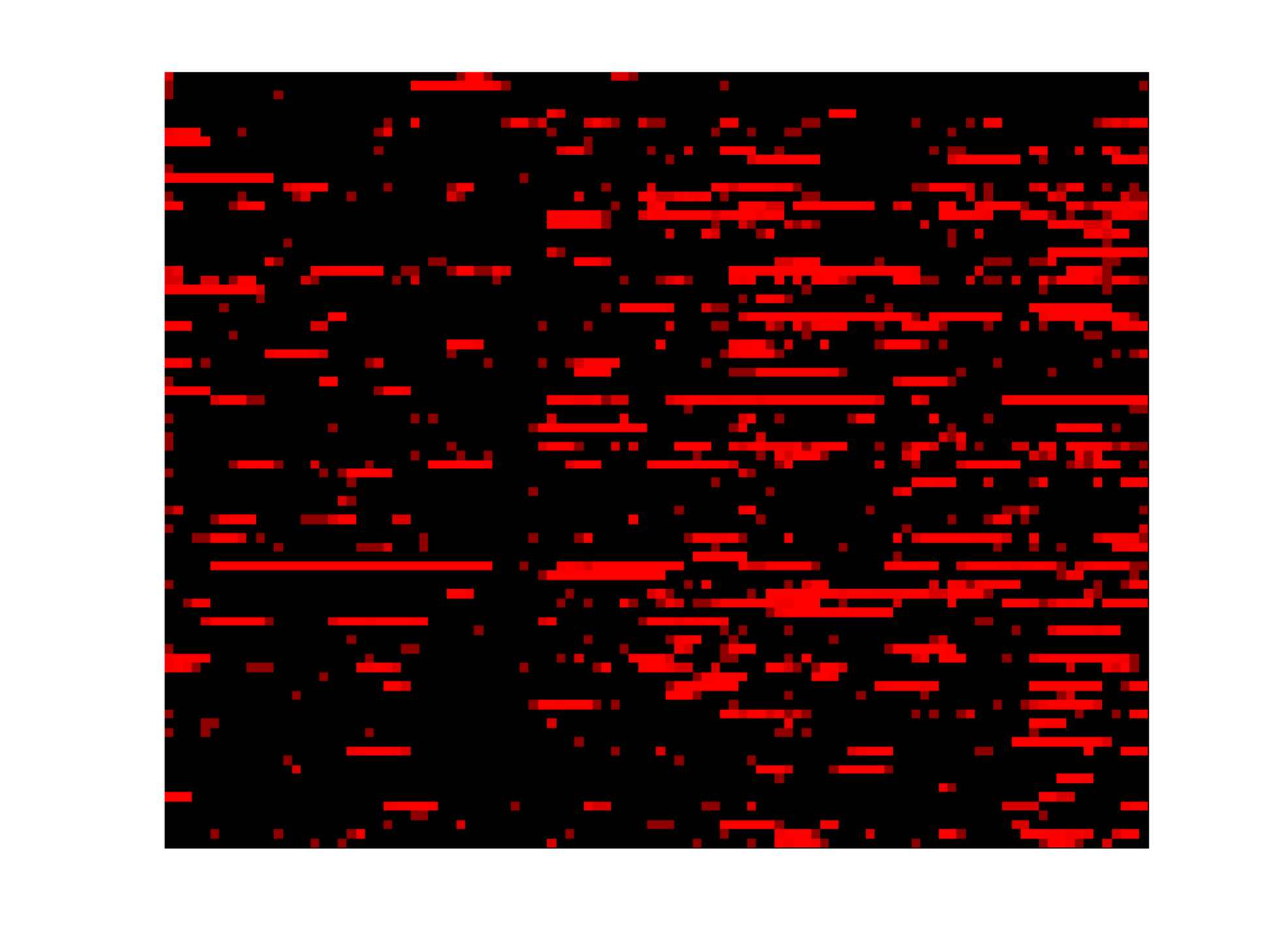}
}
\centering
\caption{Infection state prediction is visualized by a heatmap. For (b) and (c), the redder the point is, the larger the probability of getting infected is.}
\label{fig:predX}
\end{figure}

{\bf Real Data} The public MIT Social Evolution dataset contains dynamic networks $\mathbf{G}_{84\times84\times107}$ including 84 participants over 107 days, $G_t$ indexes 1 to 107, and  covariates $z_n$ exist for each participant, including 9 features, weight, height, salads per week, veggies fruits per day, healthy diet level, aerobics per week, sports per week, smoking indicator, and default feature 1. The quantity per week is frequency. Weight and height are taken as real values. Healthy diet includes 6 levels ranging from very unhealthy to very healthy based on self evaluation. The smoking indicator is literally a binary variable. Real symptoms $\mathbf{Y}$ are temporarily discarded since the true infection states $\mathbf{X}$ are unavailable for this dataset.

{\bf Synthesized Data} $\mathbf{X}$ and $\mathbf{Y}$ are then generated based on our proposed generative model. Hyperparamter $\bm\eta$ needs to be predefined, which means synthesized infection parameters $\gamma_n,\alpha_n,\beta_n$ are known because of the sigmoid function for data simulation. Only synthesized data $\mathbf{Y}$ with 6 symptoms is given to the model, but the evaluation is done on other variables. The proportion of missing values in $Y$ is set to 0.5, i.e, the observations $y_{n,t,s}$ are \texttt{NA} with probability 0.5. Our generated $\mathbf{X}_{84\times108}$ (including initial states at timestamp 0) is shown in Figure \ref{fig:predX}(a). Each row vector represents a person's infection states during the entire observation period. 

\subsubsection{Model Evaluation}
\begin{figure}[t]
\centering
\subfigure[]{
\includegraphics[width=65mm,clip,trim=5 30 20 0mm]{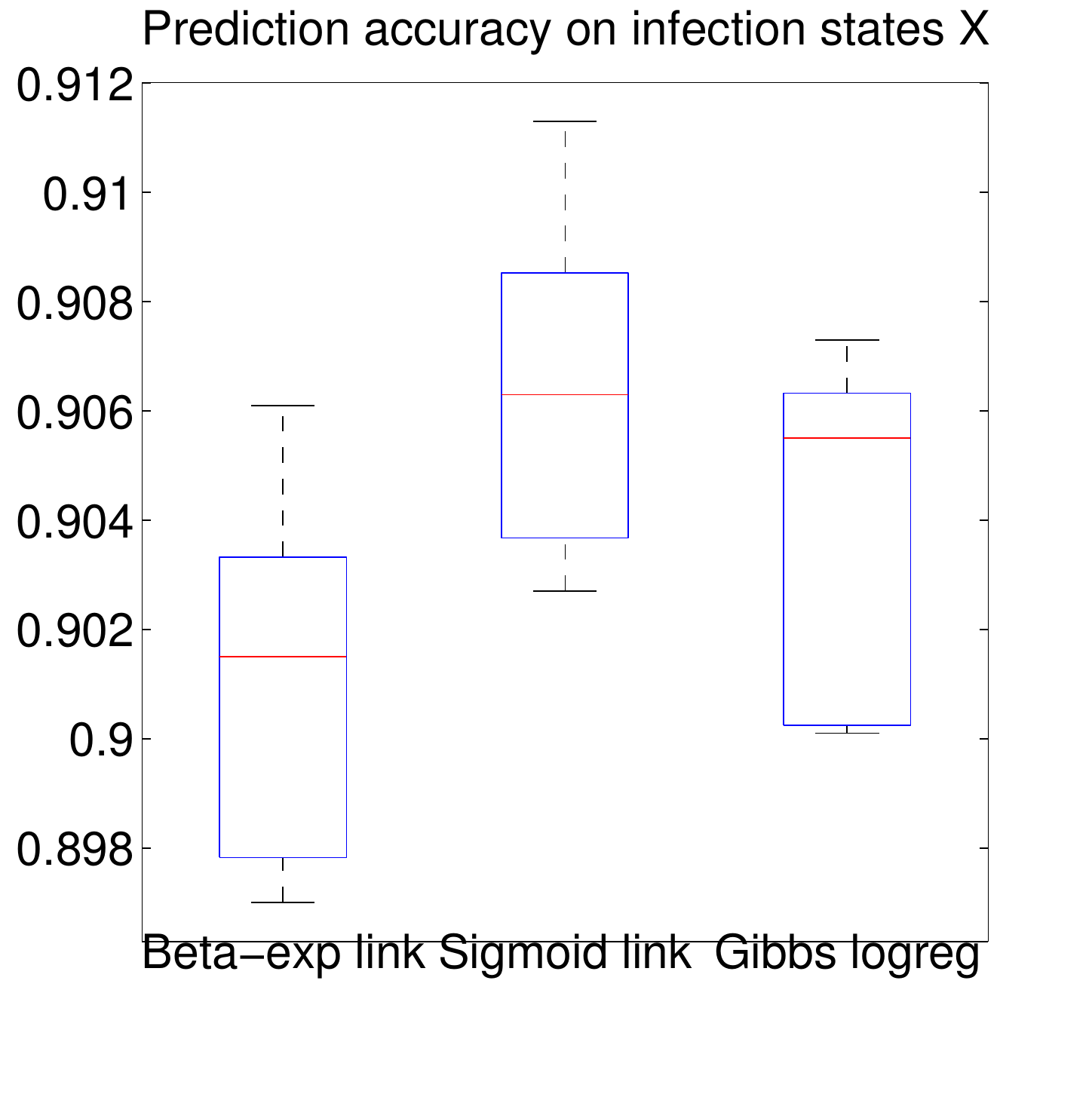}
}
\subfigure[]{
\includegraphics[width=65mm,clip,trim=5 30 20 0mm]{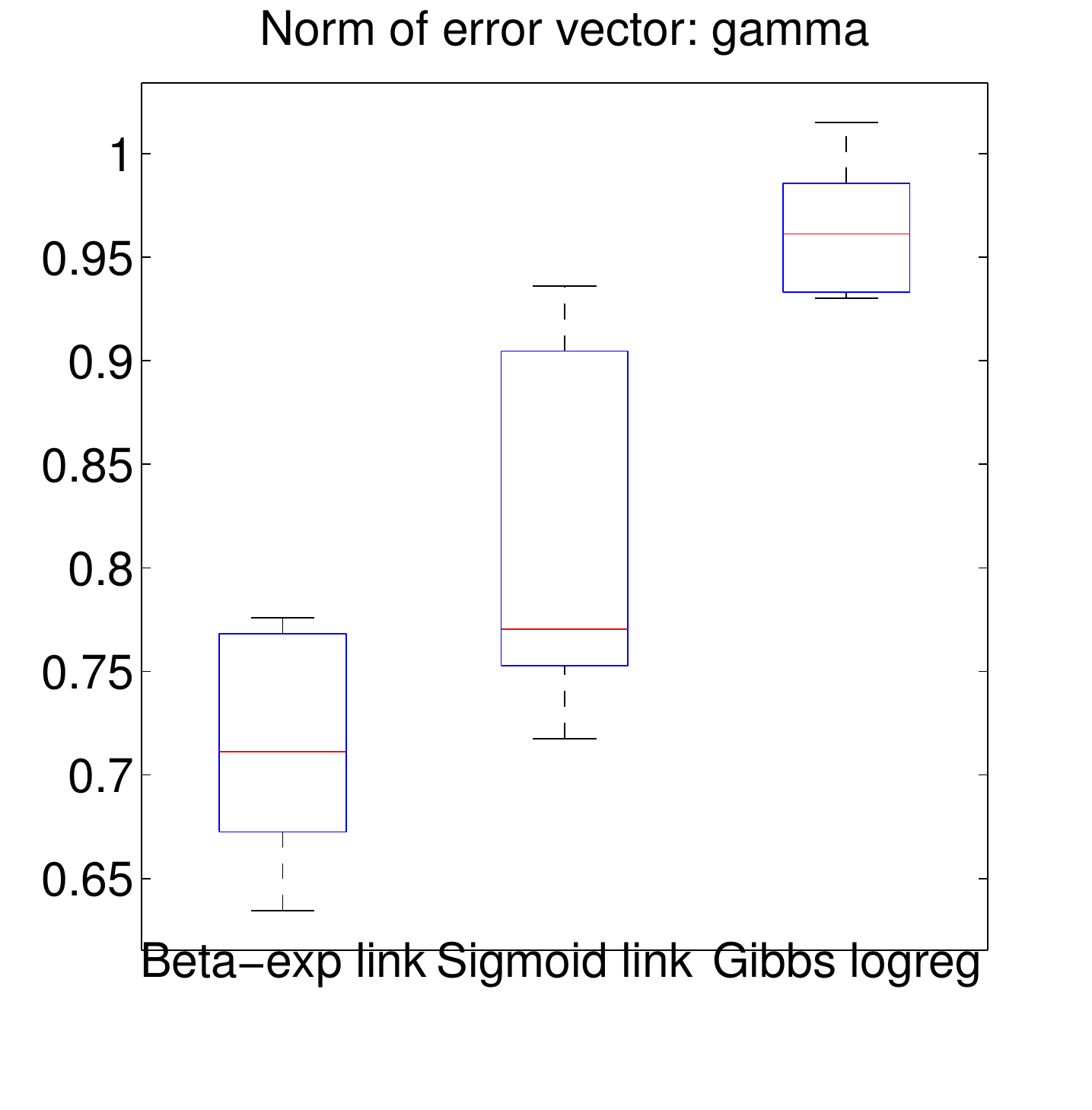}
}
\subfigure[]{
\includegraphics[width=65mm,clip,trim=5 30 20 0mm]{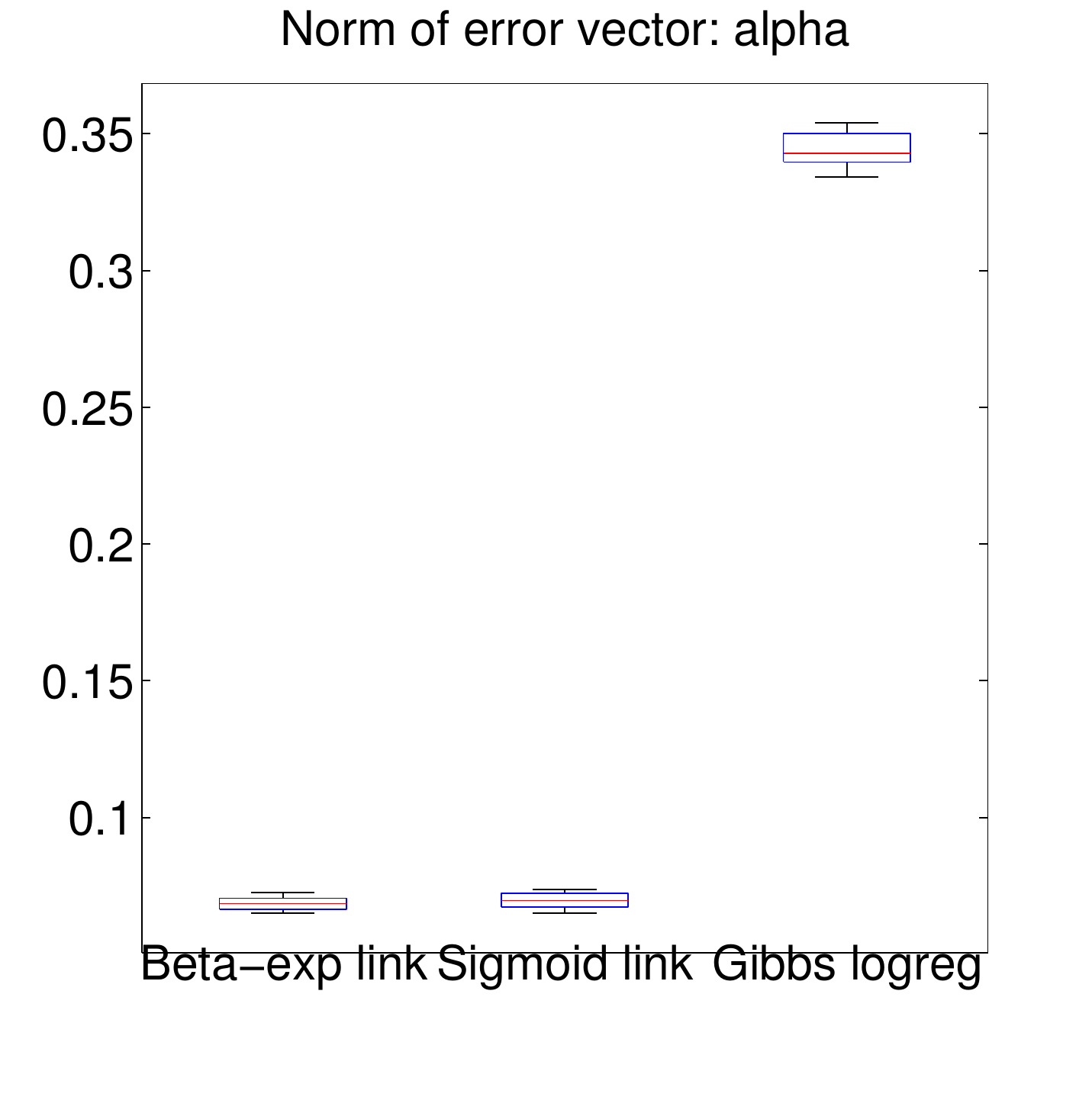}
}
\subfigure[]{
\includegraphics[width=65mm,clip,trim=5 30 20 0mm]{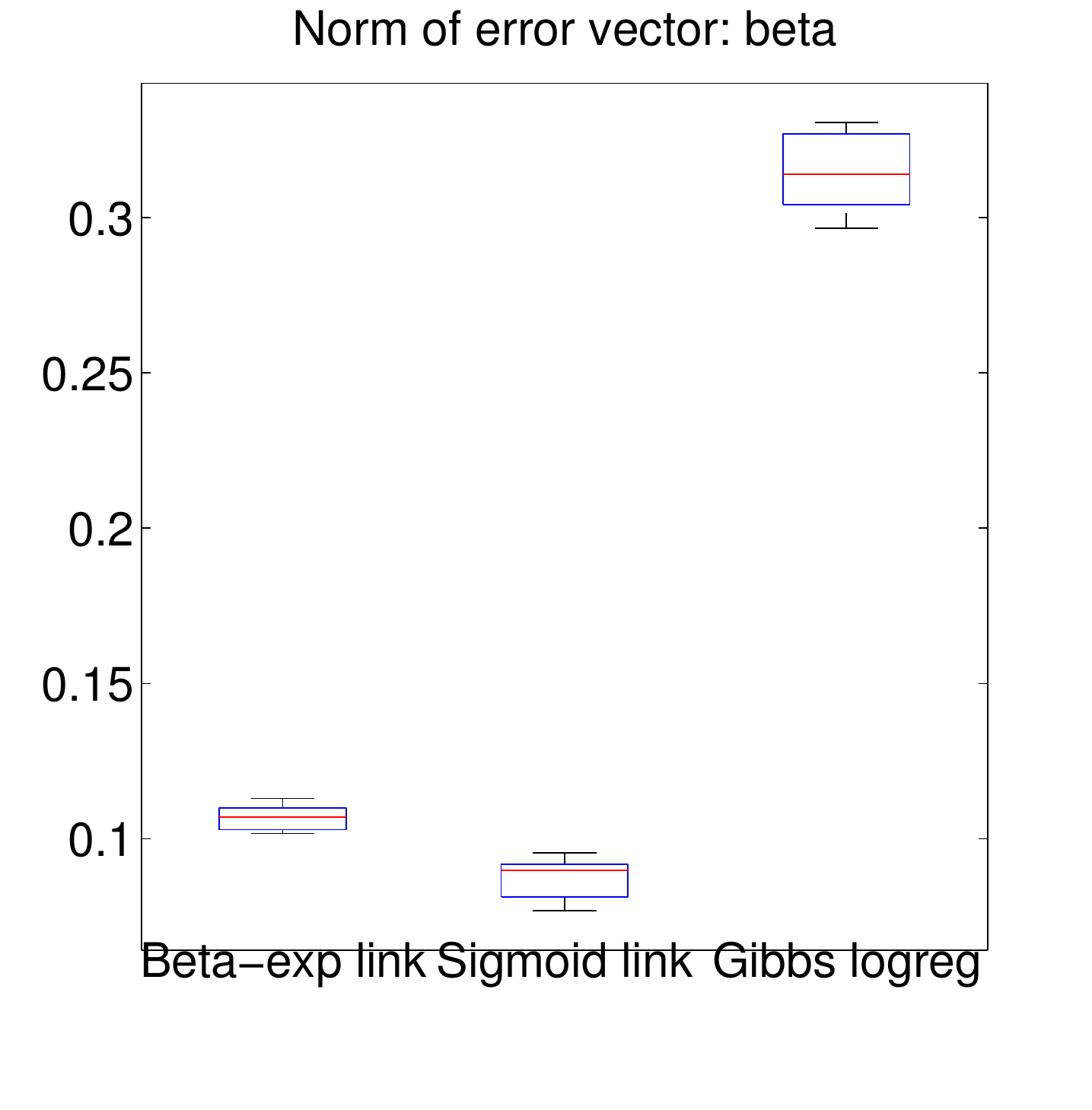}
}
\centering
\caption{(a) Accuracy comparison on latent variable prediction; (b-d) Error bar comparison on parameter estimation; Notice the $N$-dimensional error vector is normed to a scalar for statistical comparison.}
\label{fig:error}
\end{figure}

We ran the algorithm 10 times. The prediction performance on latent variable $X$ is the byproduct of the E-step, and when $x_{n,t}$ is larger than the threshold $0.5$ the person is diagnosed as being infected. Since $\mathbf{X}$ is completely unknown to the algorithm, held-out test data prediction is unnecessary, while the whole matrix $\mathbf{X}$ is used to evaluate prediction accuracy. Figure \ref{fig:predX}(a-c) shows the difference between the truth and the inferred results for each of the two linked models. The posterior mean from Gibbs sampling for prediction, in both beta-exponential and sigmoid models, leads to a real value in the interval $[0,1]$. Figure \ref{fig:error}(a) reveals a quantitative measurement on accuracy with a standard deviation. As mentioned before, the \textit{baseline} model is a two-step algorithm including the standard GCHMM and logistic regression is also implemented and compared to. The rightmost barplot in Figure \ref{fig:error}(a) shows its predictive performance. The GCHMM needs to run at least 2000 iterations of Gibbs sampling to obtain good mixing, while in our approach, we only run about 50 inner Gibbs sampling iterations in E step and less than 10 outer EM iterations. 

Figure \ref{fig:error}(b-d) displays the predictive error of the forecasted infection parameters. Since the infection parameters are individual specific, the estimation is in fact a vector of length $N$. Therefore we used the 2-norm of the error vector for statistic summarization and further comparison. It is apparent that the sigmoid model gives the best performance on latents $\mathbf{X}$ or $\gamma_n,\alpha_n,\beta_n$, in terms of the generative model. The defined Beta-exponential Model, as a probabilistic substitute to the approximate sigmoid transformation on infection parameters, proves it is competitive for parameter estimation. However, standard GCHMM with logistic regression, as two independent parts of the sigmoid model, provides an unreliable prediction on individual-specific parameters, albeit its excellent latent variable inference. All three inference methods use Gibbs sampling to infer the posterior mean of $\mathbf{X}$. This is most likely the reason why they share similar performance.

\begin{figure}[t]
\centering
\subfigure[$\gamma_n$]{
\includegraphics[width=65mm,clip,trim=0 30 0 0mm]{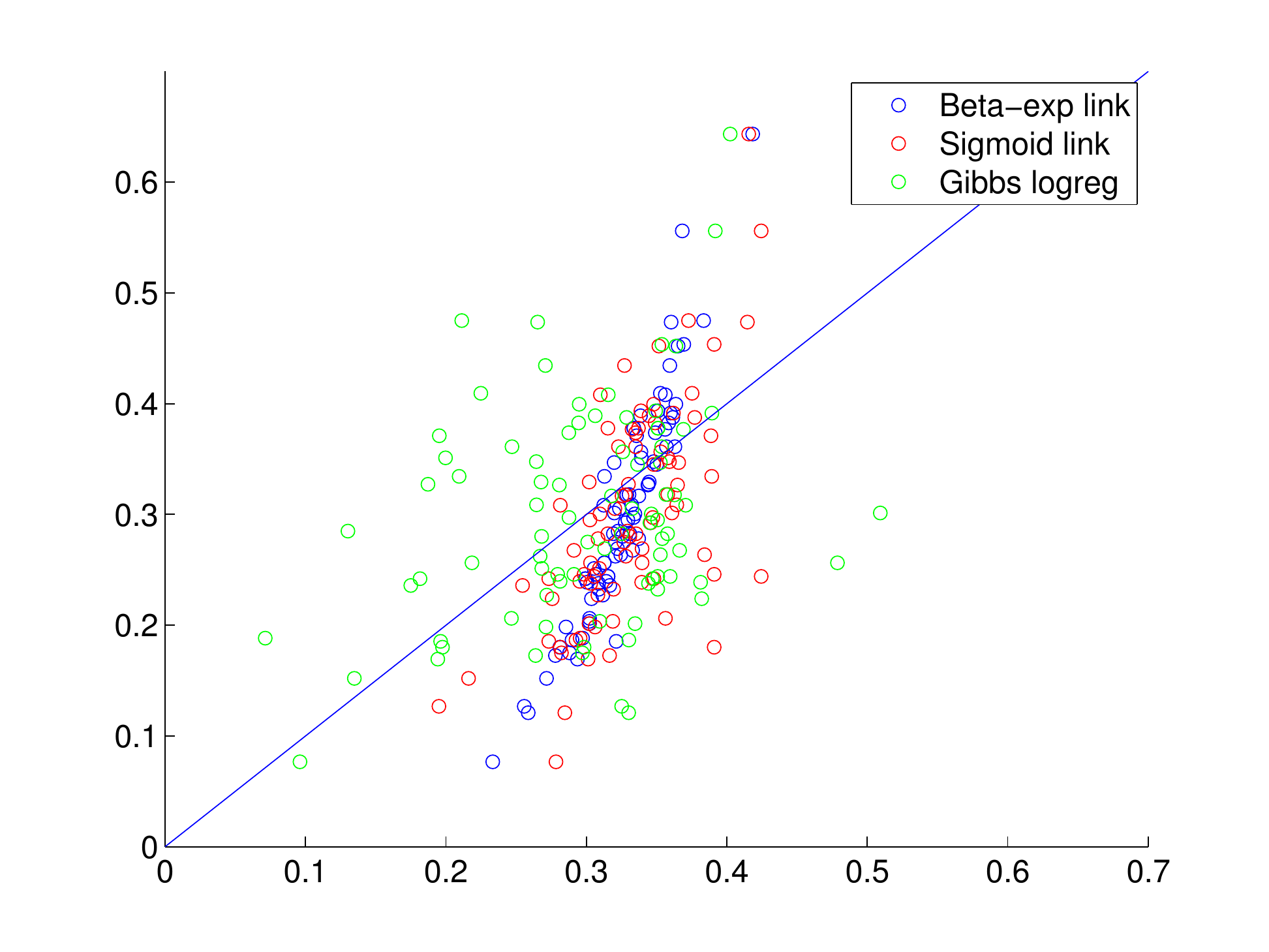}
}
\subfigure[$\alpha_n$]{
\includegraphics[width=65mm,clip,trim=0 30 0 0mm]{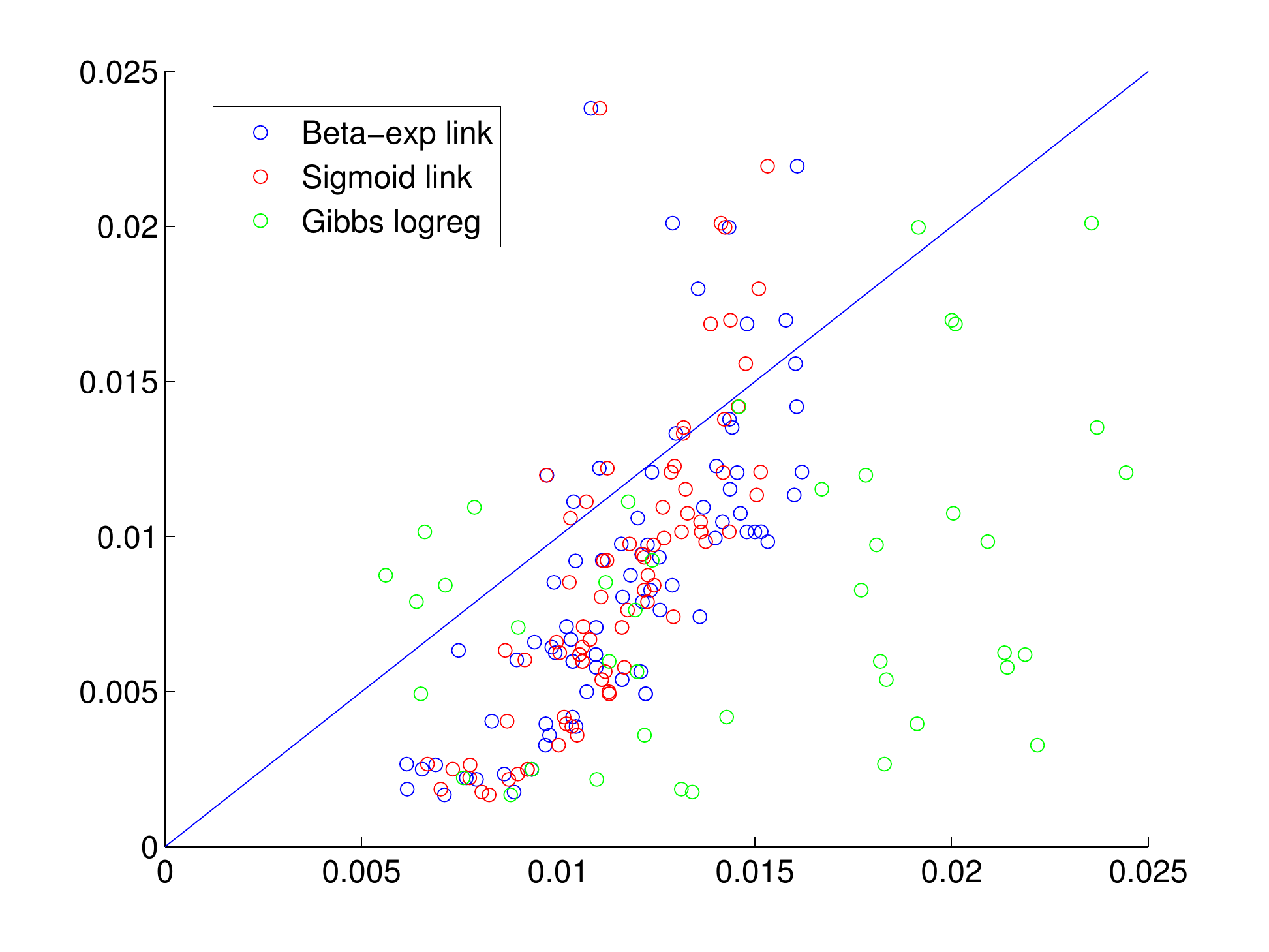}
}
\subfigure[$\beta_n$]{
\includegraphics[width=65mm,clip,trim=0 30 0 0mm]{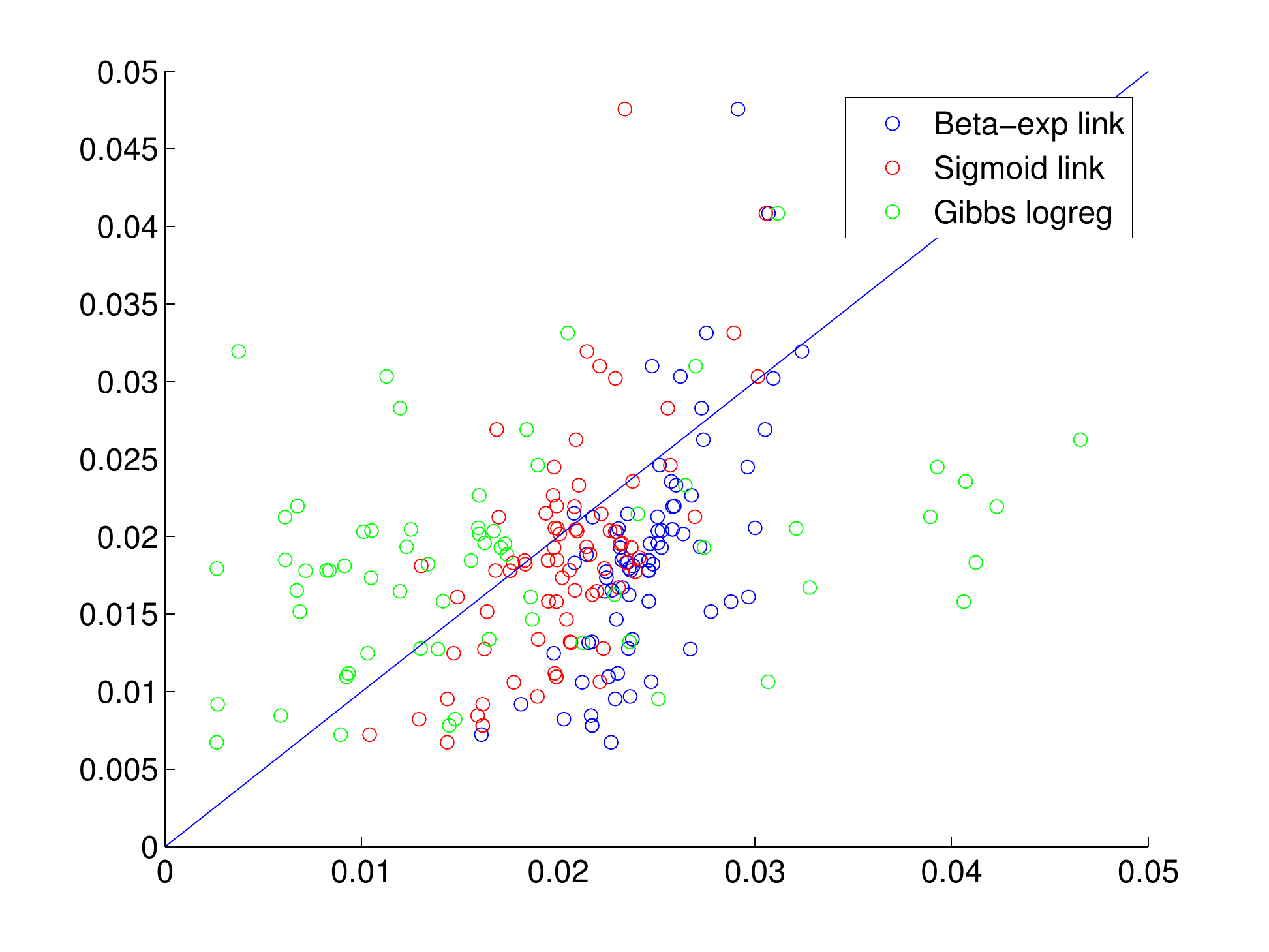}
}
\subfigure[PCA Justification]{
\includegraphics[width=65mm,clip,trim=0 30 0 0mm]{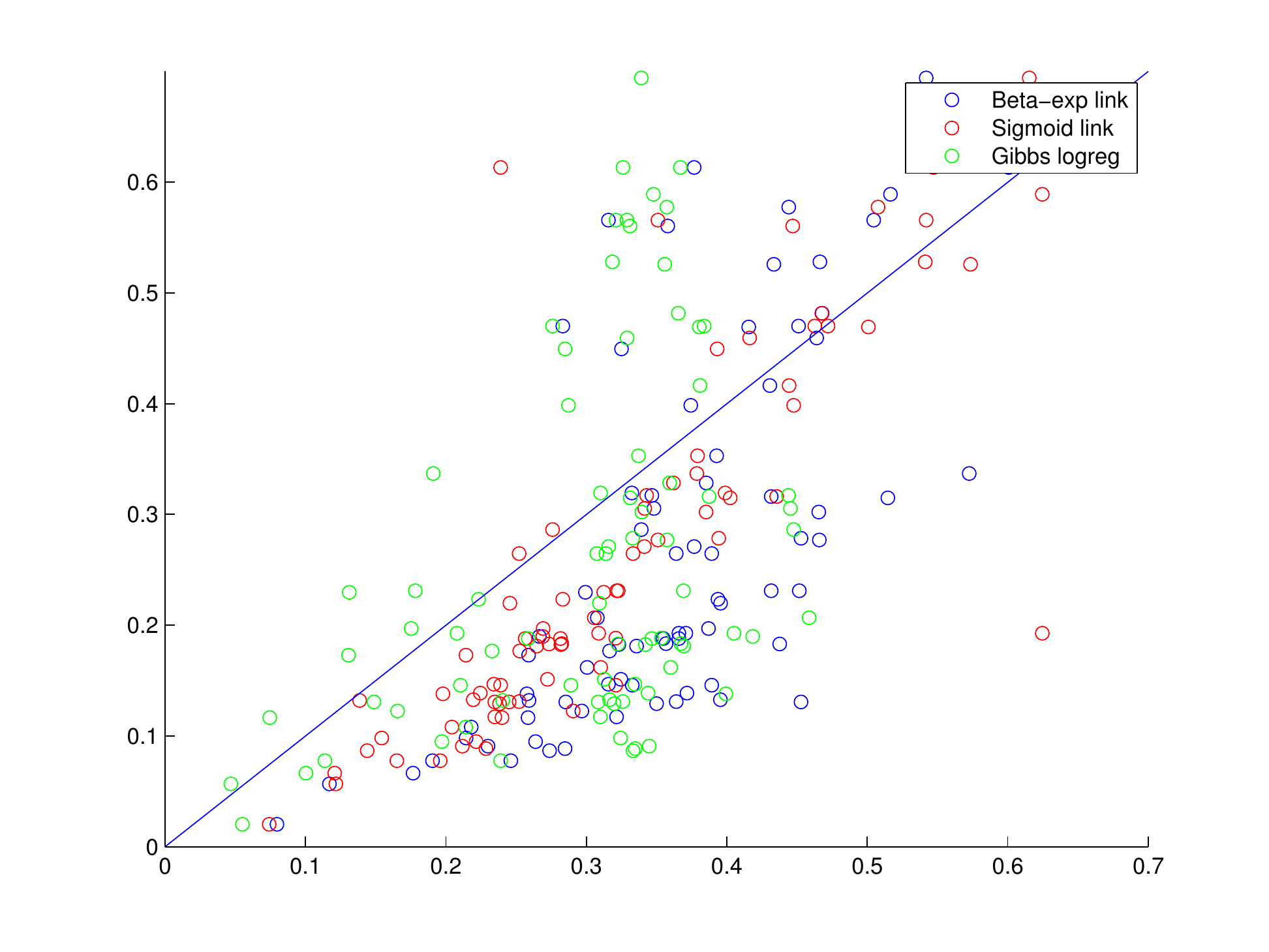}
}
\centering
\caption{ Scatter plot illustration of (e-g) Horizontal axis: estimation; Vertical axis: ground truth.Colinearity elimination on $\gamma_n$: PCA justification $\gamma$ is to obtain the regressed slope close to 1.}
\label{fig:scatter}
\end{figure}

\subsubsection{Individual Parameter Analysis}

From the perspective of general health care or disease control for large populations, $\eta$ is of concern (discussion on a real biological dataset later). However, as for individual treatment and personal medical advice, $\gamma_n,\alpha_n,\beta_n$ are more significant for physical health. Better immunity usually indicates a smaller $\alpha_n,\beta_n$ but a larger $\gamma_n$. In our model, these parameters are designed to correlate with personal health habits by using a link with influence coefficient $\bm\eta$. The prediction of the infection parameters on raw data $\mathbf{Z}$ is shown in Figure \ref{fig:scatter}(a-c). This illustration is consistent with the error bar plot in Figure \ref{fig:error}(b-d). The predicted values of our proposed models are distributed with higher concentration on the diagonal of 2D-coordinate plane, while standard GCHMM + logistic regression has relatively larger variance. The underlying linear slope for $\gamma_n$ seems inconsistent with the diagonal. This phenomenon can be blamed on the colinearity of $\mathbf{Z}$. Looking at the names of the covariates (BMI\footnote{Body mass index$=\frac{m}{h^2}$, where $m$ is mass/kg and $h$ is heigh/m.}, weight, height, salads per week, veggies and fruits per week, healthy diet or not, aerobic per week, sports per week, smoking or not), correlation obviously exists. Thus, we apply Principal component analysis (PCA) on $\mathbf{Z}$, and then select the first 4 main components (explanatory power 99.9\%) and the default feature 1. We next obtain the scatter plot of PCA justified $\gamma_n$ in Figure \ref{fig:scatter}(d). Results imply that PCA can eliminate colinearity effectively; however, the interpretability may not as effortless as Figure \ref{fig:scatter}(a).

\subsection{MIT Social Evolution Dataset}
\begin{figure}[t]
\centering
\subfigure[$y_{\cdot,5}$]{
\includegraphics[width=10mm,height=30mm,clip,trim=0 30 0 0mm]{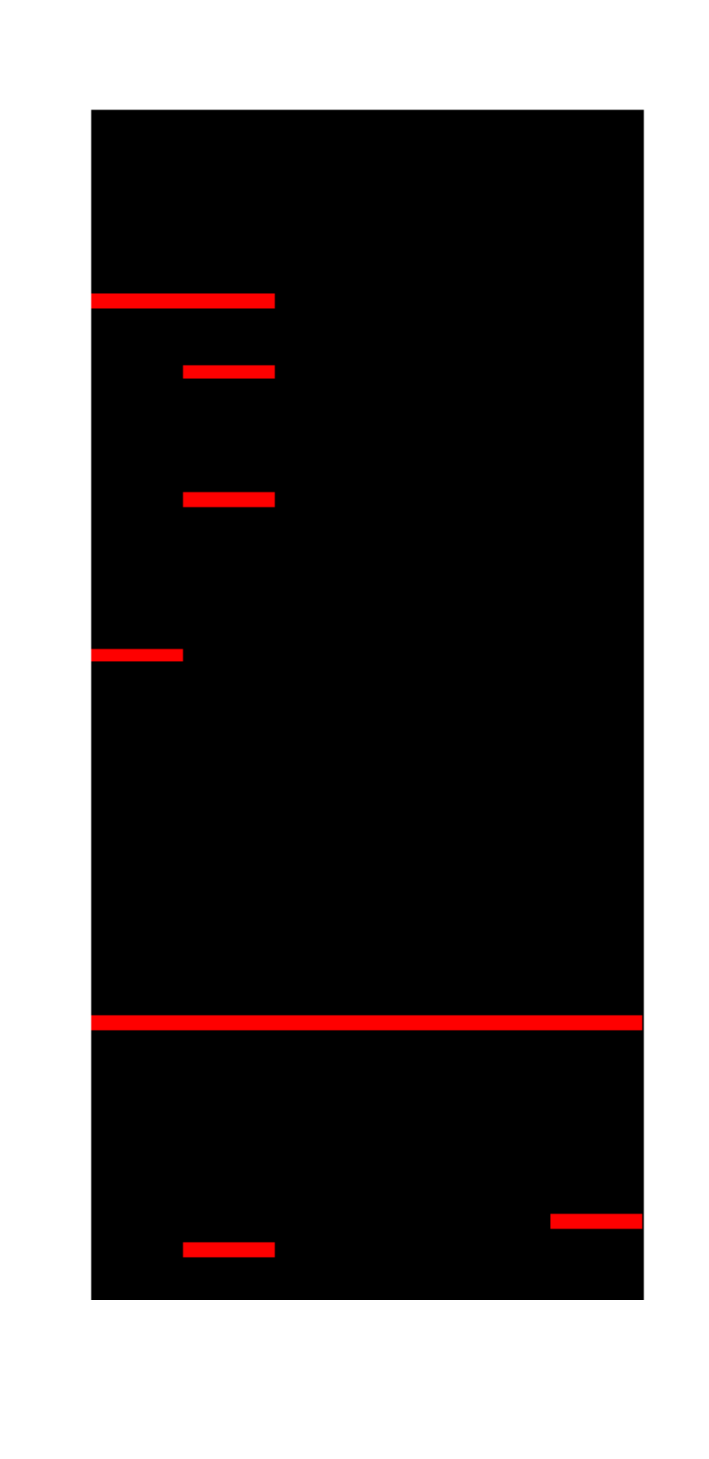}
}
\subfigure[$\hat{y}_{\cdot,5}$]{
\includegraphics[width=10mm,height=30mm,clip,trim=0 30 0 0mm]{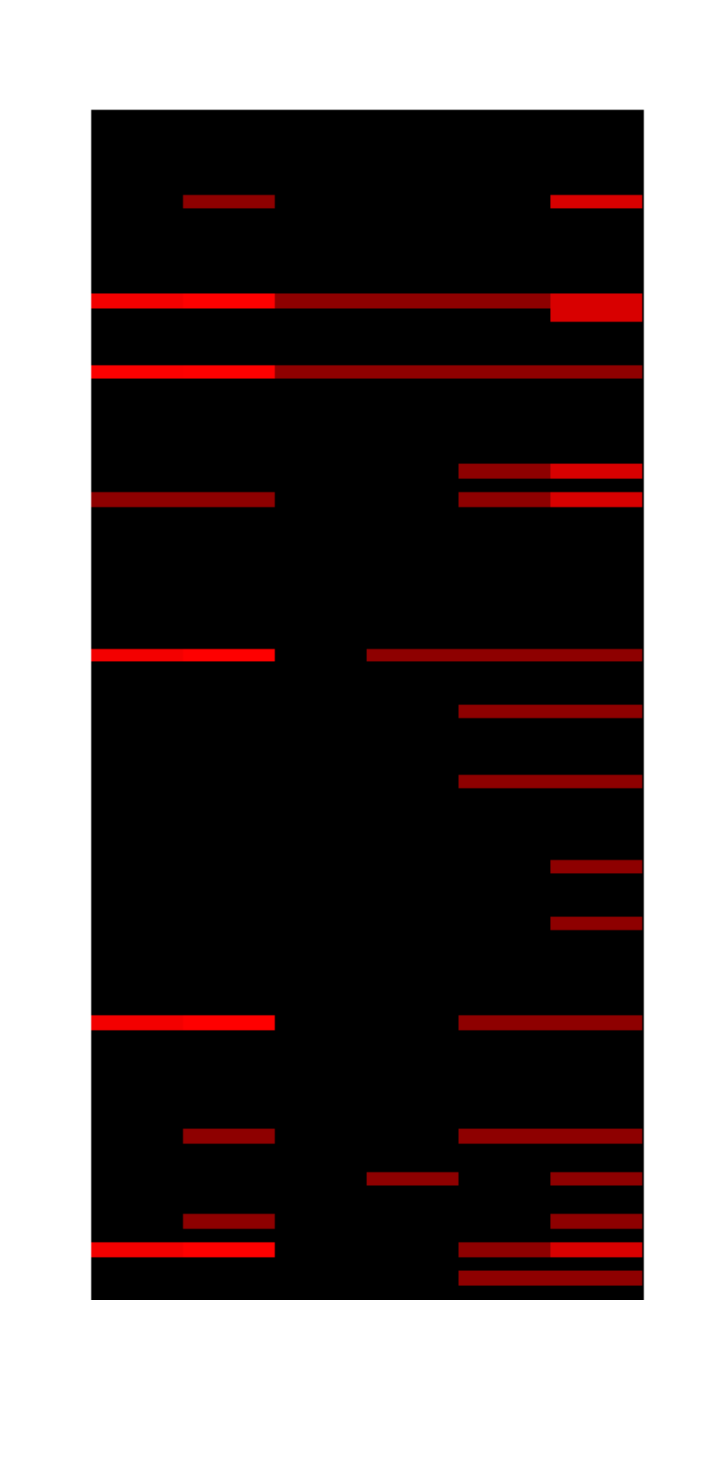}
}
\subfigure[Predicted $X$]{
\includegraphics[width=50mm,height=30mm,clip,trim=0 30 0 0mm]{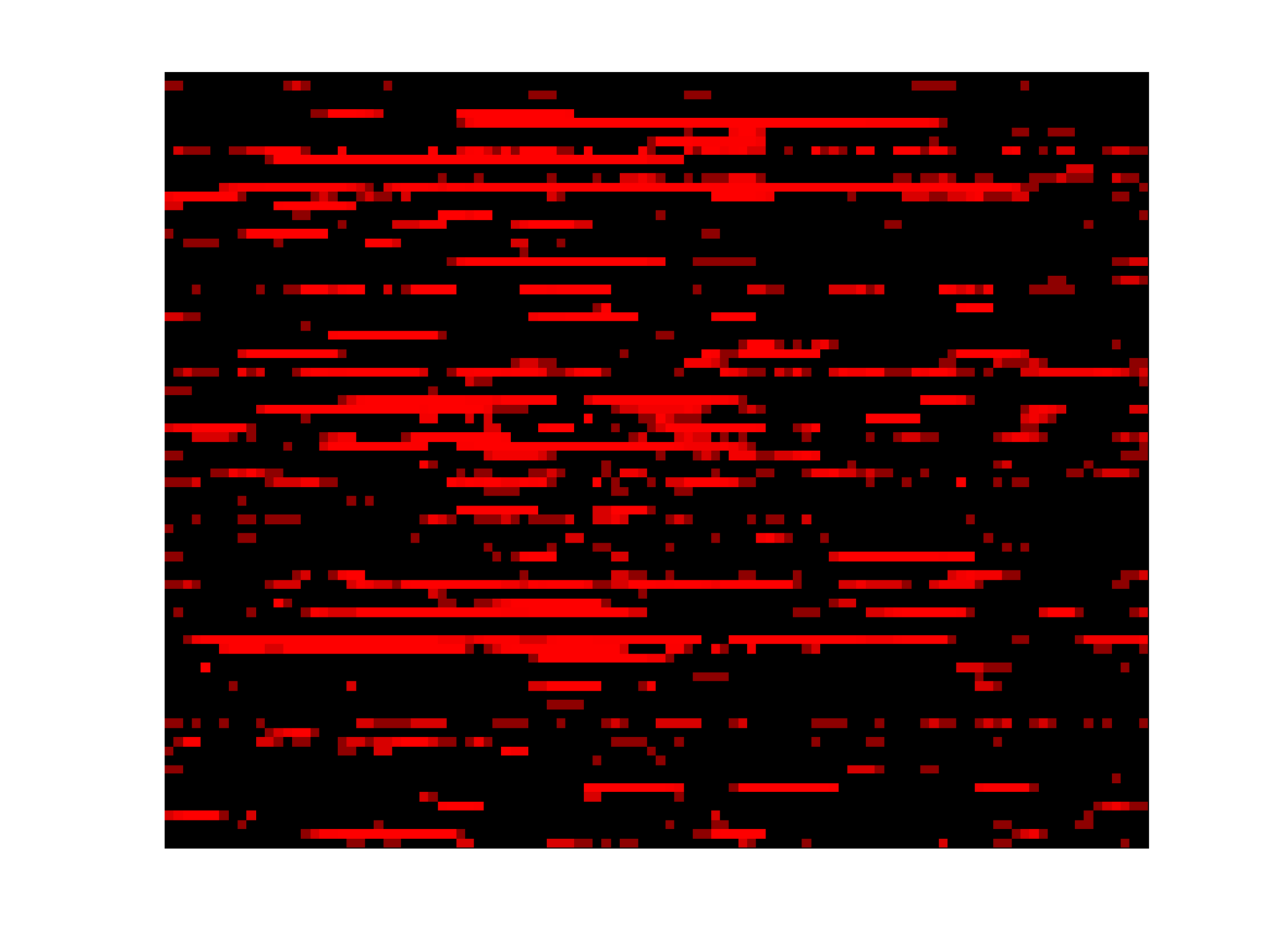}
}
\centering
\caption{Epidemic state inference on real Data: (a) shows the true reported symptoms by 84 participants at day 5; (b) gives the one step ahead prediction of (a); (c) is predicted infection.}
\label{fig:mitdata}
\end{figure}

This real world dataset \cite{madan2012sensing} is collected from a college dormitory building by web survey and contains the dynamic graphs $\mathbf{G}$, covariates $\mathbf{Z}$, and daily symptoms $\mathbf{Y}$, where $y_{n,t}$ is a 6 dimensional vector including sore throat and cough, runny nose, congestion and sneezing, fever, nausea, vomiting and diarrhea, sadness and depression, and stress. The proportion of missing values $\mathbf{Y}$ is about 0.6. The purpose is to infer latent variables $\mathbf{X}$ and infection parameters, and making tentative health suggestions to students. Even if we cannot evaluate the performance on the true $\mathbf{X}$s, the Google search of "flu" \cite{dong2012graph} implies a underlying correlation with this result. Since $\mathbf{Y}$ can be partially observed (no \texttt{NA}), one step ahead prediction on $\mathbf{Y}$ is possible, and obtains an accuracy at 92.09\% (threshold is also 0.5 for binary indicator). Results are shown in Figure \ref{fig:mitdata}.

\subsection{eX-Flu Dataset}
Evaluation on the public MIT dataset seems only partially useful, since true diagnoses are unavailable. We describe the design, study population characteristics, and social network structure of a chain referral sample of 590 students living in University of Michigan residence halls who were randomized to an intervention of isolation over a 10-week period during the 2013 influenza season. In our experiment, diagnoses are recorded immediately at flu onset.

\subsubsection{Design Description}

\begin{figure}[t]
\centering
\subfigure[]{
\includegraphics[width=75mm]{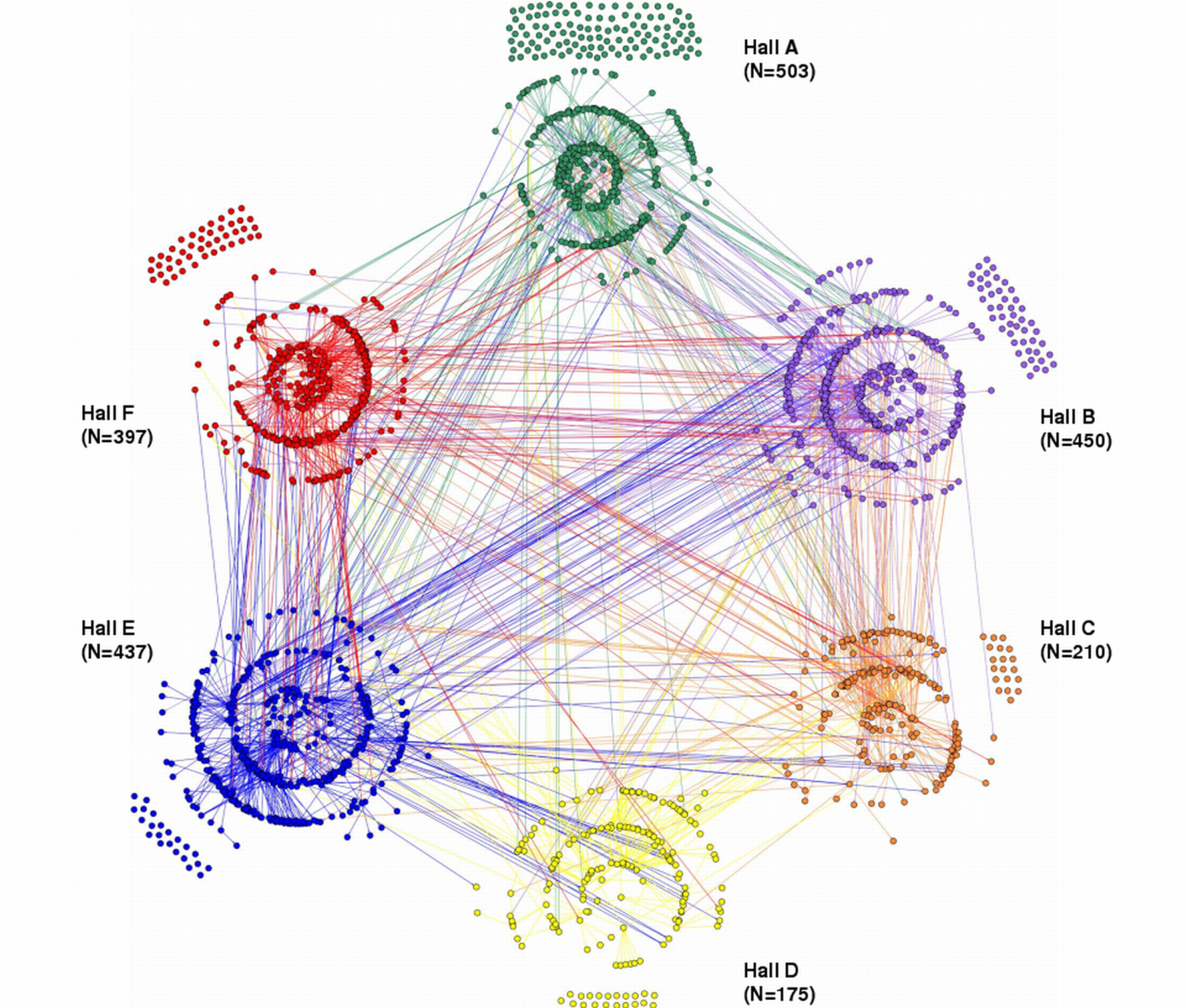}
}
\subfigure[]{
\includegraphics[width=70mm,clip=true,trim=70mm 110mm 70mm 110mm]{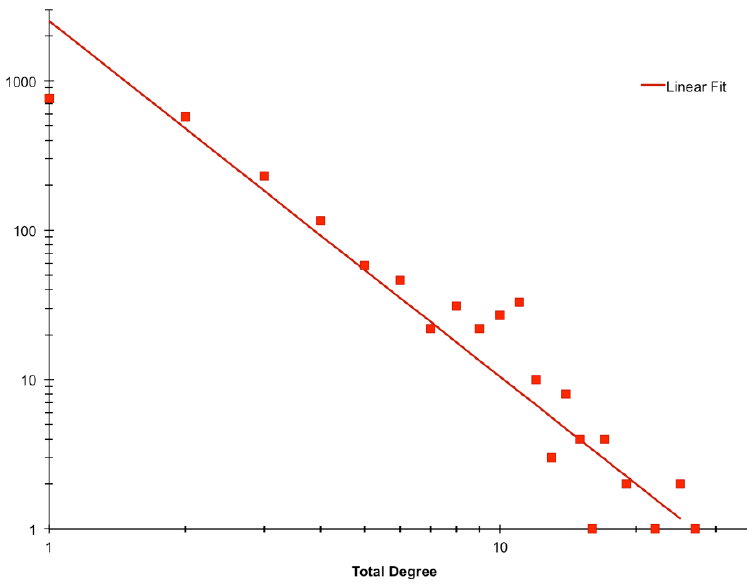}
}
\caption{Overall Social Network}
\label{fig:hall}
\end{figure}

590 students living in six eligible residence halls on the University of Michigan campus enrolled in the eX-FLU study during a chain referral recruitment process carried out from September 2012-January 2013. 262 of these, as "seed" participants, nominated their social relations to join the study as well. The rest, 328, were nominees that enrolled. Participants have to fill out weekly surveys on web apps about their health behaviors and social interactions with other participants, and a symptom indicator report of influenza-like illness (ILI). A subsample of 103 students were provided with smartphones with a mobile application, iEpi \cite{knowles2014field}, which is able to collect location sensor and contextually-dependent survey information, implying social contacts that are used in our computational model. This sub study experiment perfectly fits our proposed model, so the main evaluation will be performed on this sub dataset. Generally speaking, the underlying cumulative distribution of degrees for the overall social network on 590 students is shown in Figure \ref{fig:hall}(a). The distribution of three degree measurements (in, out, or total), were heavily right-skewed and over-dispersed. Consequently, the network appears scale-free, with a log-log plot shown in Figure \ref{fig:hall}(b) and a linear trend line ($R^2=0.91$) illustrating the approximately power-law distribution for total degree.

\subsubsection{iEpi Sub-Study and Networks Analysis}
103 (17.5\%) students of the 590 enrolled participants were equipped with provided smartphones and joined the iEpi sub-study. They were required to use their iEpi smartphone and could report their symptoms, meeting the study criteria for ILI. A total of 4843 contextually-based surveys were administered on all sub-study smartphones (mean 62.09/day), 1743 (36.0\%) of which were responded to by iEpi sub-study participants (mean 22.35/day). There were a total of 60131 Bluetooth contacts between smartphones within the iEpi sub-study, and 148,333 total Bluetooth contacts with other devices of any kind, averaging 7.48 contacts/phone/day and 20.95 contacts/person/day, respectively. 

\begin{figure}[t]
\subfigure[]{
\includegraphics[width=65mm,clip=true,trim=30mm 30mm 30mm 30mm]{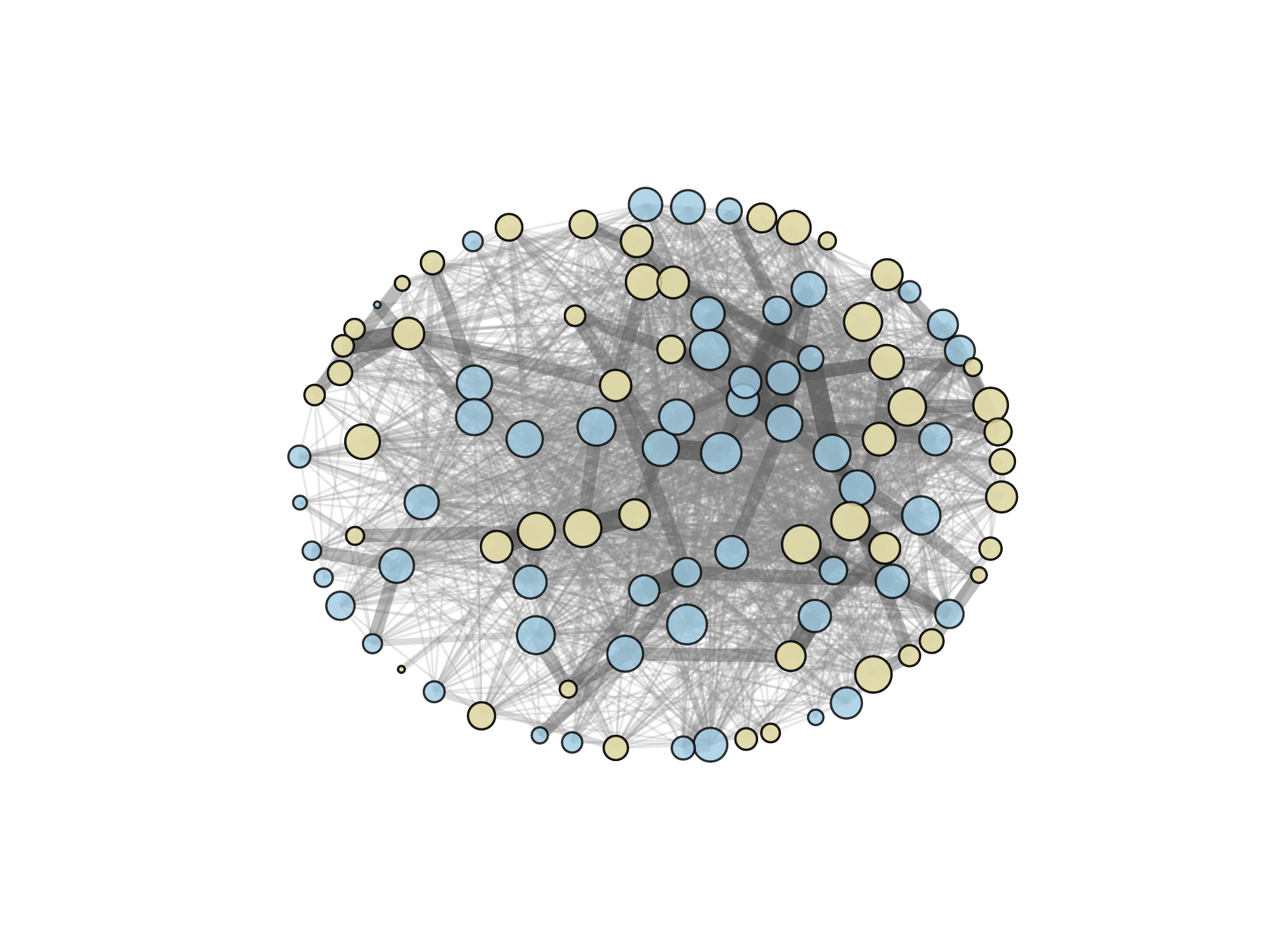}
}
\subfigure[]{
\includegraphics[width=65mm, clip=true, trim=15mm 15mm 15mm 5mm]{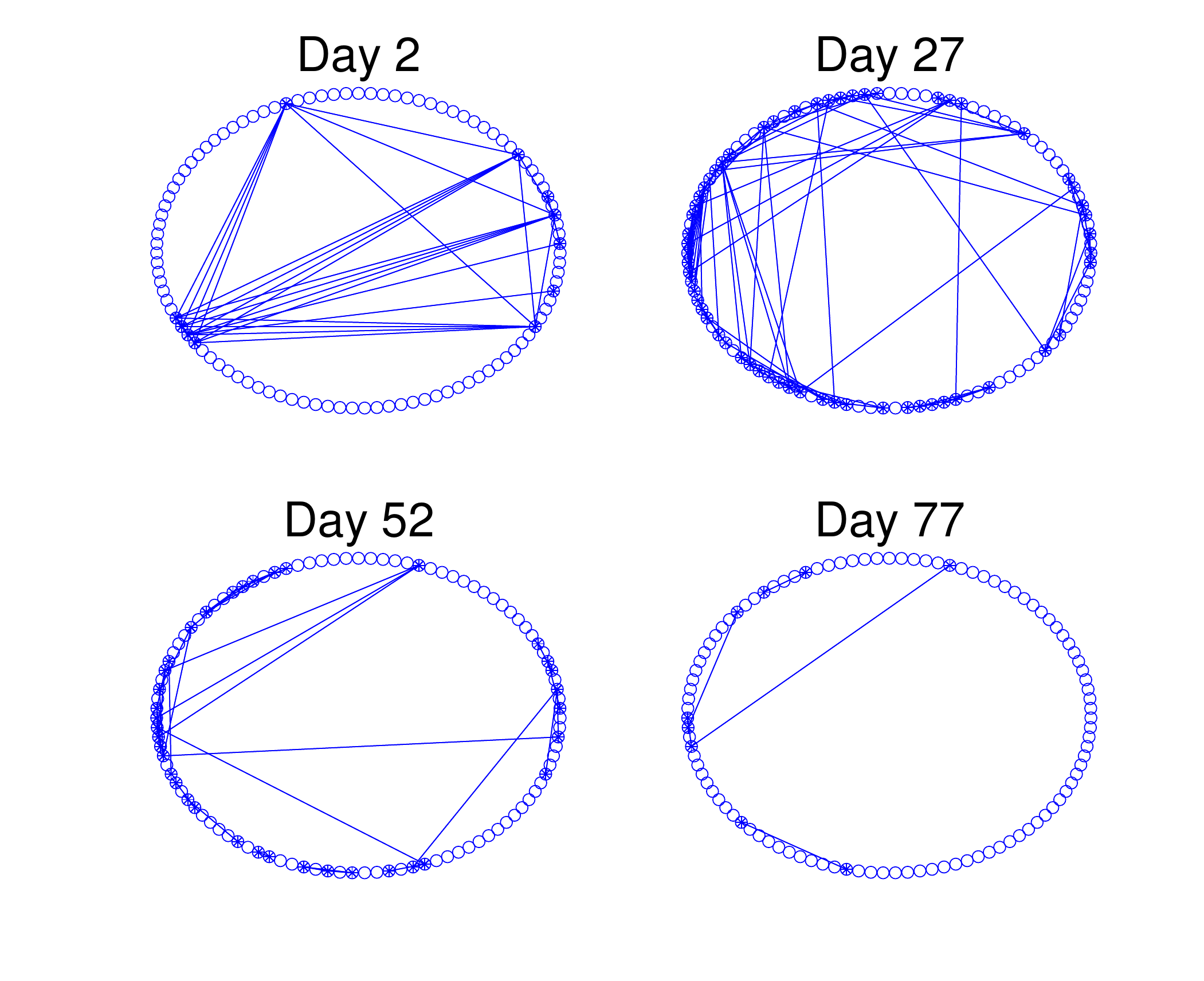}
}
\caption{(a) iEpi Bluetooth network (N=103): Network of Bluetooth contacts between smartphones in the iEpi sub-study; (b) Sampled dynamic social networks from (a): 103 dots uniformly distributed as a large circle. Contacts within the network account for edges between solid dots.}
\label{fig:network}
\end{figure}

The bluetooth detector can automatically collect contacts occurring between iEpi installed smartphones, or to other smart devices. Each node (circle) in Figure \ref{fig:network}(a) represents an individual in the sub-study, and the links (edges) between nodes represent bluetooth detections between smartphones of individuals within the sub-study networks. Node size is proportional to the total number of contacts detected in the bluetooth data (equivalent to degree), and the link thickness indicates the contact duration between the two nodes (equivalent to weight on edge). During the experiment period, we also conducted a comparison test. Some participants (yellow nodes in Figure \ref{fig:network}(a)) were isolated for three days at the onset of illness, which means no social contacts were made during this period. 

The next step is to extract daily social networks. We use the 77 days of the iEpi survey data which is relatively complete, and its corresponding bluetooth data to construct dynamic networks. Figure \ref{fig:network}(b) illustrates 4 independently sampled sub-networks, i.e. $G_t$, $t=2,27,52,77$. To make more sense of the edges, only the bluetooth data showing the total contact duration between two participants lasting more than 10 minutes will contribute to an edge on that day. The threshold of 10 minutes can be adjusted to make the graph denser or sparser, thus leading to a higher or lower computational cost.

\begin{figure}[t]
\centering
\subfigure{
\includegraphics[width=40mm,height=27mm,clip,trim=0 30 0 0mm]{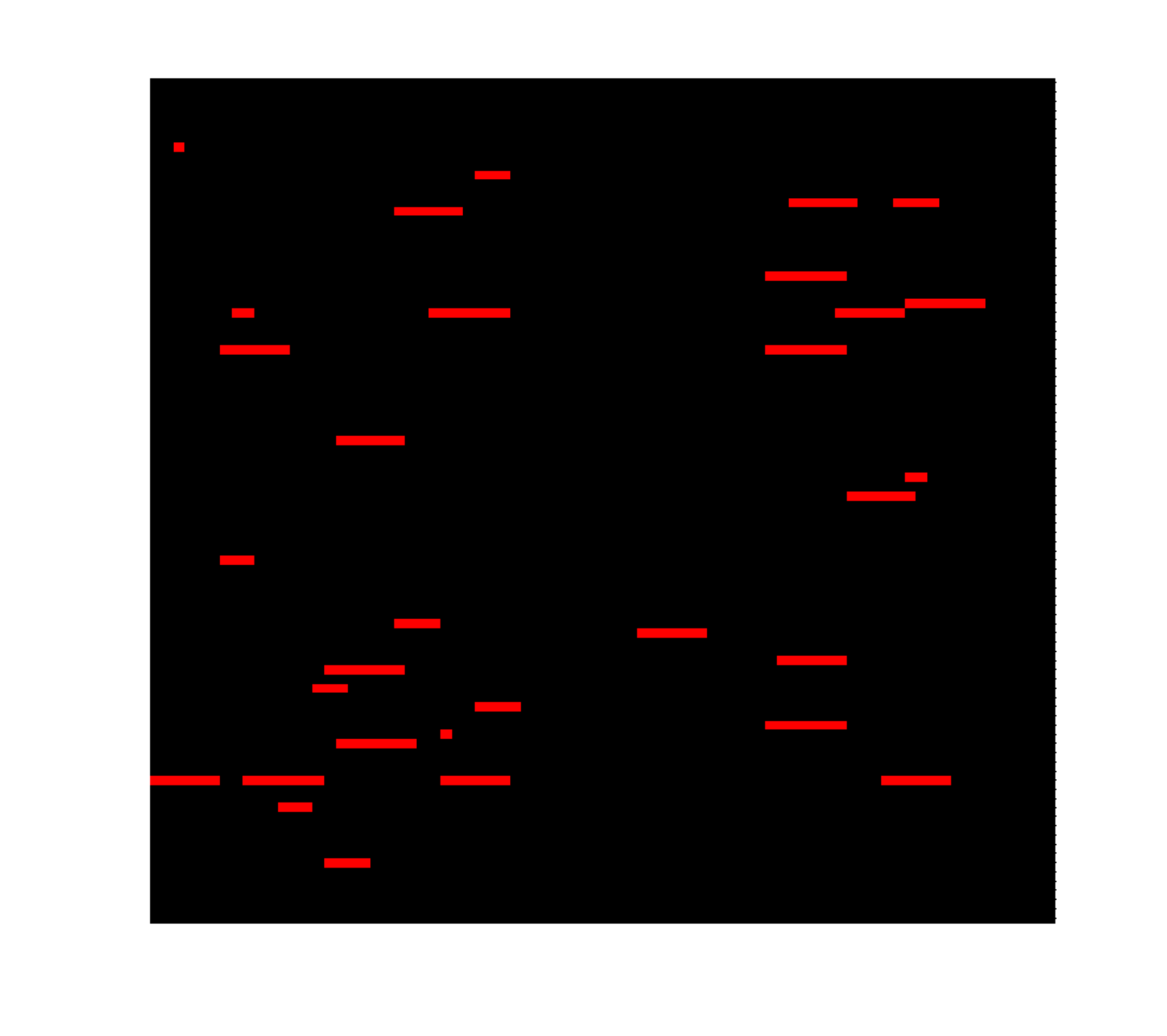}
}
\subfigure{
\includegraphics[width=40mm,height=27mm,clip,trim=0 30 0 0mm]{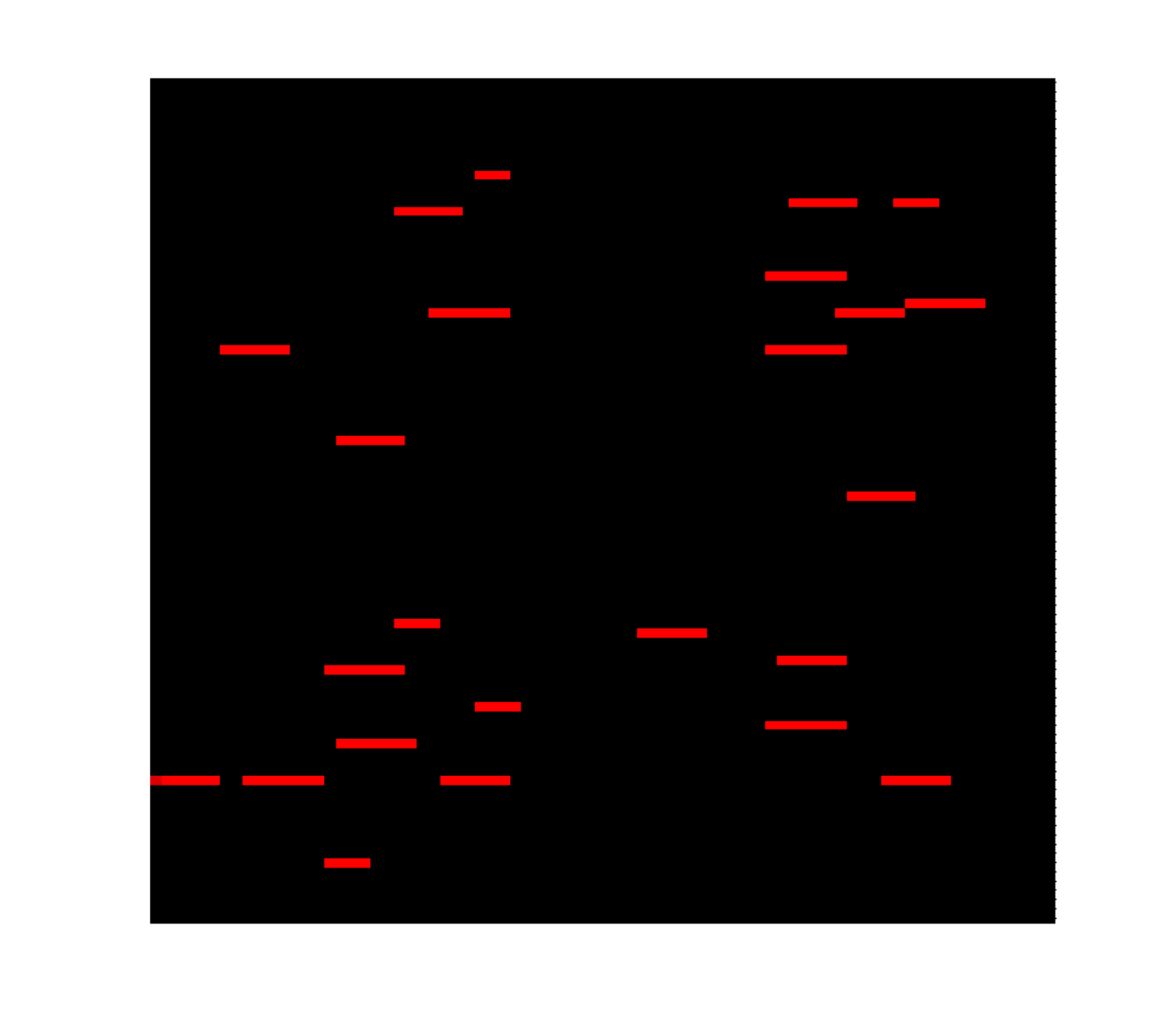}
}
\centering
\caption{Left is true Onset and its Duration. Right is predicted by Sigmoid Model.}
\label{fig:exX}
\end{figure}

\begin{table}[ht]
\centering
\caption{exFlu Epidemics State Inference Performance}
\begin{tabular}{|c|l|l|} \hline
\textbf{Model} & \textbf{Recall} & \textbf{Accuracy} \\ \hline
Sigmoid link & 0.8974 $\pm$ 0.00 & 0.9978 $\pm$ 0.00 \\
Beta-exp link & 0.7436 $\pm$ 0.00 &  0.9912 $\pm$ 0.00 \\
GCHMMs+LogReg & 0.7436  $\pm$ 0.00 & 0.9912 $\pm$ 0.00\\
\hline\end{tabular}
\label{tab:exFlu}
\end{table}

\begin{table}[t]
\centering
\caption{Coefficients Estimation on exFlu Dataset}
\begin{tabular}{l|r|r|r} \hline
\textbf{Feature\footnotemark} & \textbf{Recovery $\eta_r$} & \textbf{Outside Infect $\eta_a$} & \textbf{Inside Infect $\eta_b$} \\ \hline
Default$=$1 & -1.3022 $\pm$ 0.0146 &  -5.1517  $\pm$  0.0024 &  -4.1619  $\pm$  0.0281 \\
Gender &	-0.1575 $\pm$  0.0118 &  -0.2428  $\pm$   0.0074 &   -0.1457 $\pm$   0.0078\\
Age &	0.0074  $\pm$  0.0082 &  -0.2376  $\pm$  0.0051 &  -0.0181  $\pm$  0.0017\\
Alc\_Day&   0.1090   $\pm$ 0.0078 &  -0.1534  $\pm$  0.0003 &  -0.0410  $\pm$  0.0018\\
Vacc\_Ever&   -0.0698  $\pm$  0.0104 &   0.1092  $\pm$  0.0095 &   0.0382  $\pm$  0.0085\\
Flushot\_Yr&    0.0769 $\pm$   0.0092 &  -0.3209  $\pm$  0.0073 &   0.0837  $\pm$  0.0055\\
Smoker&   -0.1080 $\pm$   0.0029 &  -0.0536  $\pm$  0.0008 &   0.0773  $\pm$  0.0021\\
Drinker&   -0.1335 $\pm$   0.0092 &   0.0628  $\pm$ 0.0030  &  0.1408   $\pm$ 0.0029\\
Act\_Days&    0.0356 $\pm$   0.0099 &   0.0054  $\pm$  0.0063 &  -0.0622  $\pm$  0.0078\\
Sleep\_Qual&    0.0225  $\pm$  0.0069 &  -0.3686  $\pm$  0.0051 &  -0.0162  $\pm$  0.0077\\
Wash\_Opt&    0.0024 $\pm$   0.0103 &   0.0816  $\pm$  0.0132 &  -0.0714  $\pm$  0.0048\\
High\_Risk&   -0.1274 $\pm$   0.0116 &  -0.1252  $\pm$ 0.0058  & -0.0727   $\pm$ 0.0007\\
\hline
\end{tabular}
\label{tab:eta}
\end{table}
\footnotetext{Gender 1 means female; Alc\_Day: average number of times hand washing with sanitizer; Vacc\_Ever: previous vaccination; Flushot\_Yr: vaccination this year; Act\_Days: exercise in broad sense per day; Wash\_Opt: whether wash hands exceeding 20s; High\_Risk: contact with impaired immunity patient.}

\subsubsection{iEpi Flu Diffusion Analysis}

Available illness onset diagnoses in our experiment allows for the evaluation of inferred infection states. We tried all three models, Sigmoid link, Beta-exponential link and standard GCHMMs+LogReg on this dataset. Because of the specific quantized distribution of diagnosed flu onset (see red short pattern of left graph in Figure \ref{fig:exX})), the three methods perform stably, but give different results over 10 runs with no standard deviation. Though they all heavily rely on Gibbs sampling, the sigmoid link model can detect more short term patterns than the other two. Table \ref{tab:exFlu} gives both precision and recall for prediction, since the proportion of positive instances, unlike our simulation, is about one tenth. Even the sigmoid model missed some very short patterns. Two reasons may contribute to this phenomena; the first is that HMMs are a long distance dependent models; the second is that we find symptom reports for short period disease courses are always low severity.

In contrast to other models, and serving as the mainstay and novelty of this paper, we aimed to learn how personal features (first column in Table \ref{tab:eta}) were associated with individual flu vulnerability, i.e. coefficients $\eta$. A Sigmoid transform on $\eta$ will immediately give infection parameters. Larger $\gamma_n$ implies better resistance, while larger $\alpha_n,\beta_n$ indicates increased vulnerability. Because resistance or vulnerability is not an experimental quantity (difficult to measure in a real world dataset), we prefer to evaluate coefficients $\eta$ (Table \ref{tab:eta}) rather than actual infection parameters. The right three columns are the estimated $\eta$s associated with different biological meanings (indicated by their subscripts) in the Sigmoid model--possessing the best performance in both the simulation and real cases. Looking at the feature column, we can see that females seems suffer from a slower recovery but are not as likely to catch a cold. Another important factor is whether participants are addicted to alcohol. Drinkers significantly aggravate body immunity. However, whether or not one washes their hands for more than 20s, interestingly, seems not to be significant to the model, especially to the recovery rate. This may be blamed on an overly long washing duration--20s in the experimental design. Overall, the sign consistency with respect to $\eta$ makes sense, with the exception of a few relationships. For the sigmoid function, positive coefficients will enlarge infection parameters, and vice versa.

\section{Conclusion and Discussion}
\label{sec:conclusion}

We successfully extend the HMM based epidemic infection model to general and flexible hierarchical GCHMMs, enabling us to simultaneously predict individual infection and physical constitution by observing how flu spreads throughout dynamic social networks. In addition, this framework can reduce to previous models via simplifying the graphical model if less information is obtained, e.g. The model without a social network data will reduce to a standard HMM; sharing the same infection parameters will lead to a homogenous model. The hypothesis test on the necessity of social network has been researched by \cite{salathe2010high,dong2012graph}. The heterogeneity induced in our approach is validated on semi-synthetic data and epidemiological tracking data in college dormitories, based on prediction compared with previous work \cite{salathe2010high,dong2012graph}. On semi-simulation data, we evaluate our model on a number of metrics, including on the ability to correctly infer parameters. On the MIT social evolution data, we mainly focus on one step ahead prediction of the observed states (or symptoms). In our eX-Flu study, we successfully discovered the underlying social network pattern and personal feature relationships with respect to influenza vulnerability. In fact, the prediction on the infection state of the next day beyond the surveyed period is also available due to the property of Markov chain. From an application perspective, this is quite promising for health app development. 

Future research will explore learning a robust dynamic network generative model, since the bluetooth data can provide the contact duration which is a positive real value. Currently, the binary indicator determined by hard threshold is not fine-grained. One possibility is to model $\frac{N(N-1)}{2}$ stochastic processes for each edge, such as constructing a Hawkes process model with a latent space model, which requires combining the work of \cite{blundell2012modelling,palla2012infinite}. Alternatively, we may explore constructing features based on personal history (e.g., the contact frequency or duration between two nodes), and using this history feature as a predictor for the future Poisson rate \cite{gunawardana2011model,lianmultitask}. Another possible area of future research would be to implement remark (3) or investigate the identifiability of the auxiliary variable, and infection network learning by detecting the disease spread path. We can further relax the heterogeneity assumption to a cluster assumption, i.e. participants with similar health feature share the same parameters. Inspired by HDP-HMMs \cite{teh2006hierarchical}, this tradeoff can be realized by constructing a nonparametric version GCHMMs, enforcing similar HMMs to share the same parameters. 


\acks{We would like to thank Dylan Knowles, under the supervision of Dr. Nathaniel Osgood and Dr. Kevin Stanley, and the aid of other students and fellows developed the iEpi application and helped oversee the iEpi app smartphone upload, data collection procedures, and trouble shooting for the eXFLU study. This work was supported by Duke NSF grant \#3331830. Any opinions, findings and conclusions or recommendations expressed in this material are the authors' and do not necessarily reflect those of the sponsor.}


\newpage

\appendix
\section*{Appendix A. Update of Message Passing}
\label{app:mp}

\paragraph{Updating $\lambda$}
\begin{align*}
\lambda_{x_{n,t}}^{(i+1)}(x_{n,t-1})&\propto\sum_{x_{n,t}}\lambda^{(i)}(x_{n,t})\sum_{x_{n',t-1}:(n',n)\in E_{t-1}}\phi_{n,t-1,t}\prod_{n':(n',n)\in E_{t-1}}\pi_{x_{n,t}}^{(i)}(x_{\cdot,t-1})\\
\lambda_{x_{n,t}}(x_{n,t-1}=1)&\propto\sum_{x_{n,t}}\lambda(x_{n,t})\gamma^{\mathbb{I}_{x_{n,t}=0}}(1-\gamma)^{\mathbb{I}_{x_{n,t}=1}}\\
\lambda_{x_{n,t}}(x_{n,t-1}=0)&\propto\sum_{x_{n,t}}\lambda(x_{n,t})\sum_{x_{n',t-1}:(n',n)\in E_{t-1}}\left[1-(1-\alpha)(1-\beta)^{\sum_{n':(n',n)\in E_{t-1}}x_{n',t-1}}\right]^{\mathbb{I}_{x_{n,t}=1}}\\
&\left[(1-\alpha)(1-\beta)^{\sum_{n':(n',n)\in E_{t-1}}x_{n',t-1}}\right]^{\mathbb{I}_{x_{n,t}=0}}\prod_{n':(n',n)\in E_{t-1}}\pi_{x_{n,t}}(x_{\cdot,t-1})\\
\lambda_{x_{n,t}}^{(i+1)}(x_{n',t-1})&\propto\sum_{x_{n,t}}\lambda^{(i)}(x_{n,t})\sum_{x_{n,t-1},x_{n'',t-1}\neq x_{n',t-1}:(n'',n)\in E_{t-1}}\phi_{n,t-1,t}\prod_{n\cup\{n''\neq n':(n'',n)\in E_{t-1}\}}\pi_{x_{n,t}}^{(i)}(x_{\cdot,t-1})
\end{align*}

\paragraph{Updating $\pi$}
\begin{align*}
\pi_{x_{n,t+1}}^{(i+1)}(x_{n,t})&\propto \prod_{s=1}^S\lambda_{y_{n,t,s}}(x_{n,t})\prod_{n':(n,n')\in E_t}\lambda_{x_{\cdot,t+1}}^{(i)}(x_{n,t})\pi^{(i)}(x_{n,t})=\frac{BEL^{(i)}(x_{n,t})}{\lambda_{x_{n,t+1}}^{(i)}(x_{n,t})}\\
\pi_{x_{n',t+1}}^{(i+1)}(x_{n,t})&\propto \prod_{s=1}^S\lambda_{y_{n,t,s}}(x_{n,t})\prod_{n\cup\{n''\neq n':(n,n'')\in E_t\}}\lambda_{x_{\cdot,t+1}}^{(i)}(x_{n,t})\pi^{(i)}(x_{n,t})=\frac{BEL^{(i)}(x_{n,t})}{\lambda_{x_{n',t+1}}^{(i)}(x_{n,t})}\\
\pi_{y_{n,t,s}}^{(i+1)}(x_{n,t})&\propto \prod_{s'\neq s}\lambda_{y_{n,t,s'}}(x_{n,t})\prod_{n\cup\{n':(n,n')\in E_t\}}\lambda_{x_{\cdot,t+1}}^{(i)}(x_{n,t})\pi^{(i)}(x_{n,t})=\frac{BEL^{(i)}(x_{n,t})}{\lambda_{y_{n,t,s}}^{(i)}(x_{n,t})}
\end{align*}

\paragraph{Boundary Conditions}
\begin{itemize}
\item Root nodes, i.e. $x_{n,0}$. The prior distribution is $p(x_{n,0})=\pi^{x_{n,0}}(1-\pi)^{1-x_{n,0}}$. 
\item Evidence nodes, i.e. $y_{n,t,s}$
\begin{align*}
\lambda(y_{n,t,s})&=(\mathbb{I}_{y_{n,t,s}=0},\mathbb{I}_{y_{n,t,s}=1}) \\
\pi(y_{n,t,s})&=\sum_{x_{n,t}}\phi_{n,t,y|x,s}(y_{n,t,s}|x_{n,t})\pi_{y_{n,t,s}}(x_{n,t})\\
BEL(y_{n,t,s})&\propto \lambda(y_{n,t,s})\pi(y_{n,t,s})\\
\lambda_{y_{n,t,s}}(x_{n,t})&\propto \sum_{y_{n,t,s}}\lambda(y_{n,t,s})\phi_{n,t,y|x,s}(y_{n,t,s}|x_{n,t})=\mathbb{I}_{y_{n,t,s}=0}(1-\theta_{x_{n,t},s})+\mathbb{I}_{y_{n,t,s}=1}\theta_{x_{n,t},s}
\end{align*}
\end{itemize}

\section*{Appendix B. On Likelihood computation for 2nd Interpretation on $\beta_n$}
\label{sec:beta}
Notice when  $\beta_n$ to obtain (\ref{eq:PR2}), we use the following identity.
\begin{align*}
&\prod_{n=1}^N\prod_{t=1}^{T-1}\prod_{n'\in S_{n,t}}(1-\beta_{n'})^{\mathbb{I}(R_{n,t}=0)+\mathbb{I}(X_{n,t}=0,X_{n,t+1}=0)}\beta_{n'}^{\mathbb{I}(R_{n,t}=n')}\\
=&\prod_{n=1}^N\prod_{t=1}^{T-1}\prod_{n'=1}^N \left((1-\beta_{n'})^{\mathbb{I}(R_{n,t}=0)+\mathbb{I}(X_{n,t}=0,X_{n,t+1}=0)}\beta_{n'}^{\mathbb{I}(R_{n,t}=n')}\right)^{\mathbb{I}(n'\in S_{n,t})}\\
=&\prod_{n'=1}^N\prod_{t=1}^{T-1}\prod_{n=1}^N \left((1-\beta_{n'})^{\mathbb{I}(R_{n,t}=0)+\mathbb{I}(X_{n,t}=0,X_{n,t+1}=0)}\beta_{n'}^{\mathbb{I}(R_{n,t}=n')}\right)^{\mathbb{I}(n'\in S_{n,t})}\\
=&\prod_{n=1}^N\prod_{t=1}^{T-1}\prod_{n'=1}^N \left((1-\beta_n)^{\mathbb{I}(R_{n',t}=0)+\mathbb{I}(X_{n',t}=0,X_{n',t+1}=0)}\beta_n^{\mathbb{I}(R_{n',t}=n)}\right)^{\mathbb{I}(n\in S_{n',t})}
\end{align*}



%
%
%

\vskip 0.2in
\bibliography{sample}

\end{document}